\documentclass[12pt]{iopart}

\usepackage{epsfig}
\def\ss{SUSY }
\def\gs{SUSY GUTs }
\def\f{\frac}
\def\beq{\begin{equation}}
\def\eeq{\end{equation}}
\def\bea{\begin{eqnarray}}
\def\eea{\end{eqnarray}}
\def\nn{\nonumber}
\def\abs#1{\left|{#1}\right|}
\def\dltmh{$\Delta m_H^2\;$}
\def\dltmb{$\Delta m_b \;$}
\def\ch2{$\chi^2$}
\def\L{\lambda}
\newcommand{\GeV}{\,{\rm GeV}}
\begin{document}

\title{Desperately Seeking Supersymmetry [SUSY] }

\author{Stuart Raby\dag\
\footnote[3]{To whom correspondence should be addressed (raby@pacific.mps.ohio-state.edu)} }

\address{\dag\ Department of Physics, The Ohio State University, 174 W. 18th Ave., Columbus, OH 43210, USA}

\begin{abstract}
The discovery of X-rays and radioactivity in the waning years of the 19th century lead to one of the most awe
inspiring scientific eras in human history.   The 20th century witnessed a level of scientific discovery never
before seen or imagined. At the dawn of the 20th century only two forces of Nature were known -- gravity and
electromagnetism.  The atom was believed by chemists to be the elemental, indestructible unit of matter, coming
in many unexplainably different forms.  Yet J.J. Thomson, soon after the discovery of X-rays, had measured the
charge to mass ratio of the electron, demonstrating that this carrier of electric current was ubiquitous and
fundamental. All electrons could be identified by their unique charge to mass ratio.

In the 20th century the mystery of the atom was unravelled, the atomic nucleus was smashed,  and two new forces
of Nature were revealed -- the weak force [responsible for radioactive $\beta$ decay and the nuclear fusion
reaction powering the stars] and the nuclear force binding the nucleus.   Quantum mechanics enabled the
understanding of the inner structure of the atom, its nucleus and further inward to quarks and gluons [the
building blocks of the nucleus] and thence outward to an understanding of large biological molecules and the
unity of chemistry and microbiology.

Finally the myriad of new fundamental particles, including electrons, quarks, photons, neutrinos, etc. and the
three fundamental forces -- electromagnetism, the weak and the strong nuclear force -- found a unity of
description in terms of relativistic quantum field theory.   These three forces of Nature can be shown to be a
consequence of symmetry rotations in internal spaces and the particular interactions of each particle are solely
determined by their symmetry charge.  This unifying structure, describing all the present experimental
observations, is known as the standard model.   Moreover, Einstein's theory of gravity can be shown to be a
consequence of the symmetry of local translations and Lorentz transformations.

As early as the 1970s, it became apparent that two new symmetries, a grand unified theory of the strong, weak and
electromagnetic interactions in conjunction with supersymmetry,  might unify all the known forces and particles
into one unique structure. Now 30 years later, at the dawn of a new century, experiments are on the verge of
discovering (or ruling out) these possible new symmetries of Nature.   In this article we try to clarify why
supersymmetry [\ss] and supersymmetric grand unified theories [\gs] are the new standard model of particle
physics, i.e. the standard by which all other theories and experiments are measured.

\end{abstract}



\maketitle

\section{Introduction}

Supersymmetry is a space-time symmetry; an extension of the group of transformations known as the Poincar\'{e}
group including space-time translations, spatial rotations and pure Lorentz transformations.  The Poincar\'{e}
transformations act on the three space $\vec{x}$ and one time coordinate $t$.  The supersymmetric extension adds
two anti-commuting complex coordinates $\theta_\alpha,  \ \alpha = 1,2$ satisfying $\theta_\alpha \theta_\beta +
\theta_\beta \theta_\alpha = 0$. Together they make superspace ${\bf z} = \{ t, \vec{x}, \theta_\alpha \} $ and
supersymmtry transformations describe translations/rotations in superspace.  Local supersymmetry implies a
supersymmetrized version of Einstein's gravity known as supergravity.

In the standard model, ordinary matter is made of quarks and electrons.  All of these particles are Fermions with
spin $s = \frac{1}{2} \hbar$, satisfying the Pauli exclusion principle.  Hence no two identical matter particles
can occupy the same space at the same time.  In field theory, they are represented by anti-commuting space-time
fields, we generically denote by, $\psi_\alpha(\vec{x},t)$.  On the other hand, all the force particles, such as
photons, gluons, $W^\pm, \ Z^0$, are so-called gauge Bosons with spin $s = 1 \hbar$, satisfying Bose-Einstein
statistics and represented by commuting fields $\phi(\vec{x},t)$. As a result Bosons prefer to sit, one on top of
the other; thus enabling them to form macroscopic classical fields.  A Boson - Fermion pair form a supermultiplet
which can be represented by a superfield $\Phi(z) = \phi(\vec{x},t) + \theta \ \psi(\vec{x},t)$. Hence a rotation
in superspace, rotates Bosons (force particles) into Fermions (matter particles) and vice versa.

This simple extension of ordinary space into two infinitesimal directions has almost miraculous consequences
making it one of the most studied possible extensions of the standard model [SM] of particle physics.  It
provides a ``technical" solution to the so-called gauge hierarchy problem, i.e. why is $M_Z/M_{pl} \ll 1$.  In
the SM,  all matter derives its mass from the vacuum expectation value [VEV] $v$ of the Higgs field.  The $W^\pm$
and $Z^0$ mass are of order $g \ v$ where $g$ is the coupling constant of the weak force. While quarks and
leptons (the collective name for electrons, electron neutrinos and similar particles having no strong
interactions) obtain mass of order $\lambda \ v$ where $\lambda$ is called a Yukawa coupling; a measure of the
strength of the interaction between the Fermion and Higgs fields. The Higgs vacuum expectation value is fixed by
the Higgs potential and in particular by its mass $m_H$ with $v \sim m_H$. The problem is that in quantum field
theory, the Lagrangian (or bare) mass of a particle is subject to quantum corrections. Moreover for Bosons, these
corrections are typically large.  This was already pointed out in the formative years of quantum field theory by
[Weisskopf (1939)]. In particular for the Higgs we have $m_H^2 = m_0^2 + \alpha \ \Lambda^2$ where $m_0$ is the
bare mass of the Higgs, $\alpha$ represents some small coupling constant and $\Lambda$ is typically the largest
mass in the theory. In electrodynamics $\alpha$ is the fine-structure constant and $\Lambda$ is the physical
cutoff scale, i.e. the mass scale where new particles and their new interactions become relevant.  For example,
it is known that gravitational interactions become strong at the Planck scale $M_{pl} \sim 10^{19}$ GeV; hence we
take $\Lambda \sim M_{pl}$.   In order to have $M_Z \ll M_{pl}$ the bare mass must be fine-tuned to one part in
$10^{17}$, order by order in perturbation theory, against the radiative corrections in order to preserve this
hierarchy. This appears to be a particularly ``unnatural" accident or, as most theorists believe, an indication
that the SM is incomplete.   Note that neither Fermions nor gauge Bosons have this problem.  This is because
their mass corrections are controlled by symmetries.   For Fermions these chiral symmetries become exact only
when the Fermion mass vanishes.    Moreover with an exact chiral symmetry the radiative corrections to the
Fermion's mass vanish to all orders in perturbation theory.   As a consequence when chiral symmetry is broken the
Fermion mass corrections are necessarily proportional to the bare mass. Hence $m_F = m_0  + \alpha \ m_0 \
\log(\Lambda/m_0)$ and a light Fermion mass does not require any ``unnatural" fine-tuning.  Similarly for gauge
Bosons, the local gauge symmetry prevents any non-zero corrections to the gauge boson mass.   As a consequence,
massless gauge bosons remain massless to all orders in perturbation theory.  What can we expect in a
supersymmetric theory?   Since supersymmetry unifies Bosons and Fermions,  the radiative mass corrections of the
Bosons are controlled by the chiral symmetries of their Fermionic superpartners.  Moreover for every known
Fermion with spin $\frac{1}{2} \hbar$ we necessarily have a spin 0 Boson (or Lorentz scalar) and for every spin
$1 \hbar$ gauge Boson, we have a spin $\frac{1}{2} \hbar$ gauge Fermion (or gaugino).    Exact supersymmetry then
requires Boson-Fermion superpartners to have identical mass.  Thus in \ss an electron necessarily has a spin 0
superpartner, a scalar electron, with the same mass.   Is this a problem? The answer is yes, since the
interaction of the scalar electron with all SM particles is determined by \ss.   In fact, the scalar electron
necessarily has the same charge as the electron under all SM local gauge symmetries. Thus it has the same
electric charge and it would have been observed long ago.   We thus realize that \ss can only be an approximate
symmetry of Nature.   Moreover it must be broken in such a way to raise the mass of the scalar partners of all SM
Fermions and the gaugino partners of all the gauge Bosons.  This may seem like a tall order.   But what would we
expect to occur once \ss is softly broken at a scale $\Lambda_{\ss}$?   Then scalars are no longer protected by
the chiral symmetries of their Fermionic partners.  As a consequence they receive radiative corrections to their
mass of order $\delta m^2 \propto \alpha \ \Lambda_{\ss}^2 \ \log(\Lambda/m_0)$. As long as $\Lambda_{\ss} \leq
100$ TeV, the Higgs Boson can remain naturally light.  In addition, the gauge Boson masses are still protected by
gauge symmetries.   The gauginos are special, however, since even if \ss is broken, gaugino masses may still be
protected by a chiral symmetry known as R symmetry [Farrar and Fayet (1979)]. Thus gaugino masses are controlled
by both the \ss and R symmetry breaking scales.

Before we discuss SUSY theories further, let us first review the standard model [SM]in some more detail. The
standard model of particle physics is defined almost completely in terms of its symmetry and the charges (or
transformation properties) of the particles under this symmetry.  In particular the symmetry of the standard
model is $SU(3) \times SU(2) \times U(1)_Y$.   It is a local, internal symmetry, by which we mean it acts on
internal properties of states as a rotation by an amount which depends on the particular space-time point. Local
symmetries demand the existence of gauge Bosons (or spin 1 force particles) such as the gluons of the strong
$SU(3)$ interactions or the $W^\pm, \ Z^0$ or photon ($\gamma$) of the electroweak interactions $SU(2) \times
U(1)_Y$. The strength of the interactions are determined by parameters called coupling constants.  The values of
these coupling constants however are not determined by the theory, but must be fixed by experiment.

There are three families of matter particles, spin 1/2 quarks and leptons; each family carrying identical SM
symmetry charges. The first and lightest family contains the up (u) and down (d) quarks, the electron (e) and the
electron neutrino ($\nu_e$) (the latter two are leptons).   Two up quarks and one down quark bind via gluon
exchange forces to make a proton, while one up and two down quarks make a neutron. Together different numbers of
protons, neutrons bind via residual gluon and quark exchange forces to make nuclei and finally nuclei and
electrons bind via electromagnetic forces (photon exchanges) to make atoms, molecules and us. The strong forces
are responsible for nuclear interactions.  The weak forces on the other hand are responsible for nuclear beta
decay. In this process typically a neutron decays thereby changing into a proton, electron and electron neutrino.
This is so-called $\beta^-$ decay since the electron (or $beta$ particle) has negative charge, $-e$.   $\beta^+$
decays also occur where a proton (bound in the nucleus of an atom) decays into a neutron, anti-electron and
electron neutrino. The anti-electron (or positron) has positive charge, $+e$ but identical mass to the electron.
If the particle and anti-particle meet they annihilate, or disappear completely, converting their mass into pure
energy in the form of two photons.  The energy of the two photons is equal to the energy of the particle -
anti-particle pair, which includes the rest mass of both. Nuclear fusion reactions where two protons combine to
form deuterium (a p-n bound state), $e^+ + \nu_e$ is the energy source for stars like our sun and the energy
source of the future on earth. The weak forces occur very rarely because they require the exchange of the $W^\pm,
\ Z^0$ which are one hundred times more massive than the proton or neutron.

The members of the third family \{$t, \ b, \ \tau, \ \nu_\tau$\} are heavier than the second family   \{$c, \ s,
\ \mu, \ \nu_\mu$\} which are heavier than the first family members \{$u, \ d, \ e, \ \nu_e$\}.   Why there are
three copies of families and why they have the apparent hierarchy of masses is a mystery of the SM.   In addition
why each family has the following observed charges is also a mystery.   A brief word about the notation.  Quarks
and leptons have four degrees of freedom each (except for the neutrinos which in principle may only have two
degrees of freedom) corresponding to a left or right-handed particle or anti-particle.  The field labelled $e$
contains a left-handed electron and a right-handed anti-electron, while $\bar e$ contains a left-handed
anti-electron and a right-handed electron. Thus all four degrees of freedom are naturally (this is in accord with
Lorentz invariance) contained in two independent fields $e, \ \bar e$.   This distinction is a property of
Nature, since the charges of the SM particles depend on their handedness.   In fact in each family we have five
different charge multiplets given by
\begin{equation}  Q  = \left(\begin{array}{c} u \\ d \end{array}\right) \;\; \bar u \;\;\;\; \bar d \;\;\;\;\;\;
L = \left(\begin{array}{c} \nu_e \\ e \end{array}\right)  \;\; \bar e
\end{equation} where $Q$ is a triplet under color $SU(3)$, a doublet under weak $SU(2)$ and carries
$U(1)_Y$ weak hypercharge $Y = 1/3$.   The color anti-triplets $( \bar u, \ \bar d)$ are singlets under $SU(2)$
with $Y = (-\frac{4}{3}, \ \frac{2}{3})$ and finally the leptons appear as a electroweak doublet ($ L $) and
singlet ($ \bar e$) with  $Y = -1, +2 $ respectively.   Note, by definition, leptons are color singlets and thus
do not feel the strong forces.   The electric charge for all the quarks and leptons is given by the relation
$Q_{EM} = T_3 + \frac{Y}{2}$ where the (upper, lower) component of a weak doublet has $T_3 = (+1/2, \ -1/2)$.
Finally the Higgs boson multiplet,
\begin{equation} H = \left(\begin{array}{c} h^+ \\ h^0 \end{array}\right)
\end{equation} with $Y = + 1$ is necessary to give mass to the $W^\pm, Z^0$ and to all quarks and leptons.  In the
SM vacuum, the field $h^0$ obtains a non-zero vacuum expectation value $\langle h^0 \rangle = v/\sqrt{2}$.
Particle masses are then determined by the strength of the coupling to the Higgs.  The peculiar values of the
quark, lepton and Higgs charges is one of the central unsolved puzzles of the SM.  The significance of this
problem only becomes clear when one realizes that the interactions of all the particles (quarks, leptons, and
Higgs bosons), via the strong and electroweak forces, are completely fixed by these charges.

Let us now summarize the list of fundamental parameters needed to define the SM. If we do not include gravity or
neutrino masses, then the SM has 19 fundamental parameters.  These include the $Z^0$ and Higgs masses ($M_Z, \;
m_h$) setting the scale for electroweak physics.  The three gauge couplings $\alpha_i(M_Z), \ i= 1,2,3$, the 9
charged fermion masses and 4 quark mixing angles.   Lastly, there is the QCD theta parameter which violates CP
and thus is experimentally known to be less than $\approx 10^{-10}$. Gravity adds one additional parameter,
Newton's constant $G_N = 1/M_{Pl}^2$ or equivalently the Planck scale. Finally neutrino masses and mixing angles
have been definitively observed in many recent experiments measuring solar and atmospheric neutrino oscillations,
and by carefully measuring reactor or accelerator neutrino fluxes.  The evidence for neutrino masses and flavor
violation in the neutrino sector has little controversy.   It is the first strong evidence for new physics beyond
the SM. We shall return to these developments later.   Neutrino masses and mixing angles are described by 9 new
fundamental parameters -- 3 masses, 3 real mixing angles and 3 CP violating phases.

Let us now consider the minimal supersymmetric standard model [MSSM].  It is defined by the following two
properties, (i) the particle spectrum , and (ii) their interactions.  \begin{enumerate}  \item  Every matter
fermion of the SM has a bosonic superpartner. In addition, every gauge boson has a fermionic superpartner.
Finally, while the SM
has one Higgs doublet, the MSSM has two Higgs doublets.  \begin{equation}   H_u = \left(\begin{array}{c} h^+ \\
h^0
\end{array}\right), \;\; H_d = \left(\begin{array}{c} \bar h^0 \\ \bar h^- \end{array}\right) \end{equation} with
$Y = + 1, \;\; - 1$. The two Higgs doublets are necessary to give mass to up quarks, and to down quarks and
charged leptons, respectively. The vacuum expectation values are now given by $\langle h^0 \rangle = v
\sin\beta/\sqrt{2}$, $\langle \bar h^0 \rangle = v \cos\beta/\sqrt{2}$ where $\tan\beta$ is a new free parameter
of the MSSM.   \item The MSSM has the discrete symmetry called R parity.\footnote{One may give up R parity at the
expense of introducing many new interactions with many new arbitrary couplings into the MSSM.  These interactions
violate either baryon or lepton number. Without R parity the LSP is no longer stable. There are many papers which
give limits on these new couplings. The strongest constraint is on the product of couplings for the dimension
four baryon and lepton number violating operators which contributes to proton decay.  We do not discuss R parity
violation further in this review.} All SM particles are R parity even, while all superpartners are R odd. This
has two important consequences. \subitem $\bullet$ The lightest superpartner [LSP] is absolutely stable, since
the lightest state with odd R parity cannot decay into only even R parity states. Assuming that the LSP is
electrically neutral, it is a weakly interacting massive particle.  Hence it is a very good candidate for the
dark matter of the universe. \subitem $\bullet$ Perhaps more importantly, the interactions of all superpartners
with SM particles is completely determined by supersymmetry and the observed interactions of the SM.  Hence,
though we cannot predict the masses of the superpartners,  we know exactly how they interact with SM particles.
\end{enumerate}

The MSSM has some very nice properties. It is perturbative and easily consistent with all precision electroweak
data.  In fact global fits of the SM and the MSSM provide equally good fits to the data [de Boer and Sander
(2003)]. Moreover as the \ss particle masses increase, they decouple from low energy physics. On the other hand
their masses cannot increase indefinitely since one soon runs into problems of ``naturalness."    In the SM the
Higgs boson has a potential with a negative mass squared, of order the $Z^0$ mass, and an arbitrary quartic
coupling. The quartic coupling stabilizes the vacuum value of the Higgs.  In the MSSM the quartic coupling is
fixed by supersymmetry in terms of the electroweak gauge couplings.  As a result of this strong constraint, at
tree level the light Higgs boson mass is constrained to be lighter than $M_Z$.    One loop corrections to the
Higgs mass are significant. Nevertheless the Higgs mass is bounded to be lighter than about 135 GeV [Okada \etal
(1991), Ellis \etal (1991), Casas \etal (1995), Carena \etal (1995,1996), Haber \etal (1997), Zhang (1999),
Espinosa and Zhang (2000a,b), Degrassi \etal (2003)]. The upper bound is obtained in the limit of large
$\tan\beta$.

It was shown early on that, even if the tree level Higgs mass squared was positive, radiative corrections due to
a large top quark Yukawa coupling are sufficient to drive the Higgs mass squared negative [Iba\~{n}ez and Ross
(1982), Alvarez-Gaume \etal (1983), Iba\~{n}ez and Ross (1992)]. Thus radiative corrections naturally lead to
electroweak symmetry breaking at a scale determined by squark and slepton \ss breaking masses. Note, a large top
quark Yukawa coupling implies a heavy top quark. Early predictions for a top quark with mass above 50 GeV
[Iba\~{n}ez and Lopez (1983)] were soon challenged by the announcement of the discovery of the top quark by UA1
with a mass of 40 GeV. Of course, this false discovery was much later followed by the discovery of the top quark
at Fermilab with a mass of order 175 GeV.

If the only virtue of \ss is to explain why the weak scale ($M_Z, \; m_h$) is so much less than the Planck scale,
one might ponder whether the benefits outweigh the burden of doubling the SM particle spectrum.  Moreover there
are many other ideas addressing the hierarchy problem, such as Technicolor theories with new strong interactions
at a TeV scale.  One particularly intriguing possibility is that the universe has more than 3 spatial dimensions.
In these theories the fundamental Planck scale $M_*$ is near a TeV, so there is no apparent hierarchy.   I say
apparent since in order to have the observed Newton's constant $1/M_{Pl}^2$ much smaller than $1/M_*^2$ one needs
a large extra dimension such that the gravitational lines of force can probe the extra dimension.  If we live on
a 3 dimensional brane in this higher dimensional space then at large distances compared to the size of the $d$
extra dimensions we will observe an effective Newton's constant given by  $G_N = 1/M_{Pl}^2 = 1/(R^d M_*^{d +
2})$ [Arkani-Hamed \etal (1998)]. For example with $d = 2$ and $M_* = 1$ TeV we need the radius of the extra
dimension $R \approx 1$ mm. If any of these new scenarios with new strong interactions at a TeV
scale\footnote{Field theories in extra dimensions are divergent and require new non-perturbative physics, perhaps
string theory, at the TeV scale.} are true then we should expect a plethora of new phenomena occurring at the
next generation of high energy accelerators, i.e. the Large Hadron Collider [LHC] at CERN.   It is thus important
to realize that \ss does much more.  It provides a framework for understanding the 16 parameters of the SM
associated with gauge and Yukawa interactions and also the 9 parameters in the neutrino sector.  This will be
discussed in the context of supersymmetric grand unified theories [\gs] and family symmetries.   As we will see
these theories are very predictive and will soon be tested at high energy accelerators or underground detectors.
We will elaborate further on this below. Finally it is also naturally incorporated into string theory which
provides a quantum mechanical description of gravity. Unfortunately this last virtue is apparently true for all
the new ideas proposed to solve the gauge hierarchy problem.

A possible subtitle for this article could be ``A Tale of Two Symmetries: \gs."   Whereas \ss by itself provides
a framework for solving the gauge hierarchy problem, i.e. why $M_{Z} \ll M_{GUT}$ ,  \gs (with the emphasis on
GUTs) adds the framework for understanding the relative strengths of the three gauge couplings and for
understanding the puzzle of charge and mass.  It also provides a theoretical lever arm for uncovering the physics
at the Planck scale with experiments at the weak scale.  Without any exaggeration it is safe to say that \gs also
address the following problems.
\begin{itemize}
\item   They explain charge quantization since weak hypercharge ($Y$) is imbedded in a non-abelian symmetry
group.

\item   They explain the family structure and in particular the peculiar color and electroweak charges of
fermions in one family of quarks and leptons.

\item     They predict gauge coupling unification.   Thus given the experimentally determined values of two gauge
couplings at the weak scale,  one predicts the value of the third.    The experimental test of this prediction is
the one major success of \ss theories.   It relies on the assumption of \ss particles with mass in the 100 GeV to
1 TeV range.   Hence it predicts the discovery of \ss particles at the LHC.

\item  They predict Yukawa coupling unification for the third family.  In SU(5) we obtain $b - \tau$ unification,
while in SO(10) we have $t - b - \tau$ unification.   We shall argue that the latter prediction is eminently
testable at the Tevatron, the LHC or a possible Next Linear Collider.

\item   With the addition of family symmetry they provide a predictive framework for understanding the hierarchy
of fermion masses.

\item   It provides a framework for describing the recent observations of neutrino masses and mixing.  At zeroth
order the See - Saw scale for generating light neutrino masses probes physics at the GUT scale.

\item  The LSP is one of the best motivated candidates for dark matter.   Moreover back of the envelope
calculations of LSPs, with mass of order 100 GeV and annihilation cross-sections of order $1/TeV^2$, give the
right order of magnitude of their cosmological abundance for LSPs to be dark matter.    More detailed
calculations agree. Underground dark matter detectors will soon probe the mass/cross-section region in the LSP
parameter space.

\item  Finally the cosmological asymmetry of baryons vs. anti-baryons can be explained via the process known as
leptogenesis [Fukugita and Yanagida (1986)].  In this scenario an initial lepton number asymmetry, generated by
the out of equilibrium decays of heavy Majorana neutrinos,  leads to a net baryon number asymmetry today.
\end{itemize}

Grand unified theories are the natural extension of the standard model.  Ever since it became clear that quarks
are the fundamental building blocks of all strongly interacting particles, protons, neutrons, pions, kaons, etc.
and that they appear to be just as elementary as leptons, it was proposed [Pati and Salam (1973a,b, 1974)] that
the strong SU(3) color group should be extended to SU(4) color with lepton number as the fourth color.

${\bf G_{\rm Pati-Salam} \equiv \;\;\; SU_4(color) \times SU_2(L) \times SU_2(R)}$
\begin{center} \beq   \left( \begin{array}{c} u \\ d
\end{array} \ \begin{array}{c} \nu_e \\ e \end{array} \right) , \;\;\;\;
 \left( \begin{array}{c} \bar u \\ \bar d \end{array} \ \begin{array}{c} \bar \nu_e \\ \bar e
\end{array} \right) \label{eq:family} \eeq \beq  \left(  H_u \;\;  H_d  \right) \label{eq:higgsps} \eeq
\end{center}   In the Pati-Salam [PS] model, quarks and leptons of one family are united into two irreducible
representations (Eqn. \ref{eq:family}).   The two Higgs doublets of the MSSM sit in one irreducible
representation (Eqn. \ref{eq:higgsps}). This has significant consequences for fermion masses as we discuss later.
However the gauge groups are not unified and there are still three independent gauge couplings, or two if one
enlarges PS with a discrete parity symmetry where $L \leftrightarrow R$.   PS must be broken spontaneously to the
SM at some large scale  $M_G$. Below the PS breaking scale the three low energy couplings $\alpha_i, \ i= 1,2,3$
renormalize independently. Thus with the two gauge couplings and the scale $M_G$ one can fit the three low energy
couplings.

Shortly after PS, the completely unified gauge symmetry ${\bf SU_5}$ was proposed [Georgi and Glashow (1974)].
Quarks and leptons of one family sit in two irreducible representations. \beq \{  Q = \left(\begin{array}{c} u
\\ d
\end{array}\right) \;\;\;  e^c \;\;\; u^c \} \;\;
\subset \;\; \bf 10 ,\eeq  \beq \{  d^c \;\;\; L = \left(\begin{array}{c} \nu \\ e
\end{array}\right) \} \;\; \subset \;\; \bf \bar{5} .\eeq
The two Higgs doublets necessarily receive color triplet SU(5) partners filling out  $5, \ \bar 5$
representations.  \beq  \left( \begin{array}{c} H_u \\ T \end{array} \right),\;\; \left( \begin{array}{c} H_d \\
\bar T \end{array} \right) \;\; \subset \;\; {\bf 5_H},\;\; {\bf \bar{5}_H} \label{eq:higgs} \eeq   As a
consequence of complete unification the three low energy gauge couplings are given in terms of only two
independent parameters, the one unified gauge coupling $\alpha_G(M_G)$ and the unification (or equivalently the
SU(5) symmetry breaking ) scale $M_G$ [Georgi \etal (1974)].   Hence there is one prediction.  In addition we now
have the dramatic prediction that a proton is unstable to decay into a $\pi^0$ and a positron, $e^+$.

Finally complete gauge and family unification occurs in the group $ {\bf SO_{10}}$ [Georgi (1974), Fritzsch and
Minkowski (1974)] with one family contained in one irreducible representation \beq {\bf 10} + {\bf \bar{5}} +
\bar{\nu} \;\; \subset \;\;  \bf 16 \eeq  and the two multiplets of Higgs unified as well. \beq {\bf 5_H},\;\;
{\bf \bar{5}_H} \;\; \subset \;\; \bf 10_H , \eeq  (See Table \ref{t:so10}).

\begin{table}
\caption[8]{This table gives the particle spectrum for the 16 dimensional spinor representation of SO(10). The
states are described in terms of the tensor product of five spin 1/2 states with spin up ($+$) or down ($-$) and
in addition having an even number of ($-$) spins.  \{ r,b,y \} are the three color quantum numbers, and Y is weak
hypercharge given in terms of the formula $\frac{2}{3} \Sigma ({\rm C}) - \Sigma ({\rm W})$ where the sum
($\Sigma$) is over all color and weak spins with values ($\pm \ \frac{1}{2}$). Note, an SO(10) rotation
corresponds either to raising one spin and lowering another or raising (or lowering) two spins.  In the table,
the states are arranged in SU(5) multiplets.  One readily sees that the first operation of raising one spin and
lowering another is an SU(5) rotation, while the others are special to SO(10).} \label{t:so10}
$$
\begin{array}{lccc}
 \hline
 {\rm State} & { \rm Y}   & {\rm Color} & {\rm Weak} \\
 & =  \frac{2}{3} \Sigma ({\rm C}) - \Sigma ({\rm W}) &
 {\rm C \; spins} & {\rm W \; spins} \\
\hline
\hline
{\bf \bar \nu}  & { 0} &  + \, + \, +  &     + \, +  \\
\hline
\\
\hline
{\bf \bar e}  & { 2} &  + \, + \, + & - \, -  \\
\hline
{\bf u_r} &  &  - \, + \, + &   + \, - \\
{\bf d_r} &  &  - \, + \, + &   - \, + \\
{\bf u_b} & { \frac{1}{3}} &  + \, - \, + &   + \, -  \\
{\bf d_b} &  &  + \, - \, + &  - \, +  \\
{\bf u_y} &  & + \, + \, - &    + \, - \\
{\bf d_y} &  & + \, + \, - &   - \, + \\
\hline
{\bf \bar u_r} &  & + \, - \, - & + \, + \\
{\bf \bar u_b} & { -\frac{4}{3}} & - \, + \, - & + \, + \\
{\bf \bar u_y} &  & - \, - \, + & + \, + \\
\hline
\\
\hline
{\bf \bar d_r} &  & + \, - \, - & - \, - \\
{\bf \bar d_b} & { \frac{2}{3}} & - \, + \, - & - \, - \\
{\bf \bar d_y} &  & - \, - \, + & - \, - \\
\hline
{\bf \nu}  & { -1} &  - \, - \, - & + \, - \\
{\bf e}  &  & - \, - \, - & - \, + \\
\hline
\end{array}
$$
\end{table}

GUTs predict that protons decay with a lifetime $\tau_p$ of order $M_G^4/\alpha_G^2 m_p^5$. The first experiments
looking for proton decay were begun in the early 1980s. However at the very moment that proton decay searches
began, motivated by GUTs, it was shown that \gs naturally increase $M_G$, thus increasing the proton lifetime.
Hence, if \gs were correct, it was unlikely that the early searches would succeed [Dimopoulos \etal (1981),
Dimopoulos and Georgi (1981), Iba\~{n}ez and Ross (1981), Sakai (1981), Einhorn and Jones (1982), Marciano and
Senjanovic (1982)].  At the same time, it was shown that \gs did not significantly affect the predictions for
gauge coupling unification (for a review see [Dimopoulos \etal (1991), Raby (2002a)]).   At present, non- \gs are
excluded by the data for gauge coupling unification; where as \gs work quite well. So well in fact, that the low
energy data is now probing the physics at the GUT scale.  In addition, the experimental bounds on proton decay
from Super-Kamiokande exclude non-\ss GUTs, while severely testing \ss GUTs.    Moreover, future underground
proton decay/neutrino observatories, such as the proposed Hyper-Kamiokande detector in Japan or UNO in the USA
will cover the entire allowed range for the proton decay rate in \gs.

If \ss is so great,  if it is Nature, then where are the \ss particles?  Experimentalists at high energy
accelerators, such as the Fermilab Tevatron and the CERN LHC (now under construction), are desperately seeking
\ss particles or other signs of \ss.   At underground proton decay laboratories, such as Super-Kamiokande in
Japan or Soudan II in Minnesota, USA,  electronic eyes continue to look for the tell-tale signature of a proton
or neutron decay.  Finally, they are searching for cold dark matter, via direct detection in underground
experiments such as CDMS, UKDMC or EDELWEISS, or indirectly by searching for energetic gammas or neutrinos
released when two neutralino dark matter particles annihilate.   In Sect. \ref{sec:where} we focus on the
perplexing experimental/theoretical problem of where are these \ss particles.   We then consider the status of
gauge coupling unification (Sect. \ref{sec:unif}),  proton decay predictions (Sect. \ref{sec:protons}),  fermion
masses and mixing angles (including neutrinos) and the \ss flavor problem (Sect. \ref{sec:flavor}),  and \ss dark
matter (Sect. \ref{sec:darkmatter}).    We conclude with a discussion of some remaining open questions.

\section{Where are the supersymmetric particles ? \label{sec:where} }
The answer to this question depends on two interconnected theoretical issues -- \begin{enumerate} \item the
mechanism for \ss breaking, and \item the scale of \ss breaking. \end{enumerate}  The first issue is inextricably
tied to the \ss flavor problem.  While the second issue is tied to the gauge hierarchy problem.  We discuss these
issues in sections \ref{sect:ssbreaking} and \ref{sect:finetuning}.
\subsection{\ss Breaking Mechanisms \label{sect:ssbreaking}}
Supersymmetry is necessarily a local gauge symmetry, since Einstein's general theory of relativity corresponds to
local Poincar\'{e} symmetry and supersymmetry is an extension of the Poincar\'{e} group.   Hence \ss breaking
must necessarily be spontaneous, in order not to cause problems with unitarity and/or relativity.  In this
section we discuss some of the spontaneous \ss breaking mechanisms considered in the literature. However from a
phenomenological stand point, any spontaneous \ss breaking mechanism results in {\em soft \ss breaking} operators
with dimension 3 or less (such as quadratic or cubic scalar operators or fermion mass terms) in the effective low
energy theory below the scale of \ss breaking [Dimopoulos and Georgi (1981), Sakai (1981), Girardello and Grisaru
(1982)].   There are a priori hundreds of arbitrary soft \ss breaking parameters (the coefficients of the soft
\ss breaking operators) [Dimopoulos and Sutter (1995)].  These are parameters not included in the SM but are
necessary to compare with data or make predictions for new experiments.

The general set of renormalizable soft \ss breaking operators, preserving the solution to the gauge hierarchy
problem, is given in a paper by [Girardello and Grisaru (1982)].    These operators are assumed to be the low
energy consequence of spontaneous \ss breaking at some fundamental \ss breaking scale $\Lambda_S \gg M_Z$.  The
list of soft \ss breaking parameters includes squark and slepton mass matrices, cubic scalar interaction
couplings, gaugino masses, etc.  Let us count the number of arbitrary parameters [Dimopoulos and Sutter (1995)].
Left and right chiral scalar quark and lepton mass matrices are a priori independent $3 \times 3$ hermitian
matrices. Each contains 9 arbitrary parameters. Thus for the scalar partners of \{$Q, \ \bar u, \ \bar d, \ L, \
\bar e$\} we have 5 such matrices or 45 arbitrary parameters. In addition corresponding to each complex $3 \times
3$ Yukawa matrix (one for up and down quarks and charged leptons) we have a complex soft \ss breaking trilinear
scalar coupling ($A$) of left and right chiral squarks or sleptons to Higgs doublets. This adds $3 \times 18 =
54$ additional arbitrary parameters. Finally, add to these 3 complex gaugino masses ($M_i, \ i= 1,2,3$), and the
complex soft \ss breaking scalar Higgs mass ($\mu B$) and we have a total of 107 arbitrary soft \ss breaking
parameters.  In additon, the minimal supersymmetric extension of the SM requires a complex Higgs mass parameter
($\mu$) which is the coefficient of a supersymmetric term in the Lagrangian.   Therefore, altogether this minimal
extension has 109 arbitrary parameters.   Granted, not all of these parameters are physical.  Just as not all 54
parameters in the three complex $3 \times 3$ Yukawa matrices for up and down quarks and charged leptons are
observable.  Some of them can be rotated away by unitary redefinitions of quark and lepton superfields. Consider
the maximal symmetry of the kinetic term of the theory --- global $SU(3)_Q \times SU(3)_{\bar u} \times
SU(3)_{\bar d} \times SU(3)_{L} \times SU(3)_{\bar e} \times U(1)^5 \times U(1)_R$.   Out of the total number of
parameters -- 163 = 109 (new \ss parameters) + 54 (SM parameters) -- we can use the $SU(3)^5$ to eliminate 40
parameters and 3 of the $U(1)$s to remove 3 phases.   The other 3 $U(1)$s however, are symmetries of the theory
corresponding to B, L and weak hypercharge Y.  We are thus left with 120 observables, corresponding to the 9
charged fermion masses, 4 quark mixing angles and {\em 107 new, arbitrary observable \ss parameters.}

Such a theory, with so many arbitrary parameters, clearly makes no predictions.  However, this general MSSM is a
``straw man" (one to be struck down), but fear not since it is the worst case scenario. In fact, there are
several reasons why this worst case scenario cannot be correct.   First, and foremost, it is severely constrained
by precision electroweak data. Arbitrary $3 \times 3$ matrices for squark and slepton masses or for trilinear
scalar interactions maximally violate quark and lepton flavor.  The strong constraints from flavor violation were
discussed by [Dimopoulos and Georgi (1981), Dimopoulos and Sutter (1995), Gabbiani \etal (1996)]. In general,
they would exceed the strong experimental contraints on flavor violating processes, such as $b \rightarrow s
\gamma$, or $b \rightarrow s l^+ l^-$, $B_s \rightarrow \mu^+ \mu^-$, $\mu \rightarrow e \gamma$, $\mu - e$
conversion in nuclei, etc. In order for this general MSSM not to be excluded by flavor violating constraints, the
soft \ss breaking terms must be either
\begin{enumerate} \item flavor independent, \item aligned with quark and lepton masses or \item the first and
second generation squark and slepton masses ($\tilde m$) should be large (i.e. greater than a TeV).
\end{enumerate}   In the first case, squark and slepton mass matrices are proportional to the $3 \times 3$
identity matrix and the trilinear couplings are proportional to the Yukawa matrices.   In this case the squark
and slepton masses and trilinear couplings are diagonalized in the same basis that quark and lepton Yukawa
matrices are diagonalized.  This limit preserves three lepton numbers -- $L_e, \ L_\mu, \ L_\tau$ -- (neglecting
neutrino masses) and gives minimal flavor violation (due only to CKM mixing) in the quark sector [Hall \etal
(1986)]. The second case does not require degenerate quark flavors, but {\em approximately} diagonal squark and
slepton masses and interactions, when in the diagonal quark and lepton Yukawa basis.   It necessarily ties any
theoretical mechanism explaining the hierarchy of fermion masses and mixing to the hierarchy of sfermion masses
and mixing.  This will be discussed further in Sections \ref{sec:familysymmetry} and \ref{sec:flavorviol}.
Finally, the third case minimizes flavor violating processes, since all such effects are given by effective
higher dimension operators which scale as $1/\tilde m^2$. The theoretical issue is what \ss breaking mechanisms
are ``naturally" consistent with these conditions.

Several such \ss breaking mechanisms exist in the literature.  They are called minimal supergravity [mSugra]
breaking, gauge mediated \ss breaking [GMSB],  dilaton mediated [DMSB], anomaly mediated [AMSB], and gaugino
mediated [gMSB]. Consider first mSugra which has been the benchmark for experimental searches.  The minimal
supergravity model [Ovrut and Wess (1982), Chamseddine \etal (1982), Barbieri \etal (1982), Iba\~{n}ez (1982),
Nilles \etal (1983), Hall \etal (1983)] is defined to have the minimal set of soft \ss breaking parameters. It is
motivated by the simplest ([Polony (1977)] hidden sector in supergravity with the additional assumption of grand
unification. This \ss breaking scenario is also known as the constrained MSSM [CMSSM] [Kane \etal (1994)]. In
mSUGRA/CMSSM there are four soft \ss breaking parameters at $M_G$ defined by $m_0$, a universal scalar mass;
$A_0$, a universal trilinear scalar coupling; $M_{1/2}$, a universal gaugino mass; and $\mu_0 B_0$, the soft \ss
breaking Higgs mass parameter where $\mu_0$, is the supersymmetric Higgs mass parameter.  In most analyses,
$|\mu_0|$ and $\mu_0 B_0$ are replaced, using the minimization conditions of the Higgs potential, by $M_Z$ and
the ratio of Higgs VEVs $\tan\beta = \langle H_u \rangle/ \langle H_d \rangle$. Thus the parameter set defining
mSugra/CMSSM is given by
\begin{equation} m_0, \ A_0, \ M_{1/2}, \ {\rm sign}(\mu_0), \ \tan\beta . \label{eq:cmssm} \end{equation}
This scenario is an example of the first case above (with minimal flavor violation), however it is certainly not
a consequence of the most general supergravity theory and thus requires further justification. Nevertheless it is
a useful framework for experimental \ss searches.

In GMSB models, \ss breaking is first felt by messengers carrying standard model charges and then transmitted to
to the superpartners of SM particles [sparticles] via loop corrections containing SM gauge interactions. Squark
and slepton masses in these models are proportional to $\alpha_i \ \Lambda_{\ss}$ with $\Lambda_{\ss} = F/M$. In
this expression, $\alpha_i,\; i=1,2,3$ are the fine structure constants for the SM gauge interactions, $\sqrt{F}$
is the fundamental scale of \ss breaking, $M$ is the messenger mass, and $\Lambda_{\ss}$ is the effective \ss
breaking scale.  In GMSB the flavor problem is naturally solved since all squarks and sleptons with the same SM
charges are degenerate and the $A$ terms vanish to zeroth order.  In addition, GMSB resolves the formidable
problems of model building [Fayet and Ferrara (1977)] resulting from the direct tree level \ss breaking of
sparticles. This problem derives from the supertrace theorem, valid for tree level \ss breaking,
\begin{equation} \Sigma (2J + 1) (-1)^{2J} \ M^2_J = 0
\end{equation} where the sum is over all particles with spin $J$ and mass $M_J$. It generically leads to charged
scalars with negative mass squared [Fayet and Ferrara (1977), Dimopoulos and Georgi (1981)].  Fortunately the
supertrace theorem is explicitly violated when \ss breaking is transmitted radiatively.\footnote{It is also
violated in supergravity where the right-hand side is replaced by $2 (N - 1) m_{3/2}^2$ with $m_{3/2}$, the
gravitino mass and $N$ the number of chiral superfields.} Low energy \ss breaking models [Dimopoulos and Raby
(1981), Dine \etal (1981), Witten (1981), Dine and Fischler (1982), Alvarez-Gaume \etal (1982)], with $\sqrt{F}
\sim M \sim \Lambda_{\ss} \sim 100$ TeV make dramatic predictions [Dimopoulos \etal (1996)]. Following the
seminal work of [Dine and Nelson (1993), Dine \etal (1995,1996)] complete GMSB models now exist (for a review,
see [Giudice and Rattazzi (1999)]. Of course the fundamental \ss breaking scale can be much larger than the weak
scale.  Note \ss breaking effects are proportional to $1/M$ and hence they decouple as $M$ increases. This is a
consequence of \ss breaking decoupling theorems [Polchinski and Susskind (1982), Dimopoulos and Raby (1983),
Banks and Kaplunovsky (1983)]. However when $\sqrt{F} \geq 10^{10}$ GeV then supergravity becomes important.

DMSB, motivated by string theory, and AMSB and gMSB, motivated by brane models with extra dimensions, also
alleviate the \ss flavor problem.    We see that there are several possible \ss breaking mechanisms which solve
the \ss flavor problem and provide predictions for superpartner masses in terms of a few fundamental parameters.
Unfortunately we do not a priori know which one of these (or some other) \ss breaking mechanism is chosen by
nature. For this we need experiment.

\subsection{Fine Tuning or ``Naturalness" \label{sect:finetuning}}
Presently, the only evidence for supersymmetry is indirect, given by the successful prediction for gauge coupling
unification.   Supersymmetric particles at the weak scale are necessary for this to work, however it is
discouraging that there is yet no direct evidence.  Searches for new supersymmetric particles at CERN or Fermilab
have produced only lower bounds on their mass.  The SM Higgs mass bound, applicable to the MSSM when the CP odd
Higgs (A) is much heavier, is 114.4 GeV [LEP2 (2003)].  In the case of an equally light A,  the Higgs bound is
somewhat lower $\sim 89$ GeV.   Squark, slepton and gluino mass bounds are of order 200 GeV,  while the chargino
bound is 103 GeV [LEP2 (2003)].  In addition other indirect indications for new physics beyond the standard
model, such as the anomalous magnetic moment of the muon ($a_\mu$), are inconclusive. Perhaps Nature does not
make use of this beautiful symmetry?   Or perhaps the \ss particles are heavier than we once believed.

Nevertheless, global fits to precision electroweak data in the SM or in the MSSM give equally good $\chi^2/dof$
[de Boer (2003)].   In fact the fit is slightly better for the MSSM due mostly to the pull of $a_\mu$.  The real
issue among \ss enthusiasts is the problem of {\em fine tuning}.  If \ss is a solution to the gauge hierarchy
problem (making the ratio $M_Z/M_G \sim 10^{-14}$ ``naturally" small), then radiative corrections to the Z mass
should be insensitive to physics at the GUT scale, i.e. it should not require any ``unnatural" fine tuning of GUT
scale parameters. A numerical test of fine tuning is obtained by defining the {\em fine tuning parameter}
$\Delta_\alpha ( = |\frac{a_\alpha}{M_Z^2} \frac{d M_Z^2}{d a_\alpha}|$), the logarithmic derivative of the Z
mass with respect to different ``fundamental" parameters $a_\alpha$ = \{$\mu, \ M_{1/2}, \ m_0, \ A_0, \ \dots$\}
defined at $M_G$ [Ellis \etal (1986), Barbieri and Giudice (1988), de Carlos and Casas (1993), Anderson \etal
(1995)]. Smaller values of $\Delta_\alpha$ correspond to less fine tuning and roughly speaking $p =
Max(\Delta_\alpha)^{-1}$ is the probability that a given model is obtained in a random search over \ss parameter
space.

There are several recent analyses, including LEP2 data, by [Chankowski \etal (1997), Barbieri and Strumia (1998),
Chankowski \etal (1999)]. In particular [Barbieri and Strumia (1998), Chankowski \etal (1999)] find several
notable results.  In their analysis [Barbieri and Strumia (1998)] only consider values of $\tan\beta < 10$ and
soft \ss breaking parameters of the CMSSM or gauge-mediated \ss breaking. [Chankowski \etal (1999)] also consider
large $\tan\beta = 50$ and more general soft \ss breaking scenarios.  They both conclude that the value of
$Max(\Delta_\alpha)$ is significantly lower when one includes the one loop radiative corrections to the Higgs
potential as compared to the tree level Higgs potential used in the analysis of [Chankowski \etal (1997)].  In
addition they find that the experimental bound on the Higgs mass is a very strong constraint on fine tuning.
Larger values of the light Higgs mass require larger values of $\tan\beta$.  Values of $Max(\Delta_\alpha) < 10$
are possible for a Higgs mass $< 111$ GeV (for values of $\tan\beta < 10$ used in the analysis of [Barbieri and
Strumia (1998)]). However allowing for larger values of $\tan\beta$ [Chankowski \etal (1999)] allows for a
heavier Higgs. With LEP2 bounds on a SM Higgs mass of 114.4 GeV,  larger values of $\tan\beta > 4$ are required.
It is difficult to conclude too much from these results.  Note, the amount of fine tuning is somewhat sensitive
to small changes in the definition of $\Delta_\alpha$. For example, replacing $a_\alpha \rightarrow a_\alpha^2$
or $M_Z^2 \rightarrow M_Z$ will change the result by a factor of $2^{\pm 1}$. Hence factors of two in the result
are definition dependent. Let us assume that fine tuning by 1/10 is acceptable, then is fine tuning by 1 part in
100 ``unnatural." Considering the fact that the fine tuning necessary to maintain the gauge hierarchy in the SM
is at least 1 part in $10^{28}$, a fine tuning of 1 part in 100 (or even $10^3$) seems like a great success.

A slightly different way of addressing the fine tuning question says if I assign equal weight to all
``fundamental" parameters at $M_G$ and scan over all values within some a priori assigned domain,  what fraction
of this domain is already excluded by the low energy data.  This is the analysis that [Strumia (1999)] uses to
argue that 95\% of the \ss parameter space is now excluded by LEP2 bounds on the \ss spectrum and in particular
by the Higgs and chargino mass bounds. This conclusion is practically insensitive to the method of \ss breaking
assumed in the analysis which included the CMSSM, gauge-mediated or anomaly mediated \ss breaking or some
variations of these.   One might still question whether the a priori domain of input parameters,  upon which this
analysis stands, is reasonable.   Perhaps if we doubled the input parameter domain we could find acceptable
solutions in 50\% of parameter space.    To discuss this issue in more detail, let us consider two of the
parameter domains considered in [Strumia (1999)].   Within the context of the CMSSM, he considers the domain
defined by
\begin{equation} m_0, \ |A_0|, \ |M_{1/2}|, \ |\mu_0|, \ |B_0| = (0 \div 1) \ m_{\ss}  \end{equation}   where  $(a
\div b) \equiv$ a random number between $a$ and $b$ and the overall mass scale $m_{\ss}$ is obtained from the
minimization condition for electroweak symmetry breaking.   He also considered an alternative domain defined by
\begin{eqnarray} m_0 = (\frac{1}{9} \div 3) \ m_{\ss}, & \;\;\;\; |\mu_0|, \ |M_{1/2}| = (\frac{1}{3} \div 3) \ m_{\ss}, &  \\
&  \ A_0, \ B_0 = (-3 \div 3) \ m_0 & \nn
\end{eqnarray} with the sampling of $m_0, \ M_{1/2}, \ \mu_0$ using a flat distribution on a log scale.  In both
cases, he concludes that 95\% of parameter space is excluded with the light Higgs and chargino mass providing the
two most stringent constraints.  Hence we have failed to find \ss in 95\% of the {\em allowed region of parameter
space}.  But perhaps we should open the analysis to other, much larger regions, of \ss parameter space.   We
return to this issue in Sections \ref{sec:focuspt} and \ref{sec:rismh}.

For now however, let us summarize our discussion of naturalness constraints with the following quote from
[Chankowski \etal (1999)], ``We re-emphasize that naturalness is [a] subjective criterion, based on physical
intuition rather than mathematical rigour.  Nevertheless, it may serve as an important guideline that offers some
discrimination between different theoretical models and assumptions. As such, it may indicate which domains of
parameter space are to be preferred. However, one should be very careful in using it to set any absolute upper
bounds on the spectrum. We think it safer to use relative naturalness to compare different scenarios, as we have
done in this paper."   As these authors discuss in their paper,  in some cases the amount of fine tuning can be
dramatically decreased if one assumes some linear relations between GUT scale parameters.  These relations may be
due to some, yet unknown, theoretical relations coming from the fundamental physics of \ss breaking, such as
string theory.

In the following we consider two deviations from the simplest definitions of fine tuning and naturalness. The
first example, called focus point [FP] [Feng and Moroi (2000), Feng \etal (2000a,b,c), Feng and Matchev (2001)]
is motivated by infra-red fixed points of the renormalization group equations for the Higgs mass and other
dimensionful parameters.  The second had two independent motivations.  In the first case it is motivated by the
\ss flavor problem and in this incarnation it is called the radiative inverted scalar mass hierarchy [RISMH]
[Bagger \etal (1999,2000)].   More recently it was reincarnated in the context of SO(10) Yukawa unification for
the third generation quarks and leptons [YU] [Raby (2001), Derm\' \i \v sek (2001), Baer and Ferrandis (2001),
Bla\v{z}ek \etal (2002a,b), Auto \etal (2003), Tobe and Wells (2003)] scenarios.  In both scenarios the upper
limit on soft scalar masses is increased much above 1 TeV.

\subsection{The focus point region of \ss breaking parameter space \label{sec:focuspt}}

In the focus point \ss breaking scenario, Feng \etal [Feng and Moroi (2000), Feng \etal (2000a,b)] consider the
renormalization group equations [RGE] for soft \ss breaking parameters, assuming a universal scalar mass $m_0$ at
$M_G$.  This may be as in the CMSSM (Eqn. \ref{eq:cmssm}) or even a variation of AMSB.  They show that, if the
top quark mass is approximately $174$ GeV, then the RGEs lead to a Higgs mass which is naturally of order the
weak scale, independent of the precise value of $m_0$ which could be as large as 3 TeV. It was also noted that
the only fine tuning in this scenario was that necessary to obtain the top quark mass, i.e. if the top quark mass
is determined by other physics then there is no additional fine tuning needed to obtain electroweak symmetry
breaking.\footnote{For a counter discussion of fine tuning in the focus point region, see [Romanino and Strumia
(2000)].} As discussed in [Feng \etal (2000c)] this scenario opens up a new window for neutralino dark matter.
Cosmologically acceptable neutralino abundances are obtained even with very large scalar masses. Moreover as
discussed in [Feng and Matchev (2001)] the focus point scenario has many virtues.   In the limit of large scalar
masses,  gauge coupling unification requires smaller threshold corrections at the GUT scale, in order to agree
with low energy data.  In addition, larger scalar masses ameliorate the \ss flavor and CP problems.  This is
because both processes result from effective higher dimensional operators suppressed by two powers of squark
and/or slepton masses.  Finally a light Higgs mass in the narrow range from about 114 to $\sim$ 120 GeV is
predicted.  Clearly the focus point region includes a much larger range of soft \ss breaking parameter space than
considered previously.   It may also be perfectly ``natural."

The analysis of the focus point scenario was made within the context of the CMSSM.  The focus point region
extends to values of $m_0$ up to 3 TeV.   This upper bound increases from 3 to about 4 TeV as the top quark mass
is varied from 174 to 179 GeV.  On the other hand, as $\tan\beta$ increases from 10 to 50, the allowed range in
the $m_0 - M_{1/2}$ plane for $A_0 = 0$, consistent with electroweak symmetry breaking, shrinks.  As we shall see
from the following discussion, this narrowing of the focus point region is most likely an artifact of the precise
CMSSM boundary conditions used in the analysis.  In fact the CMSSM parameter space is particularly constraining
in the large $\tan\beta$ limit.

\subsection{SO(10) Yukawa unification and the radiative inverted scalar mass hierarchy [RISMH] \label{sec:rismh}}

The top quark mass $M_t \sim 174$ GeV requires a Yukawa coupling $\lambda_t \sim 1$.  In the minimal SO(10) \ss
model [MSO$_{10}$SM] the two Higgs doublets, $H_u, \ H_d$, of the MSSM are contained in one $10$. In addition the
three families of quarks and leptons are in $16_i, \ i = 1,2,3$.    In the MSO$_{10}$SM the third generation
Yukawa coupling is given by  \begin{eqnarray} \lambda \ 16_3 \ 10 \ 16_3 & = \lambda \ ( Q_3 \ H_u \ \bar t \ + \
L_3 \ H_u \ \bar \nu_\tau \ + \ Q_3 \ H_d \ \bar b \ + \ L_3 \ H_d \ \bar \tau ) & \label{eq:yukawa3} \\ & +
\lambda \ ( \frac{1}{2} Q_3 \ Q_3 \ + \ \bar t \ \bar \tau ) \ T + \lambda ( Q_3 \ L_3 \ + \ \bar t \ \bar b ) \
\bar T . & \nonumber
\end{eqnarray}  Thus we obtain the unification of all third generation Yukawa couplings with
\begin{equation} \lambda = \lambda_t = \lambda_b = \lambda_\tau = \lambda_{\nu_\tau} . \label{eq:yukunif} \end{equation}
Of course this simple Yukawa interaction, with the constant $\lambda$ replaced by a $3 \times 3$ Yukawa matrix,
does not work for all three families.\footnote{In such a theory there is no CKM mixing matrix and the down quark
and charged lepton masses satisfy the bad prediction $\frac{1}{20} \sim \frac{m_d}{m_s} = \frac{m_e}{m_\mu} \sim
\frac{1}{200}$.}  In this discussion, we shall assume that the first and second generations obtain mass using the
same $10$, but via effective higher dimensional operators resulting in a hierarchy of fermion masses and mixing.
In this case, Yukawa unification for the third family (Eqn. \ref{eq:yukunif}) is a very good approximation.  The
question then arises, is this symmetry relation consistent with low energy data given by \bea M_t = 174.3 \pm 5.1
\ {\rm GeV}, & \;\;\;  m_b(m_b) = 4.20 \pm 0.20 \ {\rm GeV}, & \\  M_{\tau} = 1.7770 \pm 0.0018, & & \nn \eea
where the error on $M_{\tau}$ is purely a theoretical uncertainty due to numerical errors in the analysis.
Although this topic has been around for a long time, it is only recently that the analysis has included the
complete one loop threshold corrections at the weak scale [Raby (2001), Derm\' \i \v sek (2001), Baer and
Ferrandis (2001), Bla\v{z}ek \etal (2002a,b), Auto \etal (2003), Tobe and Wells (2003)].   It turns out that
these corrections are very important. The corrections to the bottom mass are {\em functions of squark and gaugino
masses} times a factor of $\tan\beta \sim m_t(m_t)/m_b(m_t) \sim 50$.  For typical values of the parameters the
relative change in the bottom mass $\delta m_b/m_b$ is very large, of order 50\%.   At the same time, the
corrections to the top and tau masses are small. For the top, the same one loop corrections are proportional to
$1/\tan\beta$, while for the tau, the dominant contribution from neutralino loops is small.  These one loop
radiative corrections are determined, through their dependence on squark and gaugino masses, by the soft \ss
breaking parameters at $M_G$. In the MSO$_{10}$SM we assume the following dimensionful parameters.
\begin{equation} m_{16}, \ m_{10}, \ \Delta m_H^2, \ A_0, \ M_{1/2}, \ \mu \end{equation}
where $m_{16}$ is the universal squark and slepton mass; $m_{H_{u/d}}^2 =  m_{10}^2 (1 \mp \Delta m_H^2)$ is the
Higgs up/down mass squared; $A_0$ is the universal trilinear Higgs - scalar coupling; $M_{1/2}$ is the universal
gaugino mass and $\mu$ is the supersymmetric Higgs mass.   $\tan\beta \approx 50$ is fixed by the top, bottom and
$\tau$ mass.    Note, there are two more parameters ($m_{10}, \ \Delta m_H^2$) than in the CMSSM.  They are
needed in order to obtain electroweak symmetry breaking solutions in the region of parameter space with $m_{16}
\gg \mu, \ M_{1/2}$.   We shall defer a more detailed discussion of the results of the MSO$_{10}$SM to Sects.
\ref{sec:yukawaunif} and \ref{sec:darkmatter}.   Suffice it to say here that good fits to the data are only
obtained in a narrow region of \ss parameter space given by  \begin{eqnarray}  A_0 \approx - 2 \ m_{16}, & m_{10}
\approx \sqrt{2} \ m_{16}, & \;\; \Delta m_H^2 \approx 0.13 \label{eq:mso10sm} \\  m_{16} \ \gg \ \mu \sim
M_{1/2}, & \;\;\; m_{16} \ > \ 2 \; {\rm TeV} . & \nonumber \end{eqnarray}

Once more we are concerned about fine-tuning with $m_{16}$ so large.  However, we discover a fortuitous
coincidence. This region of parameter space (Eqn. \ref{eq:mso10sm}) ``naturally" results in a radiative inverted
scalar mass hierarchy with $\tilde m_{1,2}^2 \gg \tilde m_3^2$ [Bagger \etal (1999,2000)], i.e. first and second
generation squark and slepton masses are of order $m_{16}^2$, while the third generation scalar masses are much
lighter. Since the third generation has the largest couplings to the Higgs bosons, they give the largest
radiative corrections to the Higgs mass.  Hence with lighter third generation squarks and sleptons, a light Higgs
is more ``natural."   Although a detailed analysis of fine-tuning parameters is not available in this regime of
parameter space, the results of several papers suggest that the fine-tuning concern is minimal (see for example,
[Dimopoulos and Giudice (1995), Chankowski \etal (1999), Kane \etal (2003)]).   While there may not be any
fine-tuning necessary in the MSO$_{10}$SM region of \ss parameter space (Eqn. \ref{eq:mso10sm}),  there is still
one open problem. There is no known \ss breaking mechanism which ``naturally" satisfies the conditions of Eqn.
\ref{eq:mso10sm}. On the other hand, we conclude this section by noting that the latter two examples suggest that
there is a significant region of \ss breaking parameter space which is yet to be explored experimentally.

\section{Gauge coupling unification \label{sec:unif}}

The apparent unification of the three gauge couplings at a scale of order $3 \times 10^{16}$ GeV is, at the
moment, the only experimental evidence for low energy supersymmetry [Amaldi \etal (1991), Ellis \etal (1991),
Langacker and Luo (1991)]. In this section we consider the status of gauge coupling unification and the demise of
minimal \ss SU(5).

The theoretical analysis of unification is now at the level requiring two loop renormalization group running from
$M_G$ to $M_Z$.  Consistency then requires including one loop threshold corrections at both the GUT and weak
scales. Once GUT threshold corrections are considered, a precise definition of the GUT scale ($M_G$) is needed.
The three gauge couplings no longer meet at one scale,\footnote[1]{[Brodsky \etal (2003)] has argued that the
three gauge couplings always meet in a GUT at a scale above the largest GUT mass.  He defines this to be the GUT
scale.  Unfortunately, this scale cannot be defined in the effective low energy theory.} since
\begin{equation} \alpha_i(\mu)^{-1} = \alpha_G^{-1} + \Delta_i(\mu)  \end{equation}
where the corrections $\Delta_i(\mu)$ are logarithmic functions of mass for all states with GUT scale mass.  In
principle, the GUT scale can now be defined as the mass $M_X$ of the $X, \bar X$ gauge bosons mediating proton
decay or as the scale where any two couplings meet.  We define $M_G$ as the value of $\mu$ where
\begin{equation} \alpha_1(M_G) = \alpha_2(M_G) \equiv \tilde \alpha_G .\end{equation}
Using two loop RGE from $M_Z$ to $M_G$, we find \begin{eqnarray} M_G \approx & 3 \times 10^{16} \;\; {\rm GeV} &  \\
\alpha_G^{-1} \approx & 24 . & \nonumber \end{eqnarray}   In addition, good fits to the low energy data require
\begin{equation} \epsilon_3 \equiv \frac{(\alpha_3(M_G) - \tilde \alpha_G)}{\tilde \alpha_G} \sim - 3\% \;\; {\rm to} \; -4\% .
\end{equation}  Note the exact value of the threshold correction ($\epsilon_3$), needed to fit the data, depends
on the weak scale threshold corrections and in particular on the \ss particle spectrum.  We shall return to this
later. On the other hand, significant contributions to the GUT threshold correction $\epsilon_3$ typically arise
from the Higgs and GUT breaking sectors of the theory.  Above $M_G$ there is a single coupling constant $\alpha_G
\approx \tilde \alpha_G$ which then runs up to some fundamental scale $M_*$, such as the string scale, where the
running is cut off.   The GUT symmetry, in concert with supersymmetry,  regulates the radiative corrections.
Without the GUT, $\epsilon_3$ would naturally take on a value of order one.

Following [Lucas and Raby (1996)] we show that the allowed functional dependence of $\epsilon_3$ on GUT symmetry
breaking vacuum expectation values [VEVs] is quite restricted.  Consider a general SO(10) theory with
\begin{equation} (n_{16} + 3)\;{\bf 16} + n_{16} \; \overline{\bf 16} + n_{10} \; {\bf 10} + n_{45}  \;
{\bf 45} + n_{54} \; {\bf 54} + n_1 \; {\bf 1} . \end{equation}  Note, the superpotential for the GUT breaking
sector of the theory typically has a U(1)$^n$ $\times$ R symmetry which, as we shall see, has an important
consequence for the threshold corrections. The one loop threshold corrections are given by
\begin{equation} \alpha_i^{-1}(M_G)=\alpha_{G}^{-1}-\Delta_i \end{equation}  with  \begin{equation} \Delta_i={1
\over 2 \pi} \sum_\xi b_i^\xi \log \abs{M_\xi \over M_G}. \end{equation} The sum is over all super heavy
particles $\xi$ with mass $M_\xi$ and $b_i^\xi$ is the contribution the super heavy particle would make to the
beta function coefficient $b_i$ if the particle were not integrated out at $M_G$.

As a consequence of \ss and the U(1) symmetries, Lucas and Raby proved the following theorem:  {\it $\epsilon_3$
is only a function of U(1) and R invariant products of powers of VEVs} \{$\zeta_i$\}, i.e.
\begin{equation} {\epsilon}_3=F(\zeta_1,\ldots,\zeta_m) + {3 \tilde\alpha_G \over 5\pi}\log \abs{{ M_T^{eff} \over M_G}} +
\cdots . \end{equation}   As an example, consider the symmetry breaking sector given by the superpotential
\begin{eqnarray}
W_{sym\, breaking}=& {1\over M_*} A_1' (A_1^3+{\cal S}_3 S A_1+{\cal S}_4 A_1 A_2) \label{eq:symbreaking} \\
& + A_2 (\psi \overline{\psi} + {\cal S}_1 \tilde A) + S \tilde{A}^2 \nn \\
& + S' (S {\cal S}_2+A_1 \tilde A) + {\cal S}_3 S'^2 \nn
\end{eqnarray}  where the fields transform as follows
\{$A_1,\; A_2,\; \tilde A, \; A'_1$\} $\; \subset \; {\bf 45}$, \hspace{.2in} \{$S, \; S'$\} $\; \subset \; {\bf 54}$, \\
$\psi, \; \bar\psi$ $\; \subset \; {\bf 16, \ \overline{16}}$, \hspace{.2in} and  \{${\cal S}_1, \cdots , {\cal
S}_4$\} $\; \subset \; {\bf 1}$.  In addition we include the Lagrangian for the electroweak Higgs sector given by
\beq L_{Higgs}= [10_1 A_1 10_2 + {\cal S}_5 10^2_2 |_F + {1 \over M} [z^* 10_1^2 |_D . \eeq  $W_{sym\,breaking}$
has a [U(1)]$^4 \times$ R symmetry.  Since \ss is unbroken, the potential has many F and D flat directions.  One
in particular (Eqn.~\ref{eq:vevs}) breaks SO(10) to SU(3)$\times$SU(2)$\times$U(1)$_Y$ leaving only the states of
the MSSM massless plus some non-essential SM singlets.
\bea \langle A_1 \rangle =  a_1 \; \frac{3}{2} (B - L), &
\;\;\;  \langle A_2 \rangle =  a_2 \; \frac{3}{2} Y
 & \label{eq:vevs} \\
 \langle \tilde{A} \rangle =  \tilde{a} \; \frac{1}{2} X & \;\;\;  \langle S \rangle = s \; \ {\rm diag} ( 1 1 1
-3/2 -3/2 ) \otimes {\bf 1}_{2 \times 2} & \nn \\
 \langle \psi \rangle = v \, \, |{\rm SU(5) \, \, singlet} \rangle & \;\;\;  \langle \overline{\psi} \rangle =
\overline{v} \, \, |{\rm SU(5) \, \, singlet} \rangle & . \nn \eea  The VEVs \{$a_1, a_2, \tilde a, v, \overline
v, {\cal S}_4$\} form a complete set of independent variables along the F and D flat directions. Note since the
superpotential (Eqn.~\ref{eq:symbreaking}) contains higher dimension operators fixed by the cutoff scale $M_*
\sim M_{Planck}$ the GUT scale spectrum ranges from $10^{13} - 10^{20}$ GeV.   Nevertheless the threshold
corrections are controlled.  The only invariant under a [U(1)]$^4 \times$R rotation of the VEVs is
$\zeta={a_1^4\over a_2^2 {\cal S}_4^2}$.   By an explicit calculation we find the threshold correction \beq
{\epsilon}_3={3 \tilde\alpha_G \over 5 \pi} \biggl\{ \log{256\over 243} - \f{1}{2} \ \log\abs{(1-25 \zeta)^4
\over (1-\zeta)}+  \log \abs{{M_T^{eff} \over M_G}} \biggr\} . \eeq   Taking reasonable values of the VEVs given
by  \beq a_1 = 2 a_2 = 2 {\cal S}_4  = M_G \eeq and the effective color triplet Higgs mass \beq M_T^{eff} \sim
10^{19} \; {\rm GeV} \eeq we find \bea \zeta = 16   & \;\;\;\; \epsilon_3 \approx -0.030 . &  \eea  Note, the
large value of $M_T^{eff}$ is necessary to suppress proton decay rates as discussed in the following section.

\section{Nucleon Decay : Minimal SU(5) \ss GUT \label{sec:protons}}

Protons and neutrons [nucleons] are not stable particles; they necessarily decay in any GUT.   Super-Kamiokande
and Soudan II are looking for these decay products. The most recent (preliminary) Super-Kamiokande bounds on the
proton lifetime [Jung (2002)] are given in Table \ref{tab:proton}.  In the future, new detectors $\geq 10$ times
larger than Super-K have been proposed -- Hyper-Kamiokande in Japan and UNO in the USA.
\begin{table}
\caption{\label{tab:proton} Recent preliminary lower bounds on the proton lifetime into specific decay modes from
Super-Kamiokande [Jung (2002)].}
\begin{indented} \lineup \item[]
\begin{tabular}{@{}*{2}{l}} \br mode &$ \tau/{\rm B \;\; limit} \; [10^{33} \;{\rm yrs}]$\cr \mr
$p  \rightarrow  \pi^0 + e^+$   & $5.7 \;\; @ \; 90\% {\rm C.L.}$ \cr $p   \rightarrow  K^+ + \bar \nu$    & $1.9
\;\; @ \; 90\% {\rm C.L.}$ \cr \br
\end{tabular}
\end{indented}
\end{table}
Note, a generic, dimension 6 nucleon decay operator is given by a 4 Fermion operator of the form $\sim
(1/\Lambda^2) \;\; q \ q \ q \ l$.  Given the bound $ \tau_p
> 5 \times 10^{33} {\rm yrs} $  we find $ \Lambda > 4 \times 10^{15} {\rm GeV}$.  This is nice, since it is
roughly consistent with the GUT scale and with the See-Saw scale for neutrino masses.

In this section we consider nucleon decay in the Minimal SU(5) \ss GUT in more detail.    In minimal \ss SU(5),
we have the following gauge and Higgs sectors. The gauge sector includes the gauge bosons for SU(5) which
decompose, in the SM, to  SU(3) $\times$ SU(2) $\times$ U(1) plus the massive gauge bosons \{$X, \; \bar X$\}.
The $ \{X, \; \bar X\} $ bosons with charges \{$(3, \bar 2, -5/3), \; (\bar 3, 2, 5/3)$\} are responsible for
nucleon decay. The minimal SU(5) theory has, by definition, the minimal Higgs sector.  It includes a single
adjoint of SU(5), $\Sigma \subset$ 24, for the GUT breaking sector and the electroweak Higgs sector (Eqn.
\ref{eq:higgs})
$$ \left(
\begin{array}{c}
H_u \\ T \end{array} \right),\;\; \left( \begin{array}{c} H_d \\
\bar T \end{array} \right) \;\; \subset \;\; {\bf 5_H},\;\; {\bf \bar 5_H} .$$  The superpotential for the GUT
breaking and Higgs sectors of the model is given by [Witten (1981), Dimopoulos and Georgi (1981), Sakai (1981)]
\beq W = \frac{\lambda}{3} \Tr \Sigma^3 + \frac{\lambda V}{2} \Tr \Sigma^2 + \bar H ( \Sigma + 3V ) H . \eeq  In
general, nucleon decay can have contributions from operators with dimensions 4, 5 and 6.

\subsection{Dimension 6 operators}

\begin{figure}[t!]
\vspace*{0.50in} \hspace*{0.80in}
\begin{center}
\hspace*{0.75in}
\begin{minipage}{4.00in}
\epsfig{file=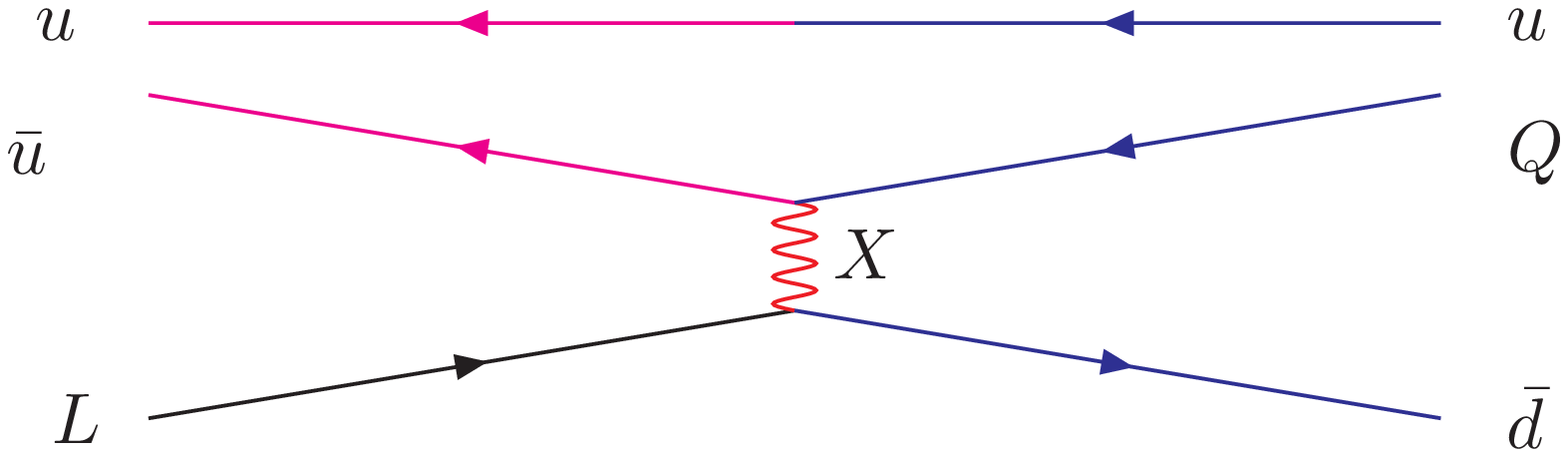,width=4.00in}
\end{minipage}
\end{center}
%\vspace*{-0.00in}
\begin{center} \caption{\label{fig:dim6op} X boson exchange diagram giving the dimension 6 four fermion operator
for proton and neutron decay.}
\end{center}
\end{figure}

The dimension 6 operators are derived from gauge boson exchange (see Fig. \ref{fig:dim6op}).  We obtain the
effective dimension 6 (four Fermion) operator given by \beq \frac{g_G^2}{2 M_X^2} \; \bar u^\dagger \ Q \ \bar
d^\dagger \ L . \eeq  Thus the decay amplitude is suppressed by one power of $1/M_X^2$.  How is $M_X$ related to
the GUT scale $M_G$ determined by gauge coupling unification?  Recall, in general we have \beq \epsilon_3 =
\frac{3 \alpha_G}{5 \pi} \ln \frac{M_T}{M_G} + \epsilon_3(M_X, M_\Sigma) . \eeq    However in minimal SU(5) we
find  \beq \epsilon_3(M_X, M_\Sigma) \equiv 0 . \eeq   Thus gauge coupling unification fixes the values of the
three parameters, \{$\alpha_G (\equiv g_G^2/4 \pi), \ M_G, \ M_T$\}.   In addition, the $\alpha_1(M_G) =
\alpha_2(M_G)$ condition gives the relation \beq (\frac{M_G}{M_T})^{1/15} = \frac{M_G}{(M_X^2 M_\Sigma)^{1/3}}
\sim \frac{M_G}{(g_G^2 \lambda)^{1/3} \langle \Sigma \rangle} .  \eeq  In the last term we used the approximate
relations \beq M_X \sim g_G \ \langle \Sigma \rangle,  \;\;\; M_\Sigma \sim \lambda \ \langle \Sigma \rangle .
\eeq   Hence the natural values for these parameters are given by
 \beq M_X \sim M_\Sigma \sim M_G \approx 3 \times 10^{16} \; {\rm GeV} . \eeq  As a result, the proton lifetime
 is given by
\beq \tau_p \sim 5 \times 10^{36} \; (\frac{M_X}{3 \times 10^{16} \ {\rm GeV}})^4 \; (\frac{0.015 \ {\rm
GeV}^3}{\beta_{lattice}})^2 \; {\rm yrs.} \eeq   and the dominant decay mode is
 \beq p \rightarrow \pi^0 + e^+ . \eeq   Note it is not possible to enhance the decay rate by taking
 $M_X \ll M_G$ without spoiling perturbativity, since this limit requires $\lambda \gg 1$.  On the other hand,
 $M_X \gg M_G$ is allowed.

\subsection{Dimension 4 \& 5 operators}

The contribution of dimension 4 \& 5 operators to nucleon decay in \gs was noted by [Weinberg (1982); Sakai and
Yanagida (1982)].   Dimension 4 operators are dangerous.   In \gs they always appear in the combination \beq (U^c
\; D^c \; D^c)  +  (Q \; L \; D^c)  +  (E^c \; L \; L) \eeq leading to unacceptable nucleon decay rates.  R
parity [Farrar and Fayet (1979)] forbids all dimension 3 and 4 (and even one dimension 5) baryon and lepton
number violating operators. It is thus a necessary ingredient of any ``natural" SUSY GUT.

Dimension 5 operators are obtained when integrating out heavy color triplet Higgs fields.   \bea W \supset & H_u
\; Q Y_u \overline U  +  H_d ( Q Y_d \overline D + L Y_e \overline E)  + & \label{eq:yukawa} \\ & T ( Q {1\over
2} c_{qq} Q + \overline U c_{ue} \overline E ) + \bar T ( Q c_{ql} L + \overline U c_{ud} \overline D ) & \nn
\eea   If the color triplet Higgs fields in Eqn. \ref{eq:yukawa} \{$T, \; \bar T$\} have an effective mass term
$M^{eff}_T \ \bar T \ T$ we obtain the dimension 5 operators \bea (1/M^{eff}_T) \; \left[ \ Q {1\over 2} c_{qq} Q
\ Q c_{ql} L + \overline U c_{ud} \overline D \ \overline U c_{ue} \overline E \ \right] , \label{eq:dim5op} \eea
denoted $L L L L$ and $R R R R$ operators, respectively.  Nucleon decay via dimension 5 operators was considered
by [Sakai and Yanagida (1982), Dimopoulos \etal (1982), Ellis \etal (1982)].
\begin{figure}[t!]
\vspace*{0.50in} \hspace*{0.80in}
\begin{center}
\hspace*{0.75in}
\begin{minipage}{4.00in}
\epsfig{file=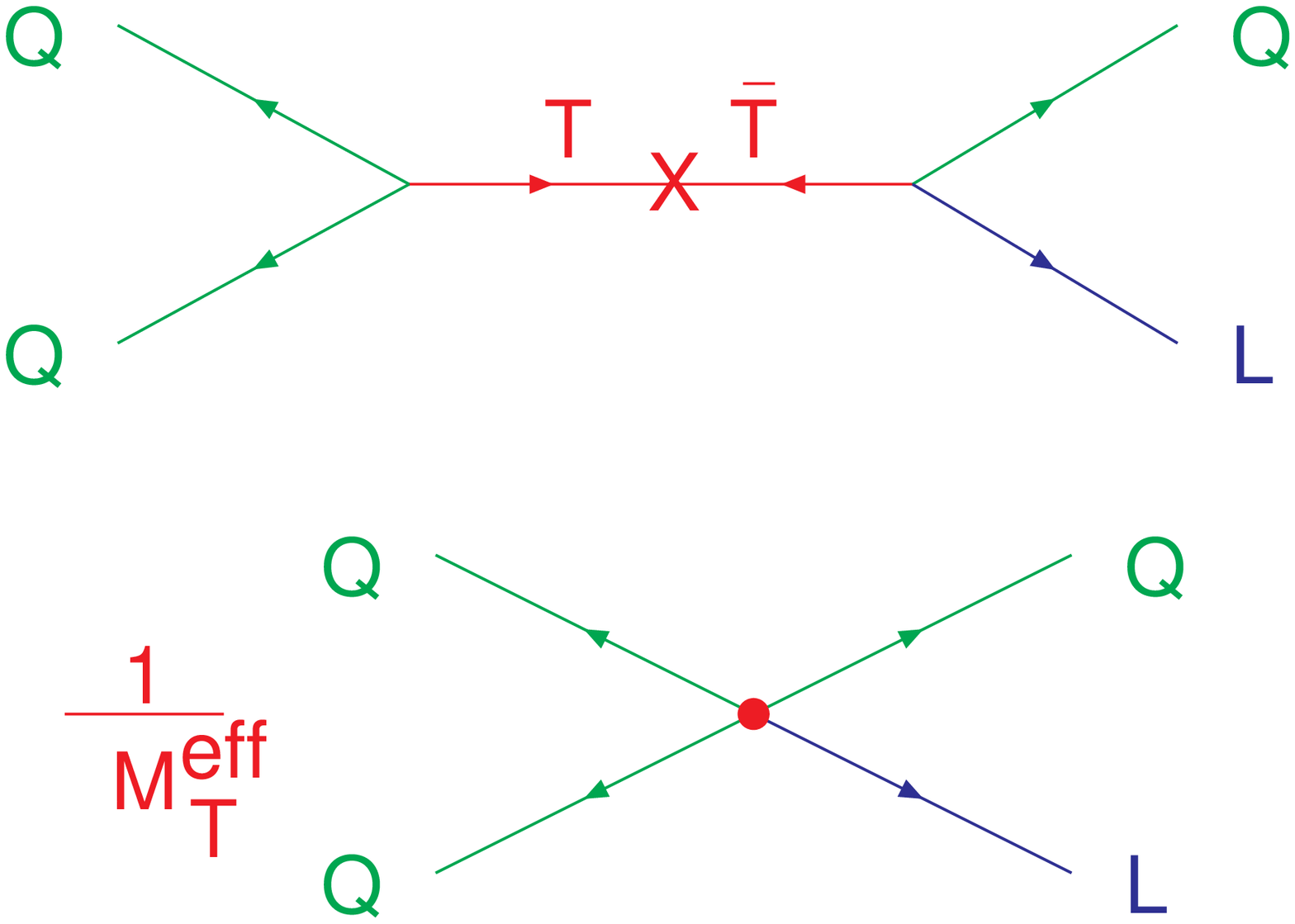,width=4.00in}
\end{minipage}
\end{center}
%\vspace*{-0.00in}
\begin{center} \caption{\label{fig:dim5op} Color triplet Higgs exchange diagram giving the dimension 5 superpotential
operator for proton and neutron decay.}
\end{center}
\end{figure}
The proton decay amplitude is then given generically by the expression \bea T(p \rightarrow K^+ + \bar \nu) &
\propto \frac{c^2}{M^{eff}_T} \; ({\rm Loop \; Factor)} \; (RG) \; \langle K^+  \bar \nu | q q q l | p \rangle &
\\ & \sim \frac{c^2}{M^{eff}_T} \; {\rm (Loop \; Factor)} \; (RG) \; \frac{\beta_{lattice}}{f_\pi} \ m_p . & \nn
\eea  The last step used a chiral Lagrangian analysis to remove the $K^+$ state in favor of the vacuum state. Now
we only need calculate the matrix element of a three quark operator between the proton and vacuum states.  This
defines the parameter $\beta_{lattice}$ using lattice QCD calculations.   The decay amplitude includes four
independent factors: \begin{enumerate} \item $\beta_{lattice}$, the three quark matrix element, \item $c^2$, a
product of two $3 \times 3$ dimensionless coupling constant matrices, \item a Loop Factor, which depends on the
SUSY breaking squark, slepton and gaugino masses, and \item $M^{eff}_T$, the effective color triplet Higgs mass
which is subject to the GUT breaking sector of the theory. \end{enumerate} Let us now consider each of these
factors in detail.

\subsubsection{$\beta_{lattice}$}

The strong interaction matrix element of the relevant three quark operators taken between the nucleon and the
appropriate pseudo-scalar meson may be obtained directly using lattice techniques.  However these results have
only been obtained recently [Aoki \etal (JLQCD) (2000), Aoki \etal (RBC) (2002)].  Alternatively, chiral
Lagrangian techniques [Chadha \etal (1983)] are used to replace the pseudo-scalar meson by the vacuum.  Then the
following three quark matrix elements are needed.   \beq \beta U({\bf k}) =\epsilon_{\alpha \beta \gamma}
\smash{<}0 | (u^\alpha d^\beta) u^\gamma |{\rm proton}({\bf k})\smash{>} ,\eeq \beq \alpha U({\bf k})
=\epsilon_{\alpha \beta \gamma} \smash{<}0 |(\overline{\vphantom{d}u}^{*\,\alpha} \overline d^{*\,\beta})
u^\gamma |{\rm proton}({\bf k})\smash{>} \eeq and $U({\bf k})$ is the left handed component of the proton's
wavefunction.  It has been known for some time that $|\beta| \approx |\alpha|$ [Brodsky \etal (1984), Gavela
\etal (1989)] and that $|\beta|$ ranges from .003 to .03 GeV$^3$ [Brodsky \etal (1984), Hara \etal (1986), Gavela
\etal (1989)]. Until quite recently, lattice calculations did not reduce the uncertainty in $|\beta|$; lattice
calculations have reported $|\beta|$ as low as .006 GeV$^3$ [Gavela \etal (1989)] and as high as .03 GeV$^3$
[Hara \etal (1986)]. Additionally, the phase between $\alpha$ and $\beta$ satisfies $\beta \approx -\alpha$
[Gavela \etal (1989)]. As a consequence, when calculating nucleon decay rates most authors have chosen to use a
conservative lower bound with $|\alpha| \sim |\beta| = 0.003$ GeV$^3$ and an arbitrary relative phase.

Recent lattice calculations [Aoki \etal (JLQCD) (2000), Aoki \etal (RBC) (2002)] have obtained significantly
improved results.   In addition, they have compared the direct calculation of the three quark matrix element
between the nucleon and pseudo-scalar meson to the indirect chiral Lagrangian analysis with the three quark
matrix element between the nucleon and vacuum.  [Aoki \etal (JLQCD) (2000)] find \beq \beta_{lattice} =  \langle
0 | q q q | N \rangle  = 0.015 (1) \; {\rm GeV}^3 . \eeq   Also [Aoki \etal (RBC) (2002)], in preliminary results
reported in conference proceedings, obtained \beq \beta_{lattice} =  0.007 (1) \; {\rm GeV}^3 . \eeq   They both
find \beq \alpha_{lattice} \approx - \beta_{lattice} . \eeq    Several comments are in order.  The previous
theoretical range $0.003 {\rm GeV}^3 < \beta_{lattice} < 0.03 {\rm GeV}^3$ has been significantly reduced and the
relative phase between $\alpha$ and $\beta$ has been confirmed.  The JLQCD central value is 5 times larger than
the previous ``conservative lower bound."   Although the new, preliminary, RBC result is a factor of 2 smaller
than that of JLQCD.  We will have to wait for further results.   What about the uncertainties?  The error bars
listed are only statistical.  Systematic uncertainties (quenched + chiral Lagrangian) are likely to be of order
$\pm 50$ \%  (my estimate).     This stems from the fact that errors due to quenching are characteristically of
order 30 \%, while the comparison of the chiral Lagrangian results to the direct calculation of the decay
amplitudes agree to within about 20 \%, depending on the particular final state meson.

\subsubsection{$c^2$ - Model Dependence}

Consider the quark and lepton Yukawa couplings in SU(5) --  \beq \lambda(\langle \Phi \rangle) \; 10 \; 10 \; 5_H
+ \lambda^\prime(\langle \Phi \rangle) \; 10 \; \bar 5 \; \bar 5_H  \eeq or in SO(10) -- \beq \lambda(\langle
\Phi \rangle) \; 16 \; 16 \; 10_H . \eeq   The Yukawa couplings \beq \lambda(\langle \Phi \rangle), \;\;
\lambda^\prime(\langle \Phi \rangle) \eeq are effective higher dimensional operators, functions of adjoint
($\Phi$) (or higher dimensional) representations of SU(5) (or SO(10)).   The adjoint representations are
necessarily there in order to correct the unsuccessful predictions of minimal SU(5) (or SO(10)) and to generate a
hierarchy of fermion masses.\footnote{Effective higher dimensional operators may be replaced by Higgs in higher
dimensional representations, such as 45 of SU(5) or 120 and 126 or SO(10).  Using these Higgs representations,
however, does not by itself address the fermion mass hierarchy.}   Once the adjoint (or higher dimensional)
representations obtain VEVs ($\langle \Phi \rangle$), we find the Higgs Yukawa couplings -- \beq H_u \; Q \ Y_u \
\overline U  + H_d \ ( \ Q \ Y_d \ \overline D + L \ Y_e \ \overline E)    \eeq  and also the effective dimension
five operators \beq (1/M^{eff}_T) \ \left[ \ Q \ {1\over 2} c_{qq} \ Q \, Q \ c_{ql} \ L + \overline U \ c_{ud} \
\overline D \ \overline U \ c_{ue} \ \overline E \ \right] . \eeq

Note, because of the Clebsch relations due to the VEVs of the adjoint representations, etc, we have \beq Y_u \neq
c_{qq} \neq c_{ue} \eeq and \beq Y_d \neq Y_e \neq c_{ud} \neq c_{ql} . \eeq  Hence, the $3 \times 3$ complex
matrices entering nucleon decay are not the same $3 \times 3$ Yukawa matrices entering fermion masses.  Is this
complication absolutely necessary and how large can the difference be?  Consider the SU(5) relation -- \beq
\lambda_b = \lambda_\tau \eeq [Einhorn and Jones (1982), Inoue \etal (1982), Iba\~{n}ez and Lopez (1984)]. It is
known to work quite well for small $\tan\beta \sim 1 - 2$ or large $\tan\beta \sim 50$ [Dimopoulos \etal (1992),
Barger \etal (1993)]. For a recent discussion see [Barr and Dorsner (2003)]. On the other hand, the same relation
for the first two families gives \bea & \lambda_s = \lambda_\mu &
\\ & \lambda_d = \lambda_e & \nn \eea leading to the unsuccessful relation \beq 20 \sim \frac{m_s}{m_d} =
\frac{m_\mu}{m_e} \sim 200 . \eeq  This bad relation can be corrected using Higgs multiplets in higher
dimensional representations [Georgi and Jarlskog (1979), Georgi and Nanopoulos (1979), Harvey \etal (1980,1982)]
or using effective higher dimensional operators [Anderson \etal (1994)].  Clearly the corrections to the simple
SU(5) relation for Yukawa and c matrices can be an order of magnitude.   Nevertheless, in predictive \gs the c
matrices are obtained once the fermion masses and mixing angles are fit [Kaplan and Schmaltz (1994), Babu and
Mohapatra (1995), Lucas and Raby (1996), Frampton and Kong (1996), Bla\v{z}ek \etal (1997), Barbieri and Hall
(1997), Barbieri \etal (1997), Allanach \etal (1997), Berezhiani (1998), Bla\v{z}ek \etal (1999,2000), Derm\' \i
\v sek and Raby (2000), Shafi and Tavartkiladze (2000), Albright and Barr (2000,2001), Altarelli \etal (2000),
Babu \etal (2000), Berezhiani and Rossi (2001), Kitano and Mimura (2001), Maekawa (2001), King and Ross (2003),
Chen and Mahanthappa (2003), Raby (2003), Ross and Velasco-Sevilla (2003), Goh \etal (2003), Aulakh \etal
(2003)].  In spite of the above cautionary remarks we still find the inexact relations \beq c_{qq} \ c_{ql}, \;\;
c_{ud} \ c_{ue} \ \propto \ m_u \ m_d \ \tan\beta . \eeq In addition, family symmetries can affect the texture of
\{$ c_{qq}, \ c_{ql}, \ c_{ud}, \ c_{ue} $\}, just as it will affect the texture of Yukawa matrices.

In order to make predictions for nucleon decay it is necessary to follow these simple steps.  Vary the GUT scale
parameters, $\tilde \alpha_G, \; M_G$, $Y_u, \; Y_d, \; Y_e$ and SOFT SUSY breaking parameters until one obtains
a good fit to the precision electroweak data.  Whereby we now explicitly include fermion masses and mixing angles
in the category of precision data.   Once these parameters are fit, then in any predictive \ss GUT the matrices $
c_{qq}, \ c_{ql}, \; c_{ud}, \ c_{ue} $ at $M_G$ are also fixed.  Now renormalize the dimension 5 baryon and
lepton number violating operators from $M_G \rightarrow M_Z$ in the MSSM;  evaluate the Loop Factor at $M_Z$ and
renormalize the dimension 6 operators from $M_Z \rightarrow 1$ GeV.  The latter determines the renormalization
constant $A_3 \sim 1.3$ [Derm\' \i \v sek \etal (2001)]. [Note, this should not be confused with the
renormalization factor $A_L \sim .22$ [Ellis \etal (1982)] which is used when one does not have a theory for
Yukawa matrices. The latter RG factor, takes into account the combined renormalization of the dimension 6
operator from the weak scale to 1 GeV and also the renormalization of fermion masses from 1 GeV to the weak
scale.]  Finally calculate decay amplitudes using a chiral Lagrangian approach or direct lattice gauge
calculation.

Before leaving this section we should remark that we have assumed that the electroweak Higgs in SO(10) models is
contained solely in a 10.  If however the electroweak Higgs is a mixture of weak doublets from $16_H, \; 10_H$
and, in addition, we include the higher dimensional operator $\frac{1}{M} (16 \ 16 \ 16_H \ 16_H)$, useful for
neutrino masses, then there are additional contributions to the dimension 5 operators considered in (Eqn.
\ref{eq:dim5op}) [Babu \etal (2000)].  However these additional terms are not required for neutrino masses
[Bla\v{z}ek \etal (1999,2000)].

\subsection{Loop factor \label{sec:loopfactor}}

The dimension 5 operators are a product of two fermion and two scalar fields.  The scalar squarks and/or sleptons
must be integrated out of the theory below the \ss breaking scale.   There is no consensus on the best choice for
an appropriate \ss breaking scale.  Moreover, in many cases there is a hierarchy of \ss particle masses.  Hence
we take the simplest assumption, integrating out all \ss particles at $M_Z$.  When integrating out the \ss
particles in loops, the effective dimension 5 operators are converted to effective dimension 6 operators.   This
results in a loop factor which depends on the sparticle masses and mixing angles.

\begin{figure}[t!]
\vspace*{0.50in} \hspace*{0.80in}
\begin{center}
\hspace*{0.75in}
\begin{minipage}{4.00in}
\epsfig{file=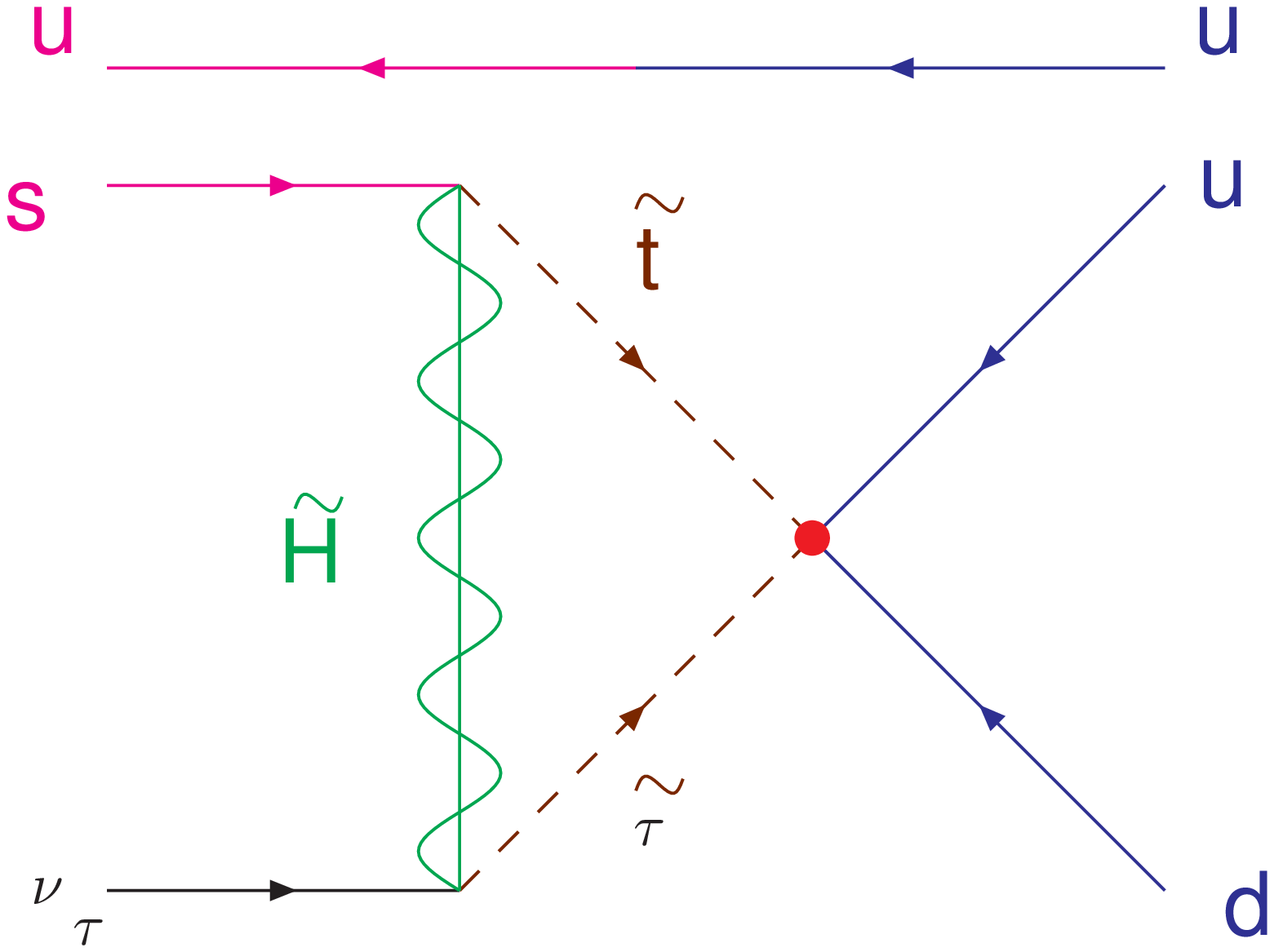,width=4.00in}
\end{minipage}
\end{center}
%\vspace*{-0.00in}
\begin{center} \caption{\label{fig:loopfactor} Higgsino and third generation stop - stau loop giving the
dominant loop contribution to dimension 5 nucleon decay.}
\end{center}
\end{figure}

Consider the contribution to the loop factor for the process $p \rightarrow K^+ + \bar \nu_\tau$ in Fig.
\ref{fig:loopfactor}.  This graph is due to the RRRR operators and gives the dominant contribution at large
$\tan\beta$ and a significant contribution for all values of $\tan\beta$ [Lucas and Raby (1997), Babu and
Strassler (1998), Goto and Nihei (1999), Murayama and Pierce (2002)].
Although the loop factor is a complicated function of the sparticle masses and mixing angles, it nevertheless has
the following simple dependence on the overall gaugino and scalar masses given by \beq \frac{\lambda_t \;
\lambda_\tau}{16 \pi^2} \frac{\sqrt{\mu^2 + M_{1/2}^2}}{m_{16}^2} . \eeq   Thus in order to minimize this factor
one needs  \beq \mu, \;  M_{1/2} \;\; \; {\rm SMALL} \eeq and \beq  m_{16} \;\;\; {\rm Large} . \eeq

\subsection{$M_t^{eff}$}

The largest uncertainty in the nucleon decay rate is due to the color triplet Higgs mass parameter $M^{eff}_T$.
As $M^{eff}_T$ increases, the nucleon lifetime increases.  Thus it is useful to obtain an upper bound on the
value of $M^{eff}_T$.  This constraint comes from imposing perturbative gauge coupling unification [Lucas and
Raby (1997), Goto and Nihei (1999), Babu \etal (2000), Altarelli \etal (2000), Derm\' \i \v sek \etal (2001),
Murayama and Pierce (2002)].  Recall, in order to fit the low energy data a GUT scale threshold correction \beq
\epsilon_3 \equiv \frac{(\alpha_3(M_G) - \tilde \alpha_G)}{\tilde \alpha_G} \sim  - 3\% \;\; {\rm to} \; - 4\%
\eeq is needed. $\epsilon_3$ is a logarithmic function of particle masses of order $M_G$, with contributions from
the electroweak Higgs and GUT breaking sectors of the theory.  \bea \epsilon_3 = & \epsilon_3^{\rm Higgs} +
\epsilon_3^{\rm GUT
\; breaking} + \cdots , &  \\
\epsilon_3^{\rm Higgs} = & \frac{3 \alpha_G}{5 \pi} \ln(\frac{M^{eff}_T}{M_G}) . &  \eea  In Table
\ref{t:threshold} we have analyzed three different GUT theories -- the minimal SU(5) model, an SU(5) model with
natural doublet-triplet splitting and minimal SO(10) (which also has natural doublet-triplet splitting).  We have
assumed that the low energy data, including weak scale threshold corrections, requires $ \epsilon_3 = - 4 \%$. We
have then calculated the contribution to $ \epsilon_3$ from the GUT breaking sector of the theory in each case.

Minimal SU(5) is defined by its minimal GUT breaking sector with one SU(5) adjoint $\Sigma$.  The one loop
contribution from this sector to $\epsilon_3$ vanishes. Hence, the $- 4$ \% must come from the Higgs sector
alone, requiring the color triplet Higgs mass $M^{eff}_T \sim 2 \times 10^{14}$ GeV. Note since the Higgs sector
is also minimal, with the doublet masses fine-tuned to vanish, we have $M^{eff}_T \equiv M_T$.  By varying the
\ss spectrum at the weak scale, we may be able to increase $\epsilon_3$ to $- 3$ \% or even $- 2$ \%, but this
cannot save minimal SU(5) from disaster due to rapid proton decay from Higgsino exchange.

In the other theories, Higgs doublet - triplet splitting is obtained without fine-tuning. This has two
significant consequences.   First, the GUT breaking sectors are more complicated, leading in these theories to
large negative contributions to $\epsilon_3$. The maximum value $|\epsilon_3| \sim 10 \%$ in minimal SO(10) is
fixed by perturbativity bounds [Derm\' \i \v sek \etal (2001)]. Secondly, the effective color triplet Higgs mass
$M^{eff}_T$ does not correspond to the mass of any particle in the theory. In fact, in both cases with ``natural"
doublet-triplet splitting, the color triplet Higgs mass is of order $M_G$ even though $M^{eff}_T \gg M_G$.   The
values for $M^{eff}_T$ in Table \ref{t:threshold} are fixed by the value of $\epsilon_3^{\rm Higgs}$ needed to
obtain $\epsilon_3 = - 4 \%$.
\begin{table}
\caption{\label{t:threshold} GUT threshold corrections in three different theories.  The first column is the
minimal SU(5) \ss theory [Dimopoulos and Georgi (1981) Sakai (1982), Goto and Nihei (1999), Murayama and Pierce
(2002)], the second column is SU(5) with ``natural" Higgs doublet-triplet splitting [Masiero \etal (1982),
Grinstein (1982), Altarelli \etal (2000)], and the third column is minimal SO(10) \ss model [Bla\v{z}ek \etal
(1999.2000), Derm\' \i \v sek \etal (2001), Bla\v{z}ek \etal (2002a,b)].}
\begin{indented}
\item[]\begin{tabular}{@{}lccc} \br
 & Minimal SU(5) & SU(5) ``Natural" D/T & Minimal SO(10)\\
\mr $\epsilon_3^{\rm GUT breaking}$ & 0 & - 7.7 \% & - 10 \% \\  & & & \\
$\epsilon_3^{\rm Higgs}$ & - 4 \% & + 3.7 \% & + 6 \% \\
\mr $M^{eff}_T \; {\rm [GeV]}$ & $2 \times 10^{14}$ & $3 \times 10^{18}$ & $6 \times 10^{19}$ \\
\br
\end{tabular}
\end{indented}
\end{table}

Before discussing the bounds on the proton lifetime due to the exchange of color triplet Higgsinos, let us
elaborate on the meaning of $M^{eff}_T$.   Consider a simple case with two pairs of SU(5) Higgs multiplets, \{
$\bar 5_H^i, \;\; 5_H^i$ \} with $i = 1,2$.   In addition, also assume that only \{ $\bar 5_H^1, \;\; 5_H^1$ \}
couples to quarks and leptons.   Then $M^{eff}_T$ is defined by the expression \beq \frac{1}{M^{eff}_T} = (
M_T^{-1} )_{11}  \eeq  where $M_T$ is the color triplet Higgs mass matrix.   In the cases with ``natural" doublet
- triplet splitting, we have \beq M_T = \left( \begin{array}{cc} 0 &  M_G \\
M_G & X
\end{array} \right) \eeq  with   \beq \frac{1}{M^{eff}_T} \equiv \frac{X}{M_G^2} . \eeq
Thus for $X \ll M_G$ we have $M^{eff}_T \gg M_G$ and {\em no} particle has mass greater than $M_G$ [Babu and Barr
(1993)]. The large Higgs contribution to the GUT threshold correction is in fact due to an extra pair of light
Higgs doublets with mass of order $X$.

\begin{figure}[t!]
\vspace*{0.50in} \hspace*{0.80in}
\begin{center}
\hspace*{-1.00in}
\begin{minipage}{4.00in}
\epsfig{file=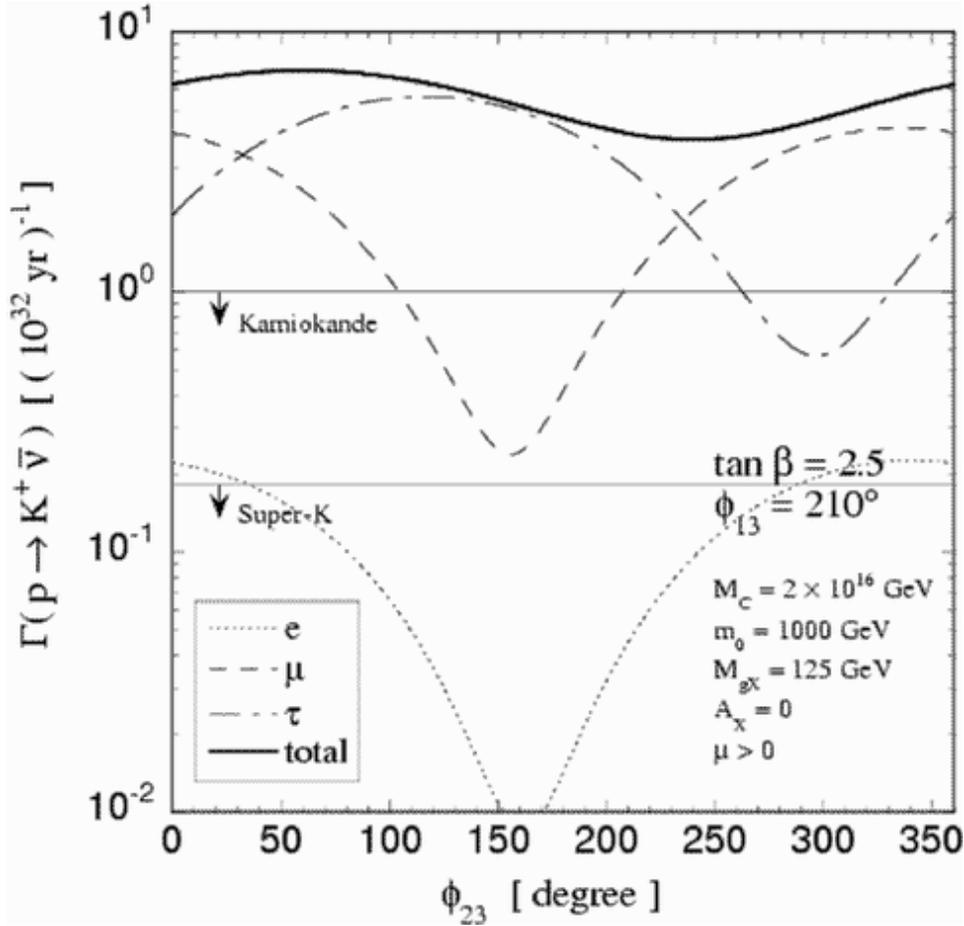,width=5.00in}
\end{minipage}
\end{center}
%\vspace*{-0.00in}
\begin{center} \caption{\label{fig:gotofig2} [Fig. 2: Goto and Nihei (1999)]  Decay rates $\Gamma(p \rightarrow K^+
\bar \nu_i) \; (i = e, \mu \; {\rm and} \; \tau)$ as functions of the phase $\phi_{23}$ for $\tan\beta = 2.5$.
The soft \ss breaking parameters are fixed at $m_0 = 1$ TeV, $M_{\tilde g} = 125$ GeV and $A_0 = 0$ with $\mu >
0$. $M_T$ and $M_\Sigma$ are taken as $M_T = M_\Sigma = 2 \times 10^{16}$ GeV.  The horizontal lower line
corresponds to the Super-Kamiokande limit $\tau(p \rightarrow K^+ \bar \nu) > 5.5 \times 10^{32}$ years, and the
horizontal upper line corresponds to the Kamiokande limit $\tau(p \rightarrow K^+ \bar \nu) > 1.0 \times 10^{32}$
years. The new Super-Kamiokande bound is $\tau(p \rightarrow K^+ \bar \nu) > 1.9 \times 10^{33}$ years.}
\end{center}
\end{figure}

\begin{figure}[t!]
\vspace*{0.50in} \hspace*{0.80in}
\begin{center}
\hspace*{-1.00in}
\begin{minipage}{4.00in}
\epsfig{file=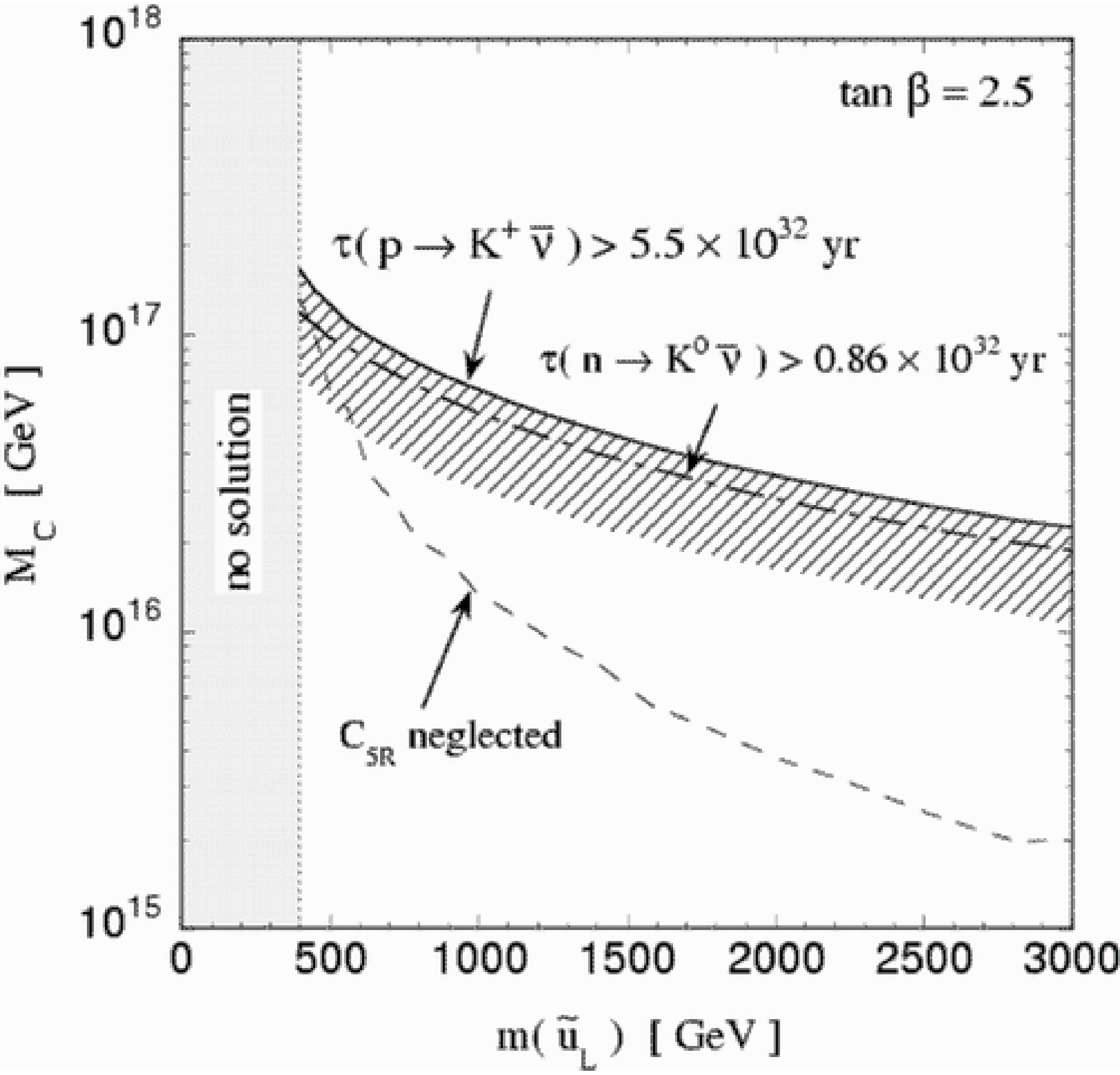,width=5.00in}
\end{minipage}
\end{center}
%\vspace*{-0.00in}
\begin{center} \caption{\label{fig:gotofig4}[Fig. 4: Goto and Nihei (1999)]  Lower bound on the colored Higgs mass $M_T$ as
a function of the left-handed scalar up quark mass $m_{\tilde u_L}$.  The soft \ss breaking parameters $m_0, \;
M_{\tilde g}, \; A_0$ are scanned within the range $0 < m_0 < 3$ TeV, $0 < M_{\tilde g} < 1$ TeV and $-5 < A_0 <
5$ with $\tan\beta$ fixed at 2.5. Both signs of $\mu$ are considered. The whole parameter region of the two
phases $\phi_{13}$ and $\phi_{23}$ is examined. The solid curve represents the bound derived from the
Super-Kamiokande limit $\tau(p \rightarrow K^+ \bar \nu) > 5.5 \times 10^{32}$ years and the dashed curve
represents the corresponding result without the RRRR effect.  The left-hand side of the vertical dotted line is
excluded by other experimental constraints.  The dash-dotted curve represents the bound derived from the
Kamiokande limit on the neutron partial lifetime $\tau(n \rightarrow K^0 \bar \nu) > 0.86 \times 10^{32}$ years.}
\end{center}
\end{figure}

Due to the light color triplet Higgsino, it has been shown that minimal SUSY SU(5) is ruled out by the
combination of proton decay constrained by gauge coupling unification [Goto and Nihei (1999), Murayama and Pierce
(2002)] !!   In Figs. \ref{fig:gotofig2} and \ref{fig:gotofig4} we reprint the figures from the paper by [Goto
and Nihei (1999)].   In Fig. \ref{fig:gotofig2} the decay rate for $p \rightarrow K^+ \bar \nu_i$ for any one of
the three neutrinos  $(i = e, \mu \; {\rm and} \; \tau)$ is plotted for fixed soft \ss breaking parameters as a
function of the relative phase $\phi_{23}$ between two LLLL contributions to the decay amplitude.  The phase
$\phi_{13}$ is the relative phase between one of the LLLL contributions and the RRRR contribution.  The latter
contributes predominantly to the $\bar \nu_\tau$ final state, since it is proportional to the up quark and
charged lepton Yukawa couplings.   As noted by [Goto and Nihei (1999)], the partial cancellation between LLLL
contributions to the decay rate is completely filled by the RRRR contribution.  It is this result which provides
the stringent limit on minimal \ss SU(5).  As one sees from Fig. \ref{fig:gotofig2}, for the color triplet Higgs
mass $M_T = 2 \times 10^{16}$ GeV ($\equiv M_C$ in the notation of [Goto and Nihei (1999)]), the universal scalar
mass $m_0 =$ 1 TeV and $\tan\beta = 2.5$, there is no value of the phase $\phi_{23}$ which is consistent with
Super-Kamiokande bounds.   Note, the proton decay rate scales as $\tan\beta^2$; hence the disagreement with data
only gets worse as $\tan\beta$ increases.   In Fig. \ref{fig:gotofig4} the contour of constant proton lifetime is
plotted in the $M_T (\equiv M_C)$ -- $m(\tilde u_L)$ plane, where $m(\tilde u_L)$ is the mass of the left-handed
up squark for $\tan\beta = 2.5$.  Again, there is no value of $m(\tilde u_L) < 3$ TeV for which the color triplet
Higgs mass is consistent with gauge coupling unification.   In [Goto and Nihei (1999)] the up squark mass was
increased by increasing $m_0$.   Hence all squarks and slepton masses increased.

One may ask whether one can suppress the proton decay rate by increasing the mass of the squarks and sleptons of
the first and second generation, while keeping the third generation squarks and sleptons light (in order to
preserve ``naturalness"). This is the question addressed by [Murayama and Pierce (2002)].   They took the first
and second generation scalar masses of order 10 TeV, with the third generation scalar masses less than 1 TeV.
They showed that since the RRRR contribution does not decouple in this limit, and moreover since any possible
cancellation between the LLLL and RRRR diagrams vanishes in this limit,  one finds that minimal \ss SU(5) cannot
be saved by decoupling the first two generations of squarks and sleptons.

Thus minimal \ss SU(5) is dead.  Is this something we should be concerned about.  In my opinion, the answer is
no, although others may disagree [Bajc \etal (2002)].   Minimal \ss SU(5) has two a priori unsatisfactory
features:
\begin{itemize} \item It requires fine-tuning for Higgs doublet-triplet splitting, and \item renormalizable
Yukawa couplings due to $5_H, \; \bar 5_H$ alone are not consistent with fermion masses and mixing angles.
\end{itemize}    Thus it was clear from the beginning that two crucial ingredients of a realistic theory were
missing.    The theories which work much better have ``natural" doublet-triplet splitting and fit fermion masses
and mixing angles.

\subsection{Summary of Nucleon Decay in 4D}

Minimal \ss SU(5) is excluded by the concordance of experimental bounds on proton decay and gauge coupling
unification.   We discussed the different factors entering the proton decay amplitude due to dimension 5
operators. \beq T(p \rightarrow K^+ + \bar \nu) \sim  \frac{c^2}{M^{eff}_T} {\rm (Loop \; Factor)} \;
\frac{\beta_{lattice}}{f_\pi} \ m_p . \eeq We find
\begin{itemize}
\item $c^2$: model dependent but constrained by fermion masses and mixing angles;

\item $\beta_{lattice}$:  JLQCD central value is 5 times larger than the previous ``conservative lower bound."
However one still needs to reduce the systematic uncertainties of quenching and chiral Lagrangian analyses.
Moreover, the new RBC result is a factor of 2 smaller than JLQCD;

\item Loop Factor: $\propto \frac{\lambda_t \ \lambda_\tau}{16 \pi^2} \frac{\sqrt{\mu^2 + M_{1/2}^2}}{m_{16}^2}$.
It is minimized by taking gauginos light and the 1st and 2nd generation squarks and sleptons heavy ($>$ TeV).
However, ``naturalness" requires that the stop, sbottom and stau masses remain less than of order 1 TeV;

\item  $M^{eff}_T$:  constrained by gauge coupling unification and GUT breaking sectors.

\end{itemize}

The {\em bottom line} we find for dimension 6 operators [Lucas and Raby (1997),  Murayama and Pierce (2002)] \beq
\tau_(p \rightarrow \pi^0 + e^+) \approx 5 \times 10^{36} \; (\frac{M_X}{3 \times 10^{16} \ {\rm GeV}})^4 \;
(\frac{0.015 \ {\rm GeV}^3}{\beta_{lattice}})^2 \;\; {\rm years} . \eeq   Note, it has been recently shown
[Klebanov and Witten (2003)] that string theory can possibly provide a small enhancement of the dimension 6
operators.  Unfortunately the enhancement is very small.   Thus it is very unlikely that these dimension 6 decay
modes $p \rightarrow \pi^0 + e^+$ will be observed.

On the other hand for dimension 5 operators in realistic \gs we obtain rough upper bounds on the proton lifetime
coming from gauge coupling unification and perturbativity [Babu \etal (2000), Altarelli \etal (2000), Derm\' \i
\v sek \etal (2001)] \beq \tau(p \rightarrow K^+ + \bar \nu) < (\frac{1}{3} - 3) \times 10^{34} \; (\frac{0.015 \
{\rm GeV}^3}{\beta_{lattice}})^2 \;\; {\rm years} . \eeq  Note in general \beq \tau(n \rightarrow K^0 + \bar \nu)
< \tau(p \rightarrow K^+ + \bar \nu) . \eeq   Moreover other decay modes may be significant,  {\em but they are
very model dependent}, for example [Carone \etal (1996), Babu \etal (2000)] \beq p \rightarrow  \pi^0 + e^+, \;\;
K^0 + \mu^+ . \eeq

\subsection{Proton decay in more than four dimensions \label{sec:extra}}

We should mention that there has been a recent flurry of activity on \gs in extra dimensions beginning with the
work of [Kawamura (2001a,b)].  However the study of extra dimensions on orbifolds goes back to the original work
of [Dixon \etal (1985,1986)] in string theory.  Although this interesting topic would require another review, let
me just mention some pertinent features here. In these scenarios, grand unification is only a symmetry in extra
dimensions which are then compactified at scales of order $1/M_G$. The effective four dimensional theory,
obtained by orbifolding the extra dimensions, has only the standard model gauge symmetry or at most a Pati-Salam
symmetry which is then broken by the standard Higgs mechanism.   In these theories, it is possible to completely
eliminate the contribution of dimension 5 operators to nucleon decay. This may be a consequence of global
symmetries as shown by [Witten (2002), Dine \etal (2002)] or a continuous U(1)$_R$ symmetry (with R parity a
discrete subgroup)[Hall and Nomura (2002)]. [Note, it is also possible to eliminate the contribution of dimension
5 operators in 4 dimensional theories with extra symmetries [Babu and Barr (2002)], but these 4 dimensional
theories are quite convoluted. Thus it is difficult to imagine that nature takes this route. On the other hand,
in one or more small extra dimensions the elimination of dimension 5 operators is very natural.] Thus at first
glance, nucleon decay in these theories may be extremely difficult to see.   However this is not necessarily the
case.   Once again we must consider the consequences of grand unification in extra dimensions and gauge coupling
unification.

Extra dimensional theories are non-renormalizable and therefore require an explicit cutoff scale $M_*$, assumed
to be larger than the compactification scale $M_c$.  The Ka\l uza-Klein excitations above $M_c$ contribute to
threshold corrections to gauge coupling unification evaluated at the compactification scale.  The one loop
renormalization of gauge couplings is given by \bea \f{2\pi}{\alpha_i (\mu)} & = & \f{2\pi}{\alpha (M_*)} + b_i
\log ( \frac{M_c}{\mu}) + \Delta_i \label{eq:rge} \eea where $\Delta_i$ are the threshold corrections due to all
the KK modes from $M_c$ to $M_*$ and can be expressed as $\Delta_i = b^{eff}_i \log (\frac{M_*}{M_c})$ [Hall and
Nomura (2001,2002a), Nomura \etal (2001), Contino \etal (2002), Nomura (2002)]. In a 5D SO(10) model, broken to
Pati-Salam by orbifolding and to the MSSM via Higgs VEVs on the brane, it was shown [Kim and Raby (2003)] that
the KK threshold corrections take a particularly simple form \bea \f{2\pi}{\alpha_i (\mu)} & = & \f{2\pi}{\alpha
(M_*)} +
b^{MSSM}_i \log ( \frac{\hat M_c}{\mu}) + \hat \Delta_i \\
& = & \f{2\pi}{\alpha (M_*)} + b^{MSSM}_i \log ( \frac{\hat M_c}{\mu}) + \f{2}{3} b^{SM}_i (V) \log (\f{M_*}{\hat
M_c}). \label{5d} \eea  with \bea \hat {\Delta}_{\rm gauge} & =
& \f{2}{3} b^{SM}_i (V) \log (\f{M_*}{\hat M_c}) \\
\hat {\Delta}_{Higgs}  & = & 0, \eea and $\hat M_c \equiv (\pi R/2)^{-1}$.  Here $b^{MSSM}$ includes the gauge
sector and the Higgs sector together and $b^{SM}(V)$ includes the gauge sector only. The running equation is very
simple and permits us to directly compare with well known 4D SUSY GUTs. In the minimal 4D SU(5) \ss model, the
running equation is given by \bea \f{2\pi}{\alpha_i (\mu)} & = & \f{2\pi}{\alpha (M_{GUT})} + b^{MSSM}_i \log (
\frac{M_T}{\mu}) + b^{SM}_i (V) \log (\f{M_{GUT}}{M_T}) . \label{4d} \eea where $M_T$ is the color triplet Higgs
mass.  As discussed earlier, we can achieve unification by adjusting the color triplet Higgs mass $M_T = 2 \times
10^{14} \GeV$ (see Table \ref{t:threshold}).

Comparing Eqn. (\ref{5d}) and (\ref{4d}), we observe that if \bea
 \hat M_c =  M_T, \,\,\,\,\,
 \f{M_*}{\hat M_c}  =
(\f{M_{GUT}}{M_T})^{\f{3}{2}}, \nn \eea the same unification is achieved here. Therefore we get $\hat M_c = 2
\times 10^{14} \GeV$ and $M_* = 3.7 \times 10^{17} \GeV$ by using $M_{GUT} = 3 \times 10^{16} \GeV$.   A few
remarks are in order. There is no problem with proton decay due to dimension 5 operators, even though the color
triplet Higgs mass is of order $10^{14} \GeV$, since these operators are excluded by R symmetry.  In a 5D model,
the 4D GUT scale has no fundamental significance.   The couplings unify at the cutoff scale and there is no
scale, above which we have a perturbative SO(10) GUT.

\begin{figure}[t!] \vspace*{0.50in} \hspace*{0.80in}
\begin{center}
\hspace*{-1.00in}
\begin{minipage}{4.00in}
\epsfig{file=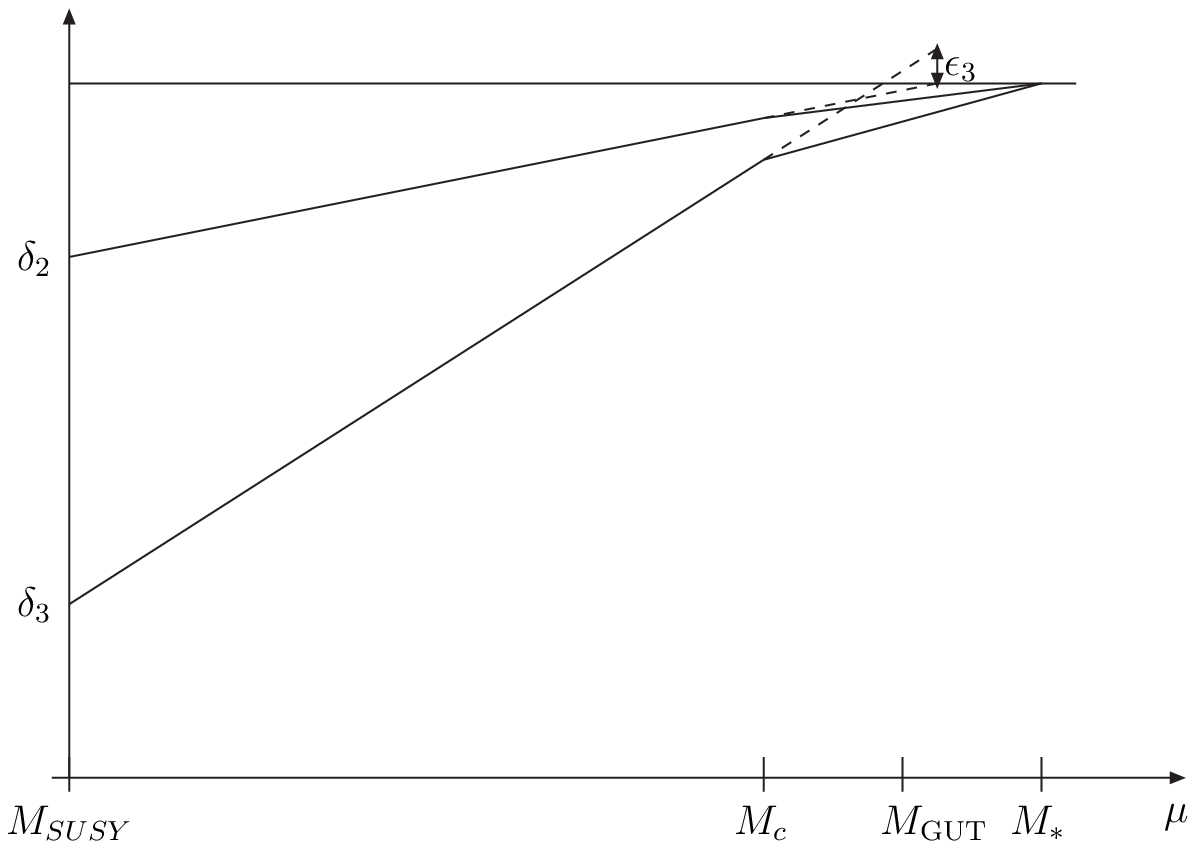,width=5.00in}
\end{minipage}
\end{center}
%\vspace*{-0.00in}
\begin{center} \caption{\label{fig:running}
Differential running of $\delta_2 = 2\pi (\frac{1}{\alpha_2} -\frac{1}{\alpha_1})$ and $\delta_3 = 2\pi
(\frac{1}{\alpha_3} -\frac{1}{\alpha_1})$.}
\end{center}
\end{figure}

In Fig. \ref{fig:running} we show the running in the difference of couplings for two independent cases.
\begin{enumerate} \item We show the couplings for four dimensional gauge theories with GUT scale thresholds, in
which the GUT scale is defined as the point where $\alpha_1$ and $\alpha_2$ meet and $\epsilon_3$ is the relative
shift in $\alpha_3$ due to threshold corrections, and \item for a five dimensional SO(10) model where the three
couplings meet at the cutoff scale and the threshold corrections due to the KK tower is defined at the
compactification scale. \end{enumerate}   In both cases, the running of the gauge couplings below the
compactification scale must be the same.   Thus we can use the low energy fits from 4D theories to constrain a 5D
theory.

Note, since the the KK modes of the baryon number violating \{$X, \ \bar X$\} gauge bosons have mass starting at
the compactification scale $M_c \approx 10^{14} \GeV$ we must worry whether proton decay due to dimension 6
operators is safe.  It has been shown [Altarelli and Feruglio (2001), Hall and Nomura (2002b), Hebecker and
March-Russell (2002)] that this depends on where in the extra dimensions the quarks and leptons reside.  If they
are on symmetric orbifold fixed points, i.e. symmetric under the GUT symmetry, then this leads to the standard
dimension 6 proton decay operators which is ruled out for the first and second families of quarks and leptons.
Hence the first two families must be either in the bulk or on broken symmetry fixed points.  If they are in the
bulk, then the \{$X, \ \bar X$\} mediated processes take massless modes to KK excitations which is not a problem.
Otherwise, if the first two families are on the broken symmetry fixed point, the wave functions for the \{$X, \
\bar X$\} bosons vanish there. However, certain effective higher dimensional operators on the broken symmetry
orbifold fixed points can allow the \{$X, \ \bar X$\} bosons to couple to the first two families. These operators
are allowed by symmetries and they naturally lead to proton lifetimes for $p \rightarrow e^+ \ \pi^0$ of order
$10^{34 \pm 2}$ years.   The large uncertainty is due to the order of magnitude uncertainty in the coefficient of
these new effective operators.

Before closing this section, we should make a few comments on theories with large extra dimensions of order 1/TeV
or as large as 1 mm.  In some of these theories, only gravity lives in higher dimensions while the ordinary
matter and gauge interactions typically reside on a three dimensional brane [Arkani-Hamed \etal (1998)].  While
if the extra dimensions are no larger than 1/TeV, all matter may live in the bulk.  Such theories replace SUSY
with new and fundamental non-perturbative physics at the 1 - 10 TeV scale. These theories must address the
question of why dimension 6 proton decay operators, suppressed only by 1/TeV$^2$, are not generated. There are
several ideas in the literature with suggested resolutions to this problem.  They include:
\begin{itemize} \item conserved baryon number on the brane with anomaly cancelling Chern-Simons terms in the
bulk, \item or displacing quarks from leptons on a ``fat" brane with a gaussian suppression of the overlap of the
quark/lepton wave functions. \end{itemize}   A novel solution to the problem of proton decay is found in 6
space-time dimensional theories with all matter spanning the two extra dimensions.   It has been shown that in
such theories [Appelquist \etal (2001)] a Z(8) remnant of the 6 dimensional Lorentz group is sufficient to
suppress proton decay to acceptable levels.   These theories predict very high dimensional proton decay operators
with multi-particle final states.

\section{Fermion masses and mixing \label{sec:flavor}}

Low energy \ss provides a natural framework for solving the gauge hierarchy problem, while \gs make the
successful prediction for gauge coupling unification and the, still un-verified, prediction for proton decay. But
these successes affect only a small subset of the unexplained arbitrary parameters in the standard model having
to do with the Z and Higgs masses (i.e. the weak scale) and the three gauge coupling constants.  On the other
hand, the sector of the standard model with the largest number of arbitrary parameters has to do with fermion
masses and mixing angles.  Grand unification also provides a natural framework for discussing the problem of
fermion masses, since it naturally arranges quarks and leptons into a few irreducible multiplets, thus explaining
their peculiar pattern of gauge charges, i.e. charge quantization and the family structure.   Moreover, it has
been realized for some time that the masses and mixing angles of quarks and leptons are ordered with respect to
their generation (or family) number. The first generation of quarks and leptons, \{$u, \ d, \ e, \ \nu_e$\}, are
the lightest; the second generation, \{$c, \ s, \ \mu, \ \nu_\mu$\}, are all heavier than the first, and the
third generation, \{$t, \ b, \ \tau, \ \nu_\tau$\}, are the heaviest (see Table \ref{t:masses}).
\begin{table}
\caption{\label{t:masses} Quark and lepton masses in units of MeV/c$^2$.}
\begin{indented}
\item[]\begin{tabular}{@{}lcccc} \br
 & $\nu_e$ & e & u & d \\
\mr 1st &  $\leq 10^{-7}$   & 1/2 &  2 &  5 \\
\mr 2nd & $\leq 10^{-7}$ & 105.6 & 1,300 & 120 \\
\mr 3rd & $\leq 10^{-7}$ & 1,777 & 174,000 &  4,500 \\
\br
\end{tabular}
\end{indented}
\end{table}
In addition, the first two generations have a weak mixing angle given by the Cabibbo angle, $\theta_C$. If we
define $\lambda \equiv \sin\theta_C \sim .22$, then the mixing of the second and third generation is of order
$\lambda^2$ and the first and third is the weakest mixing of order $\lambda^3$. This pattern is very elegantly
captured in the Wolfenstein representation of the CKM matrix given by
\begin{eqnarray}  V_{CKM}  & =  \left(  \begin{array}{ccc} V_{ud} & V_{us} & V_{ub} \\
V_{cd} & V_{cs} & V_{cb} \\
V_{td} & V_{ts} & V_{tb}  \end{array}  \right) &  \\
& \approx \left(\begin{array}{ccc} 1 - \frac{1}{2} \lambda^2 &  \lambda & ( \rho - i  \eta) \ A \ \lambda^3 \\
-  \lambda & 1 - \frac{1}{2} \lambda^2 &  A \ \lambda^2 \\
(1 -  \rho - i  \eta) \ A \ \lambda^3 &  - A \ \lambda^2 & 1  \end{array}\right) & \nonumber
\end{eqnarray}
with $ \lambda \sim 1/5$, $\; A \sim 1$ and $|\rho - i \eta| \sim 1/2$.

Although the fundamental explanation for three families is still wanting, it is natural to assume that the
families transform under some family symmetry.  Such a possibility is consistent with weakly coupled heterotic
string theory where, for example, $E(8) \times E(8)$ and SO(32) are both large enough to contain a GUT group
$\times$ a family symmetry.   Such a family symmetry has many potential virtues.
\begin{itemize}  \item A spontaneously broken family symmetry can explain the hierarchy of fermion masses
and mixing angles.   \item  In a \ss theory, the family symmetry acts on both fermions and sfermions, thus
aligning the quark and squark, and the lepton and slepton mass matrices.   This suppresses flavor violating
processes.  \item  The combination of a family and a GUT symmetry can reduce the number of fundamental parameters
in the theory, hence allowing for a predictive theory.
\end{itemize}

In the following sections, we consider several important issues.   In section \ref{sec:yukawaunif} we discuss the
simplest case of the third generation only.  Here we discuss the status of SU(5) ($\lambda_b = \lambda_\tau$) and
SO(10) ($\lambda_t = \lambda_b = \lambda_\tau = \lambda_{\nu_\tau}$) Yukawa unification.   In section
\ref{sec:familysymmetry} we consider several different analyses in the literature for three family models with
either U(1) or non-abelian family symmetry.  In section \ref{sec:neutrinos} we study the relation between charged
fermion and neutrino masses in \gs and consider some examples giving bi-large neutrino mixing consistent with the
data.  Finally, in section \ref{sec:flavorviol} we discuss some experimental consequences of \ss theories of
fermion masses.   In particular, we consider $b \rightarrow s \gamma$, $(g - 2)_\mu$, $B_s \rightarrow \mu^+ \
\mu^-$, $\mu \rightarrow e \gamma$ and the electric dipole moments $d_n, \ d_e$.

\subsection{Yukawa unification \label{sec:yukawaunif}}

Let us first discuss the most stringent case of SO(10) Yukawa unification.  It has been shown [Raby (2001),
Derm\' \i \v sek (2001), Baer and Ferrandis (2001), Bla\v{z}ek \etal (2002a,b), Auto \etal (2003), Tobe and Wells
(2003)] that SO(10) boundary conditions at the GUT scale, for soft SUSY breaking parameters as well as for the
Yukawa couplings of the third generation, are consistent with the low energy data, including $M_t, \ m_b(m_b), \
M_\tau$, ONLY in a narrow region of SUSY breaking parameter space. Moreover, this region is also preferred by
constraints from CP and flavor violation, as well as by the non-observation of proton decay. Finally we discuss
the consequences for the Higgs and SUSY spectrum.

Recall, in SO(10) we have the compelling unification of all quarks and leptons of one family into one irreducible
representation such that  $ {\bf 10} + {\bf \bar{5}} + \bar{\nu}_{sterile} \ \subset \ { \bf 16}$ and the two
Higgs doublets are also unified with ${\bf 5_H},\; {\bf \bar{5}_H} \ \subset \ {\bf 10_H} $.  Hence, minimal
SO(10) also predicts Yukawa unification for the third family of quarks and leptons with $\lambda_b = \lambda_t =
\lambda_{\tau} = \lambda_{\nu_\tau} = \lambda$ at the GUT scale [Banks (1988), Olechowski and Pokorski (1988),
Pokorski (1990), Shafi and Ananthanarayan (1991);Ananthanarayan \etal (1991,1993,1994), Anderson \etal
(1993,1994)].

Ignoring threshold corrections, one can use the low energy value for $m_b/m_{\tau}$ to fix the universal Yukawa
coupling $\lambda$. RG running from $M_G$ to $M_Z$ then gives $\lambda_\tau(M_Z)$. Then given $m_\tau =
\lambda_\tau \frac{v}{\sqrt{2}} cos\beta$ we obtain $\tan\beta \approx 50$. Finally, a prediction for the top
quark mass is given with $m_t = \lambda_t \frac{v}{\sqrt{2}} sin\beta \sim 170 \pm 20 \; {\rm GeV}$ (see
[Anderson \etal (1993)]).

\begin{figure}[t!] \vspace*{0.50in} \hspace*{0.80in}
\begin{center}
%\hspace*{-1.00in}
\begin{minipage}{4.00in}
\epsfig{file=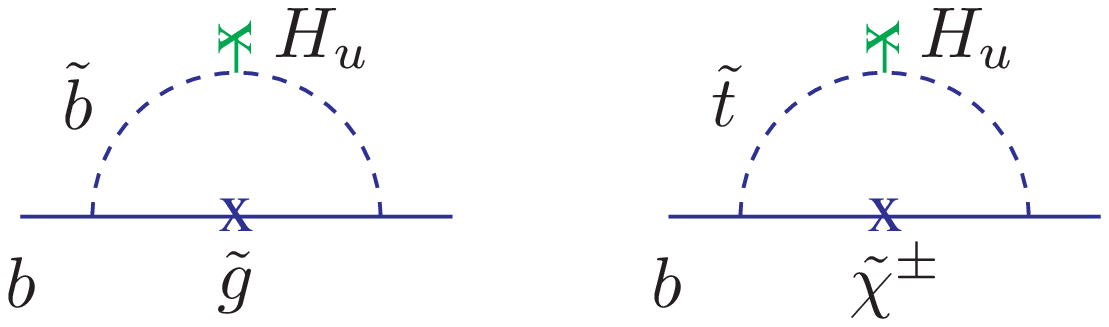,width=4.00in}
\end{minipage}
\end{center}
%\vspace*{-0.00in}
\begin{center} \caption{\label{fig:delta_mb} The one loop gluino (left) and chargino (right) corrections to the
bottom quark mass proportional to $\alpha_s$ (left) and $\lambda_t$ (right) and to $\tan\beta$.}
\end{center}
\end{figure}

Note, in this case there are insignificant GUT threshold corrections from gauge and Higgs loops.  Nevertheless,
the previous discussion is essentially a {\it straw man}, since there are {\em huge} threshold corrections at the
weak scale [Hall \etal (1994), Hempfling (1994), Carena \etal (1994), Bla\v{z}ek \etal (1995)].   The dominant
contributions are from gluino and chargino loops plus an overall logarithmic contribution due to finite wave
function renormalization given by $\delta m_b/ m_b = \Delta m_b^{\tilde g} + \Delta m_b^{\tilde \chi} + \Delta
m_b^{\log} + \cdots$ (see Fig. \ref{fig:delta_mb}). These contributions are approximately of the form
\begin{equation} \Delta m_b^{\tilde g} \approx \frac{2 \alpha_3}{3 \pi} \; \frac{\mu m_{\tilde g}}{m_{\tilde
b}^2} \; tan\beta  , \label{eq:gluino} \end{equation}
 \begin{equation} \Delta m_b^{\tilde \chi^+} \approx  \frac{\lambda_t^2}{16 \pi^2}
\; \frac{\mu A_t}{m_{\tilde t}^2} \; tan\beta  \;\;\; {\rm and} \label{eq:chargino} \end{equation}
\begin{equation}  \Delta m_b^{\log} \approx \frac{\alpha_3}{4
\pi} \log(\frac{\tilde m^2}{M_Z^2}) \sim 6 \% \label{eq:log}
\end{equation} with $\Delta m_b^{\tilde g} \sim - \Delta
m_b^{\tilde \chi} > 0$ {\em for {\bf $\mu > 0$}} [with our conventions].   These corrections can easily be of
order $\sim 50$ \%.   However good fits require\ $- 4\% < \delta m_b/ m_b <  - 2\%$.

Note, the data favors $\mu > 0$.  First consider the process {$b \rightarrow s \gamma$}.  The chargino loop
contribution typically dominates and has opposite sign to the standard model and charged Higgs contributions for
$\mu > 0$, thus reducing the branching ratio.   This is desirable, since the standard model contribution is a
little too large.  Hence $\mu < 0$ is problematic when trying to fit the data.  Secondly, the recent measurement
of the anomalous magnetic moment of the muon suggests a contribution due to NEW physics given by $a_\mu^{NEW} =
22.1 (11.3) \times 10^{-10}$ or $ 7.4 (10.5) \times 10^{-10}$ [Muon g - 2 Collaboration (2002), Davier \etal
(2003)] depending on whether one uses $e^+ e^-$ or $\tau$ hadronic decay data to evaluate the leading order
hadronic contributions.  For other recent theoretical analyses and references to previous work see [Hagiwara
\etal (2003), Melnikov and Vainshtein (2003)]. However in SUSY the sign of $a_\mu^{NEW}$ is correlated with the
sign of $\mu$ [Chattopadhyay and Nath (1996)]. Once again the data favors $\mu > 0$.

Before discussing the analysis of Yukawa unification, specifically that of [Bla\v{z}ek \etal (2002a,b)], we need
to consider one important point. $SO(10)$ Yukawa unification with the minimal Higgs sector necessarily predicts
large $\tan\beta \sim 50$. In addition, it is much easier to obtain EWSB with large $\tan\beta$ when the Higgs
up/down masses are split ($m_{H_u}^2 < m_{H_d}^2$) [Olechowski and Pokorski (1995), Matalliotakis  and Nilles
(1995), Polonsky and Pomarol (1995), Murayama, Olechowski and Pokorski (1996), Rattazzi and Sarid (1996)]. In the
following analysis we consider two particular Higgs splitting schemes we refer to as Just So and D term
splitting.\footnote{Just So Higgs splitting has also been referred to as non universal Higgs mass splitting or
NUHM [Berezinsky \etal (1996), Bla\v{z}ek \etal (1997a,b), Nath and Arnowitt (1997)].} In the first case the
third generation squark and slepton soft masses are given by the universal mass parameter $m_{16}$, and only
Higgs masses are split: $m_{(H_u, \; H_d)}^2 = m_{10}^2 \;( 1 \mp \Delta m_H^2)$. In the second case we assume D
term splitting, i.e. that the D term for $U(1)_X$ is non-zero, where $U(1)_X$ is obtained in the decomposition of
$SO(10) \rightarrow SU(5) \times U(1)_X$.  In this second case, we have $ m_{(H_u,\; H_d)}^2 = m_{10}^2 \mp 2 D_X
$, $m_{(Q,\; \bar u,\; \bar e)}^2 = m_{16}^2 + D_X$, $m_{(\bar d,\; L)}^2 = m_{16}^2 - 3 D_X $. The Just So case
does not at first sight appear to be very well motivated. However we now argue that it is quite natural
[Bla\v{z}ek \etal (2002a,b)]. In $SO(10)$, neutrinos necessarily have a Yukawa term coupling active neutrinos to
the ``sterile" neutrinos present in the {\bf 16}. In fact for $\nu_\tau$ we have $\L_{\nu_\tau} \; \bar \nu_\tau
\; L \; H_u$ with $\L_{\nu_\tau} = \L_t = \L_b = \L_\tau \equiv \; {\bf \L}$. In order to obtain a tau neutrino
with mass $m_{\nu_\tau} \sim 0.05$ eV (consistent with atmospheric neutrino oscillations), the ``sterile" $\bar
\nu_\tau$ must obtain a Majorana mass $M_{\bar \nu_\tau} \geq 10^{13}$ GeV. Moreover, since neutrinos couple to
$H_u$ (and not to $H_d$) with a fairly large Yukawa coupling (of order 0.7), they naturally distinguish the two
Higgs multiplets. With $\L = 0.7$ and $M_{\bar \nu_\tau} = 10^{14}$ GeV, we obtain a significant GUT scale
threshold correction with $\Delta m_H^2 \approx 7$\%, about 1/2 the value needed to fit the data.   At the same
time, we obtain a small threshold correction to Yukawa unification $\approx 1.75$\%.

\subsubsection{$\chi^2$ Analysis [Bla\v{z}ek \etal (2002a,b)]}

Our analysis is a top-down approach with 11 input parameters, defined at $M_G$, varied to minimize a $\chi^2$
function composed of 9 low energy observables. The 11 input parameters are: $M_G, \; \alpha_G(M_G),$ $
\epsilon_3$; the Yukawa coupling $\L$, and the 7 soft SUSY breaking parameters $\mu,\; M_{1/2},\; A_0, \;
\tan\beta$, $m_{16}^2, \; m_{10}^2$,  \dltmh ($D_X $) for Just So (D term) case. We use two (one)loop
renormalization group [RG] running for dimensionless (dimensionful) parameters from $M_G$ to $M_Z$ and complete
one loop threshold corrections at $M_Z$ [Pierce \etal (1997)]. We require electroweak symmetry breaking using an
improved Higgs potential, including $m_t^4$ and $m_b^4$ corrections in an effective 2 Higgs doublet model below
$M_{stop}$ [Haber and Hempfling (1993), Carena \etal (1995,1996)]. Note, in the figures we have chosen to keep
three input parameters $\mu,\; M_{1/2},\; m_{16}$ fixed, minimizing $\chi^2$ with respect to the remaining 8
parameters only.  The \ch2 function includes the 9 observables; 6 precision electroweak data $\alpha_{EM},\;
G_\mu, \;  \alpha_s(M_Z) = 0.118 \ (0.002),\; M_Z, \; M_W, \; \rho_{NEW}$ and the 3 fermion masses $M_{top} =
174.3 \ (5.1),\; m_b(m_b) = 4.20 \ (0.20), \; M_\tau$.
\begin{figure}[t!] \vspace*{0.50in} \hspace*{0.80in}
\begin{center}
\hspace*{-1.00in}
\begin{minipage}{4.00in}
\epsfig{file=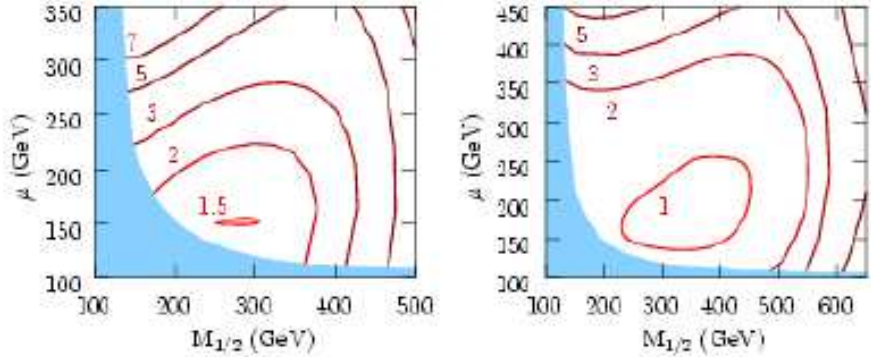,width=5.00in}
\end{minipage}
\end{center}
%\vspace*{-0.00in}
\begin{center} \caption{\label{fig:chi2} $\chi^2$ contours for $m_{16} = 1.5$ TeV (Left) and $m_{16} = 2$ TeV
(Right). The shaded region is excluded by the chargino mass limit $m_{\tilde \chi^+} > 103$ GeV.}
\end{center}
\end{figure}
Fig. \ref{fig:chi2} (Left) shows the constant $\chi^2$ contours for $m_{16} = 1.5$ TeV in the case of Just So
squark and slepton masses. We find acceptable fits ($\chi^2 < 3$) for $A_0 \sim - 1.9 \; m_{16}$, $m_{10} \sim
1.4 \; m_{16}$ and $ m_{16} \geq 1.2$ TeV. The best fits are for $m_{16} \geq 2$ TeV with $\chi^2 < 1$. Fig. 1
(Right) shows the constant $\chi^2$ contours for $m_{16} = 2$ TeV.
\begin{figure}[t!] \vspace*{0.50in} \hspace*{0.80in}
\begin{center}
\hspace*{-1.00in}
\begin{minipage}{4.00in}
\epsfig{file=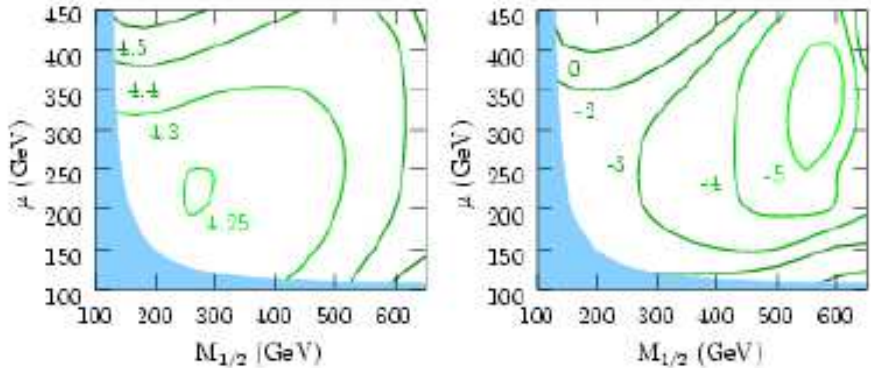,width=5.00in}
\end{minipage}
\end{center}
%\vspace*{-0.00in}
\begin{center} \caption{\label{fig:mbandcorr} Contours of constant $m_b(m_b)$[GeV] (Left) and \dltmb in \%
(Right) for $m_{16} = 2$ TeV.}
\end{center}
\end{figure}
Fig. \ref{fig:mbandcorr} gives the constant $m_b(m_b)$ and $\delta m_b/ m_b$ contours for $m_{16} = 2$ TeV. We
see that the best fits, near the central value, are found with $- 4\% \leq \delta m_b/ m_b \leq - 2$\%. The
chargino contribution (Eqn. \ref{eq:chargino}) is typically opposite in sign to the gluino (Eqn.
\ref{eq:gluino}), since $A_t$ runs to an infrared fixed point $\propto - M_{1/2}$ (see for example, [Carena \etal
(1994)]).  Hence in order to cancel the positive contribution of both the log (Eqn. \ref{eq:log}) and gluino
contributions, a large negative chargino contribution is needed. This can be accomplished for $- A_t > m_{\tilde
g}$ and $m_{\tilde t_1} \ll m_{\tilde b_1}$. The first condition can be satisfied for $A_0$ large and negative,
which helps pull $A_t$ away from its infrared fixed point.   The second condition is also aided by large $A_t$.
However in order to obtain a large enough splitting between $m_{\tilde t_1}$ and $m_{\tilde b_1}$, large values
of $m_{16}$ are needed. Note, that for Just So scalar masses, the lightest stop is typically lighter than the
sbottom. We typically find $m_{\tilde b_1} \sim 3 \; m_{\tilde t_1}$. On the other hand, D term splitting with
$D_X > 0$ gives $m_{\tilde b_1} \leq m_{\tilde t_1}$.   As a result in the case of Just So boundary conditions
excellent fits are obtained for top, bottom and tau masses; while for D term splitting the best fits give
$m_b(m_b) \geq 4.59$ GeV.\footnote{Note, [Auto \etal (2003), Tobe and Wells (2003)] use a bottom-up approach in
their analysis. The results of [Auto \etal (2003)] are in significant agreement with [Bla\v{z}ek \etal
(2002a,b)], except for the fact that they only find Yukawa unification for larger values of $m_{16}$ of order 8
TeV and higher.   The likely reason for this discrepancy has been explained by [Tobe and Wells (2003)].   They
show that [and I quote them] ``Yukawa couplings at the GUT scale are very sensitive to the low-energy SUSY
corrections. An O(1\%) correction at low energies can generate close to a O(10\%) correction at the GUT scale.
This extreme IR sensitivity is one source of the variance in conclusions in the literature. For example,
course-grained scatter plot methods, which are so useful in other circumstances, lose some of their utility when
IR sensitivity is so high. Furthermore, analyses that use only central values of measured fermion masses do not
give a full picture of what range of supersymmetry parameter space enables third family Yukawa unification, since
small deviations in low-scale parameters can mean so much to the high-scale theory viability."  It should also be
noted that [Tobe and Wells (2003)] suggest a different soft \ss breaking solution consistent with Yukawa
unification.  In particular, they suggest an extension of AMSB with the addition of a large universal scalar mass
$m_0 \geq 2$ TeV.}

The bottom line is that Yukawa unification is only possible in a narrow region of SUSY parameter space with
\begin{equation} A_0 \sim - 1.9 \;  m_{16}, \;\; m_{10} \sim 1.4 \; m_{16},\;\;\; \end{equation}
\begin{equation} (\mu,\ M_{1/2}) \sim 100 - 500 \; {\rm GeV \;\; and} \;\;  m_{16}
\geq 1.2 \; {\rm TeV} . \end{equation} It would be nice to have some a priori reason for the fundamental SUSY
breaking mechanism to give these soft SUSY breaking parameters. However, without such an a priori explanation, it
is all the more interesting and encouraging to recognize two additional reasons for wanting to be in this narrow
region of parameter space.

\begin{enumerate}
\item One mechanism for suppressing large flavor violating processes in SUSY theories is to demand heavy first
and second generation squarks and sleptons (with mass $\gg$ TeV) and the third generation scalars lighter than a
TeV. Since the third generation scalars couple most strongly to the Higgs, this limit can still leave a
``naturally" light Higgs [Dimopoulos and Giudice (1995)].  It was shown that this inverted scalar mass hierarchy
can be obtained ``naturally," i.e. purely as a consequence of renormalization group running from $M_G$ to $M_Z$,
with suitably chosen soft SUSY breaking boundary conditions at $M_G$ [Bagger \etal (1999,2000)]. All that is
needed is $SO(10)$ boundary conditions for the Higgs mass (i.e. $m_{10}$), squark and slepton masses (i.e.
$m_{16}$) and a universal scalar coupling $A_0$. In addition, they must be in the ratio [Bagger \etal (2000)]
\beq A_0^2 = 2 \ m_{10}^2 = 4 \ m_{16}^2, \;\;\; {\rm with} \;\;\; m_{16} \gg \; {\rm TeV}. \eeq

\item In order to suppress the rate for proton decay due to dimension 5 operators one must also demand [Derm\' \i
\v sek \etal (2001)] \beq (\mu,\; M_{1/2}) \ll m_{16}, \;\;\; {\rm with} \;\;\; m_{16} > \ {\rm few \; TeV} .
\eeq
\end{enumerate}

\subsubsection{Consequences for Higgs and SUSY Searches}

\begin{figure}[t!] \vspace*{0.50in} \hspace*{0.80in}
\begin{center}
\hspace*{-1.00in}
\begin{minipage}{4.00in}
\epsfig{file=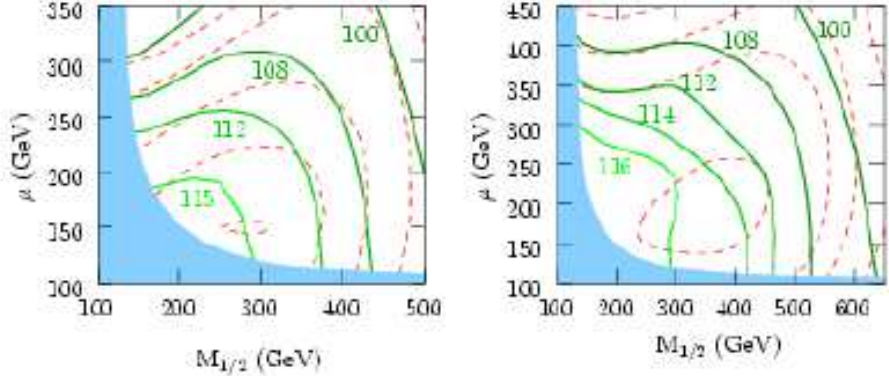,width=5.00in}
\end{minipage}
\end{center}
%\vspace*{-0.00in}
\begin{center} \caption{\label{fig:h0andchi2} Contours of constant $m_h$ [GeV] (solid lines) with $\chi^2$ contours from Fig. 1 (dotted lines) for
$m_{16} = 1.5$ TeV (Left) and $m_{16} = 2$ TeV (Right).}
\end{center}
\end{figure}

In Fig. \ref{fig:h0andchi2} we show the constant light Higgs mass contours
 for $m_{16} = 1.5$ and $2$ TeV
(solid lines) with the constant $\chi^2$ contours overlayed (dotted lines).  Yukawa unification for $\chi^2 \leq
1$  clearly prefers a light Higgs with mass in a narrow range,  112 -  118 GeV.

\begin{figure}[t!] \vspace*{0.50in} \hspace*{0.80in}
\begin{center}
\hspace*{-1.00in}
\begin{minipage}{4.00in}
\epsfig{file=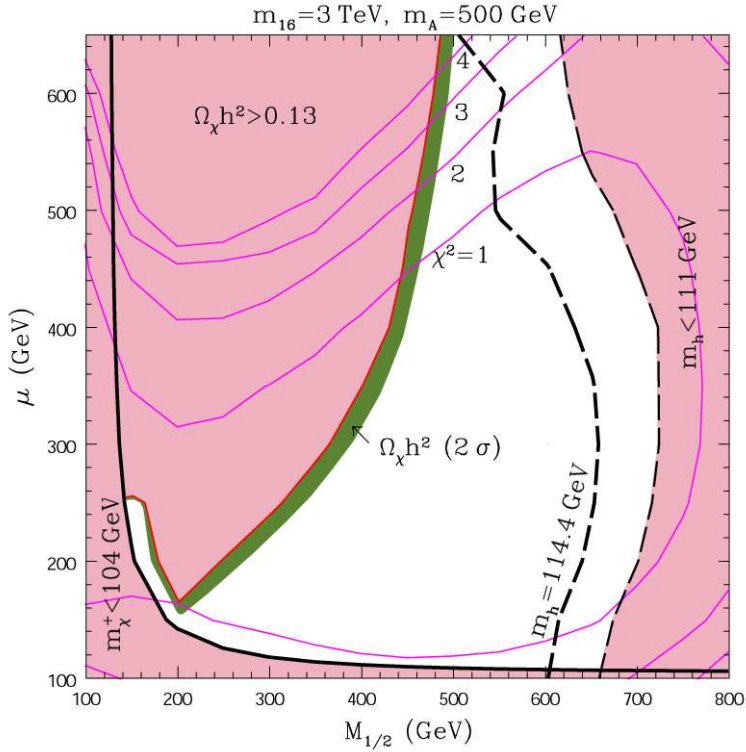,width=4.00in}
\end{minipage}
\end{center}
%\vspace*{-0.00in}
\begin{center} \caption{\label{fig:oh2} Constant $\chi^2$ contours as a function of $\mu, \ M_{1/2}$
for $m_{16} = 3$ TeV.   Note the much larger range of parameters with $\chi^2 < 1$ for this larger value of
$m_{16}$. The green shaded region is consistent with the recent WMAP data for dark matter abundance of the
neutralino LSP. The light shaded region in the lower left hand corner (separated by the solid line) is excluded
by chargino mass limits, while the light shaded region in the upper left (right side) is excluded by a
cosmological dark matter abundance which is too large (Higg mass which is too light).  To the left of the
vertical contour for a light Higgs with mass at the experimental lower limit, the light Higgs mass increases up
to a maximum value of about 121 GeV at the lower left-hand acceptable boundary.}
\end{center}
\end{figure}

In this region the CP odd $A$, the heavy CP even Higgs $H$ and the charged Higgs bosons $H^\pm$ are also quite
light.  In addition we find the mass of $\tilde t_1 \sim (150 - 250)$ GeV, $\tilde b_1 \sim (450 - 650)$ GeV,
$\tilde \tau_1 \sim (200 - 500)$ GeV, $\tilde g \sim (600 - 1200)$ GeV, $\tilde \chi^+ \sim (100 - 250)$ GeV, and
$\tilde \chi^0 \sim (80 - 170)$ GeV.  All first and second generation squarks and sleptons have mass of order
$m_{16}$.  The light stop  and chargino may be visible at the Tevatron.  With this spectrum we expect $\tilde t_1
\rightarrow \tilde \chi^+ \;b $ with $\tilde \chi^+ \rightarrow \tilde \chi^0_1 \; \bar l \; \nu$ to be dominant.
Lastly $\tilde \chi^0_1$ is the LSP and possibly a good dark matter candidate (see for example, [Roszkowski \etal
(2001)] and Fig. \ref{fig:oh2}).

Our analysis thus far has only included third generation Yukawa couplings; hence no flavor mixing.   If we now
include the second family and 2-3 family mixing, consistent with $V_{c b}$, we obtain new and significant
constraints on $m_{\tilde t_1}$ and $m_{A}$. The stop mass is constrained by $B(b \rightarrow s \gamma)$ to
satisfy $m_{\tilde t}^{MIN} > 450$ GeV (unfortunately increasing the bottom quark mass). In addition, as shown by
[Choudhury and Gaur (1999), Babu and Kolda (2000), Dedes \etal (2001), Isidori and Retico (2001)] the one loop
SUSY corrections to CKM mixing angles (see Bla\v{z}ek \etal (1995)) result in flavor violating neutral Higgs
couplings. As a consequence the CDF bound on the process $B_s \rightarrow \mu^+ \mu^-$ places a lower bound on
$m_{A} \geq 200$ GeV [Choudhury and Gaur (1999), Babu and Kolda (2000), Dedes \etal (2001), Isidori and Retico
(2001)]. \ch2, on the other hand, increases as $m_{A^0}$ increases.   However the increase in \ch2 is less than
60\% for $m_{A} < 400$ GeV. Note, the $H^\pm, \ H^0$ masses increase linearly with $m_{A}$.

\subsubsection{\label{sec:su5yukawa} SU(5) Yukawa unification}

Now consider Yukawa unification in SU(5).  In this case we only have the GUT relation $\lambda_b = \lambda_\tau$.
The RG running of the ratio $\lambda_b/\lambda_\tau$ to low energies then increases (decreases) due to QCD
(Yukawa) interactions. In addition, neglecting Yukawa interactions, this ratio is too large at the weak scale.
For a top quark mass $M_t \sim 175$ GeV, a good fit is obtained for small $\tan\beta \sim 1$ [Dimopoulos \etal
(1992), Barger \etal (1993)]. In this case, only the top quark Yukawa coupling is important.  While for large
$\tan\beta \sim 50$ we recover the results of SO(10) Yukawa unification.  For a recent analysis, see [Barr and
Dorsner (2003)].

\subsection{Fermion mass hierarchy \& Family symmetry \label{sec:familysymmetry}}

In both the standard model and the MSSM, the observed pattern of fermion masses and mixing angles has its origin
in the Higgs-quark and Higgs-lepton Yukawa couplings. In the standard model these complex $3 \times 3$ matrices
are arbitrary parameters which are under constrained by the 13 experimental observables (9 charged fermion masses
and 4 quark mixing angles). [We consider neutrino masses and mixing angles in the following section.]  In the
MSSM more of the Yukawa parameters are in principle observable, since left and right-handed fermion mixing angles
affect squark and slepton masses and mixing.  [We consider this further in Section \ref{sec:flavorviol}.] What
can \ss say about fermion masses?   \ss alone constrains the Yukawa sector of the theory simply by requiring that
all terms in the superpotential are holomorphic.   Combined with flavor symmetries, the structure of fermion
masses can be severely constrained. On the other hand, the only information we have about these flavor symmetries
is the fermion masses and mixing angles themselves, as well as the multitude of constraints on flavor violating
interactions. There are, perhaps, many different theories with different family symmetries that fit the {\em
precision low energy data} (including fermion masses and mixing angles). The goal is to find a set of predictive
theories, i.e with fewer arbitrary parameters than data, that fit this data. The more predictive the theory, the
more testable it will be.\footnote{Of course, if \ss or proton decay is observed then there will be much more low
energy data available to test these theories.} Within the context of the MSSM, theories have been constructed
with U(1) family symmetries [Binetruy \etal (1996), Elwood \etal (1997,1998), Irges \etal (1998), Faraggi and
Pati (1998), Kakizaki and Yamaguchi (2002), Dreiner \etal (2003)], with discrete family symmetries [Frampton and
Kephart (1995a,b), Hall and Murayama (1995), Carone \etal (1996), Carone and Lebed (1999), Frampton and Rasin
(2000), Aranda \etal (2000)] or non-abelian family symmetries [Hall and Randall (1990), Dine \etal (1993), Nir
and Seiberg (1993), Pouliot and Seiberg (1993), Leurer \etal (1993,1994), Pomarol and Tommasini (1995), Hall and
Murayama (1995), Dudas \etal (1995,1996), Barbieri \etal (1996), Arkani-Hamed \etal (1995,1996), Barbieri \etal
(1997), Eyal (1998)]. However, the most predictive theories combine both grand unified and family symmetries
[Kaplan and Schmaltz (1994), Babu and Mohapatra (1995), Lucas and Raby (1996), Frampton and Kong (1996),
Bla\v{z}ek \etal (1997), Barbieri and Hall (1997), Barbieri \etal (1997), Allanach \etal (1997), Berezhiani
(1998), Bla\v{z}ek \etal (1999,2000), Derm\' \i \v sek and Raby (2000), Shafi and Tavartkiladze (2000), Albright
and Barr (2000,2001), Altarelli \etal (2000), Babu \etal (2000), Berezhiani and Rossi (2001), Kitano and Mimura
(2001), Maekawa (2001), King and Ross (2003), Chen and Mahanthappa (2003), Raby (2003), Ross and Velasco-Sevilla
(2003), Goh \etal (2003), Aulakh \etal (2003)].  The Yukawa couplings in a predictive theory are completely
defined in terms of the states and symmetries of the theory. The ultimate goal of this program is to construct
one (or more) of these predictive theories, providing good fits to the data, in terms of a more fundamental
theory, such as M theory. Only then will higher order corrections to the theory be under full control. It is
important to remark at this stage that any theory, derived from some fundamental theory, includes
non-renormalizable higher dimension operators.   The higher dimension operators are suppressed by the fundamental
scale (for example, the string scale $M_S$) which is assumed to be greater than the GUT scale $M_G$. As we shall
now see, these higher dimension operators are useful in explaining the hierarchy of fermion masses.

The $3 \times 3$ up, down and charged lepton mass matrices are given by the mass terms:
\begin{eqnarray} {\cal L}_{mass} =  u \ Y_u \ \bar u \ \langle H_u \rangle \ + & \;\;\;  d \
Y_d \ \bar d \ \langle H_d \rangle \ + & \;\;\;   e \  Y_e \ \bar e \ \langle H_d \rangle .
\end{eqnarray}
Empirical descriptions of the quark mass matrices have been discussed in all the papers referenced above. As an
example, consider the following theory incorporating the hierarchy of masses and mixing angles in an SU(5) \ss
GUT with U(1) family symmetry by [Altarelli \etal (2000)]
\begin{eqnarray}   Y_u =  \left( \begin{array}{ccc} \lambda^6 & \lambda^5 & \lambda^3 \\
\lambda^{5} & \lambda^{4} & \lambda^{2} \\  \lambda^{3} & \lambda^{2} & 1 \end{array} \right), & \;\;\;\; Y_d =
Y_e^T =  \left(
\begin{array}{ccc} \lambda^5 & \lambda^3 & \lambda^3 \\ \lambda^{4} & \lambda^{2} & \lambda^{2} \\  \lambda^2 & 1
& 1 \end{array} \right) \lambda^4 .&  \label{eq:altarelli} \end{eqnarray}  Order 1 coefficients of the matrix
elements are implicit. We then obtain the rough empirical relations. \begin{eqnarray}  m_c/m_t \ \sim \ m_u/m_c &
\; \sim \ V_{cb}^2 & \ \approx \ \lambda^4  \\ m_s/m_b \ \sim \ m_d/m_s & \; \sim \ V_{us}^2 & \ \approx \
\lambda^2 . \nonumber
\end{eqnarray}
In addition, the Yukawa matrices for down quarks and charged leptons satisfy the SU(5) relations
\begin{eqnarray} \lambda_b = \lambda_\tau, & \;\; \lambda_s = \lambda_\mu, & \;\;  \lambda_d = \lambda_e
. \end{eqnarray}   This works for the third generation, as discussed in Section \ref{sec:su5yukawa}, however it
clearly doesn't work for the first and second generations, where it gives the unacceptable prediction \beq 20 \;
\approx \;  m_s/m_d \;  = \;  m_\mu/m_e  \; \approx \; 200 . \label{eq:su5yukawa} \eeq  Hence in this SU(5) model
with U(1) family symmetry,  an additional Higgs in the 75 dimensional representation (with U(1) charge zero) also
contributes to down quark and charged lepton Yukawa matrices [Altarelli \etal (2000)].   The arbitrary, order one
coefficients for each term in the Yukawa matrix are then fit to the quark masses and mixing angles and charged
lepton masses. Note in this theory there are more arbitrary parameters, than fermion mass observables; hence
there are no predictions for fermion masses and mixing angles. Nevertheless, predictions for proton decay are now
obtained. Moreover, given a model for soft \ss breaking terms like the CMSSM, one can also predict rates for
flavor violating processes.

The structure for these Yukawa matrices are determined by the U(1) family symmetry, spontaneously broken by a
scalar field $\phi$ with U(1) charge $-1$. The symmetry breaking field $\phi$ is inserted into each element of
the Yukawa matrix in order to obtain a U(1) invariant interaction.  This results in effective higher dimensional
operators suppressed by a scale $M$ with $\lambda \sim \langle \phi \rangle/M$.  This is the Froggatt-Nielsen
mechanism [Froggatt and Nielsen (1979), Berezhiani (1983,1985), Dimopoulos (1983), Bagger \etal (1984)]. For a
review see [Raby (1995)]. Given the U(1) charge assignments in Table \ref{t:U1} [Altarelli \etal (2000)] we
obtain the Yukawa matrices in Eqn. \ref{eq:altarelli}.
\begin{table}
\caption{\label{t:U1} U(1) charge $Q$ of Higgs and matter fields in the (1st,2nd,3rd) generation.}
\begin{indented}
\item[]\begin{tabular}{@{}lccccc} \br
field & $H_u$ & $H_d$ & 10 = \{$Q$, \ $\bar u$, \ $\bar e$\} & $\bar 5$ = \{$\bar d$, \ L\} & $\bar \nu$ \\
\mr $Q$ &  -2   & 1 &  (4,3,1) &  (4,2,2) & (1,-1,0) \\
\br
\end{tabular}
\end{indented}
\end{table}
Similar mass matrices using analogous U(1) family symmetry arguments have also been considered.

\begin{figure}[t!] \vspace*{0.50in} \hspace*{0.80in}
\begin{center}
%\hspace*{-1.00in}
\begin{minipage}{4.00in}
\epsfig{file=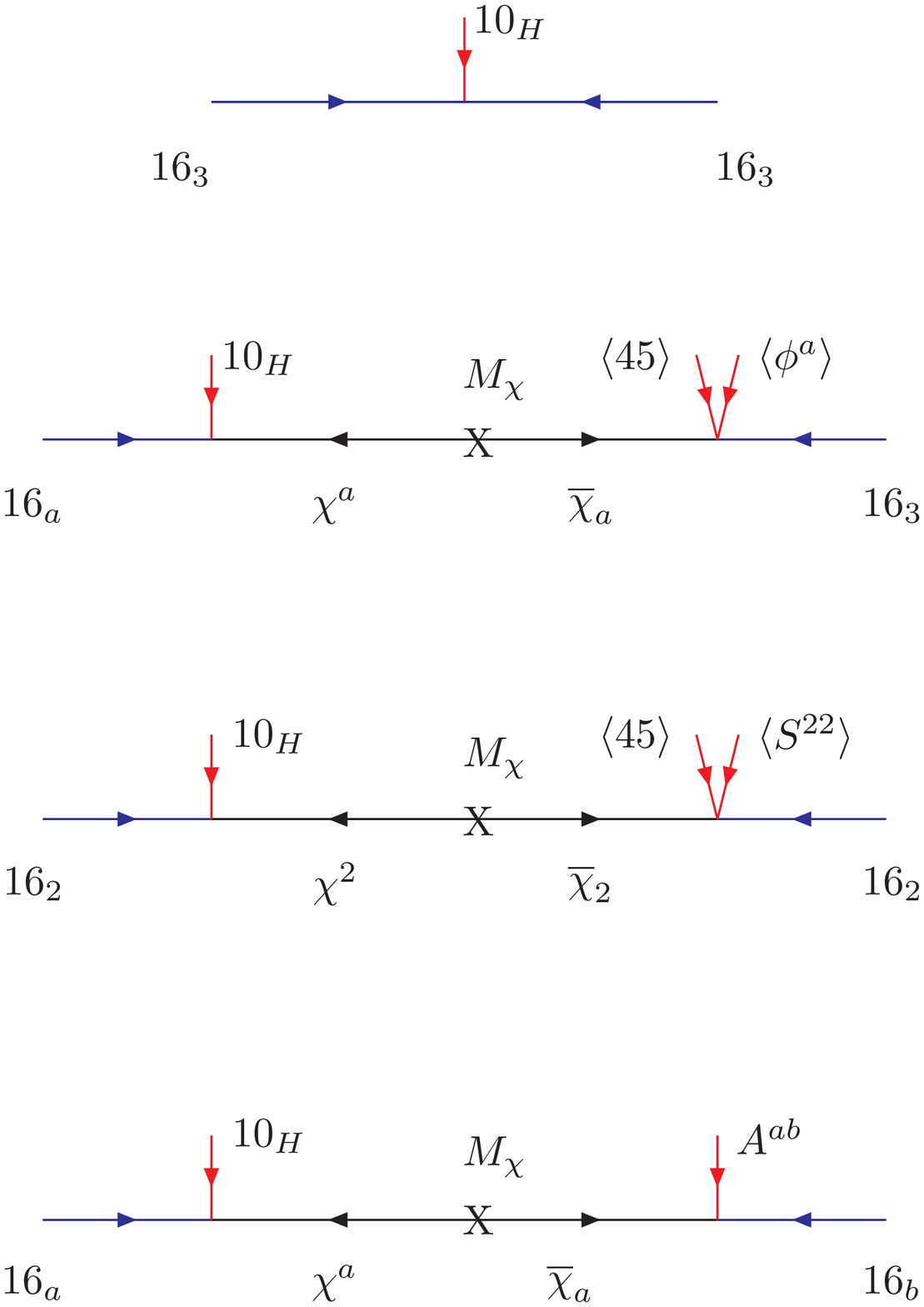,width=4.00in}
\end{minipage}
\end{center}
%\vspace*{-0.00in}
\begin{center} \caption{\label{fig:diagram} Effective fermion mass operators.  The fields $\phi^a, \ S^{a b} = S^{b a},
A^{a b} = - A^{b a}$ spontaneously break the SU(2) $\times$ U(1)$^n$ family symmetry with $\epsilon \propto
\langle \phi^2 \rangle$, $\tilde \epsilon \propto \langle S^{2 2} \rangle$, and $\epsilon^\prime \propto \langle
A^{1 2} \rangle$.}
\end{center}
\end{figure}

Using a non-abelian family symmetry, such as SU(2) $\times$ U(1) or SU(3) or discrete subgroups of SU(2), models
with fewer arbitrary parameters in the Yukawa sector have been constructed.  An example of a very predictive
SO(10) \ss GUT with SU(2) $\times$ U(1)$^n$ family symmetry is given by [Barbieri \etal (1997(a,b),1999),
Bla\v{z}ek \etal (1999,2000)].  An analogous model can be obtained by replacing the SU(2) family symmetry with a
discrete subgroup D(3) [Derm\' \i \v sek and Raby (2000)].  The model incorporates the Froggatt-Nielsen mechanism
with a hierarchy of symmetry breaking VEVs explaining the hierarchy of fermion masses. The effective fermion mass
operators are given in Fig. \ref{fig:diagram}. In particular, the family symmetry breaking pattern  \beq  SU_2
\times U_1 \;\;\; \longrightarrow \;\;\; U_1 \;\;\; \longrightarrow \;\;\; {\rm nothing} \eeq with small
parameters $\epsilon \approx \tilde \epsilon$ and $\epsilon^\prime$, respectively, gives the hierarchy of masses
with the 3rd family $\gg$ 2nd family $\gg$ 1st family.   It includes the Georgi - Jarlskog [Georgi and Jarlskog
(1979)] solution to the unacceptable SU(5) relation (Eqn. \ref{eq:su5yukawa}) with the improved relation \bea m_s
\; \sim \; \frac{1}{3} m_\mu, \;\; & m_d \;  \sim \; 3 m_e . & \eea  This is obtained naturally using the VEV
\beq \langle 45 \rangle = (B - L) \ M_G . \eeq In addition, it gives the SO(10) relation for the third generation
\beq \lambda_t = \lambda_b = \lambda_\tau = \lambda_{\nu_\tau} = \lambda  \eeq   and it uses symmetry arguments
to explain why $m_u < m_d$ even though $m_t \gg m_b$.  Finally the $SU_2$ family symmetry suppresses flavor
violation such as $\mu \rightarrow e \gamma$.  When SO(10) $\times$ SU(2) $\times$ U(1)$^n$ is broken to the MSSM
the effective Yukawa couplings (Eqn. \ref{eq:yukawa3x3}) are obtained. The superpotential for this simple model
is given by
 \begin{eqnarray} W \supset & 16_3 \ 10 \ 16_3 +  16_a \ 10 \ \chi^a &  \\
 & +  \bar \chi_a \ ( M_{\chi} \ \chi^a + \ 45 \ \frac{\phi^a}{\hat M} \  16_3 \ + \ 45 \
\frac{S^{a b}}{\hat M} \  16_b + A^{a b} \ 16_b )  & \nn
\end{eqnarray}
where $\phi^a, \; S^{a b} = S^{b a}, \; A^{a b} = - A^{b a}$ are the familon fields whose VEVs break the family
symmetry,  $M_\chi = \hat M \ (1 + \alpha \ X + \beta \ Y)$ with $X, \ Y$ charges associated with U(1)$_{X, \
Y}$, the orthogonal U(1) subgroups of SO(10), and \{ $\chi^a, \bar \chi_a$ \} are the heavy Froggatt-Nielsen
fields. After the heavy $\chi$ states are integrated out of the theory we obtain the effective fermion mass
operators given in Fig. \ref{fig:diagram}.
These four Feynman diagrams lead to the following Yukawa matrices for quarks and charged leptons.
\begin{eqnarray}
Y_u =&  \left(\begin{array}{ccc}  0 & \epsilon' \ \rho & - \epsilon \ \xi \\
             - \epsilon' \ \rho &  \tilde \epsilon \ \rho & - \epsilon  \\
      \epsilon \ \xi & \epsilon & 1 \end{array} \right) \; \lambda & \nonumber \\
Y_d =&  \left(\begin{array}{ccc}  0 & \epsilon'  & - \epsilon \ \sigma \ \xi \\
- \epsilon'   &  \tilde \epsilon  & - \epsilon \ \sigma \\
\epsilon \ \xi  & \epsilon  & 1 \end{array} \right) \; \lambda & \label{eq:yukawa3x3} \\
Y_e =&  \left(\begin{array}{ccc} 0 & - \epsilon'  & 3 \ \epsilon \ \xi \\
          \epsilon'  &  3 \ \tilde \epsilon  & 3 \ \epsilon  \\
  - 3 \ \epsilon \ \xi & - 3 \ \epsilon  & 1 \end{array} \right) \; \lambda &
 \nonumber
\end{eqnarray}
with  \begin{eqnarray}  \xi \;\; =  \;\; \langle \phi^1 \rangle/\langle \phi^2 \rangle; & \;\;
\tilde \epsilon  \;\; \propto   \;\; \langle S^{2 2} \rangle/\hat M;  & \label{eq:omega} \\
\epsilon \;\; \propto  \;\; \langle \phi^2 \rangle/\hat M; &  \;\;
\epsilon^\prime \;\; \sim  \;\;  \langle A^{1 2} \rangle/\hat M; \nonumber \\
  \sigma \;\; =   \;\; \frac{1+\alpha}{1-3\alpha}; &  \;\; \rho \;\; \sim   \;\;
  \beta \, \ll \, \alpha . & \nonumber
\end{eqnarray}

The model has only 9 arbitrary Yukawa parameters (6 real parameters \{$|\lambda|, \ |\epsilon|, \ |\tilde
\epsilon|, \ |\rho|, \ |\sigma|, \ |\epsilon^\prime|$\} and three phases \{$\Phi_\epsilon = \Phi_{\tilde
\epsilon}, \ \Phi_\rho, \ \Phi_\sigma$\}) to fit the 13 fermion masses and mixing angles (we have taken $\xi =
0$). The fit to the low energy data is given in Table \ref{t:u2fit}.\footnote{Note, some of the data, used in
this fit, have significantly improved in recent years.} More details of this fit are found in [Bla\v{z}ek \etal
(1999)] and the predictions for proton decay are found in [Derm\' \i \v sek \etal (2001)]. Note, the model fits
most of the precision electroweak data quite well.
\begin{table}
\caption{\label{t:u2fit} Fit to fermion masses and mixing angles for SO(10) GUT with SU(2) $\times$ U(1)$^n$
family symmetry [Bla\v{z}ek \etal (1999)].}
\begin{indented}
\item[]\begin{tabular}{@{}lcc} \br
Observable & Data($\sigma$)\small{masses in GeV} & Theory  \\
\mr
$M_Z$            &  91.187 \ (0.091)  &  91.17          \\
$M_W$             &  80.388 \ (0.080)    &  80.40       \\
$G_{\mu}\cdot 10^5$   &  1.1664 \ (0.0012) &  1.166     \\
$\alpha_{EM}^{-1}$ &  137.04 \ (0.14)  &  137.0         \\
$\alpha_s(M_Z)$    &  0.1190 \ (0.003)   &  0.1174       \\
$\rho_{new}\cdot 10^3$  & -1.20 \ (1.3) & +0.320   \\
\mr
$M_t$              &  173.8 \ (5.0)   &  175.0       \\
$m_b(M_b)$          &    4.260 \ (0.11)  &    4.328                  \\
$M_b - M_c$        &    3.400 \ (0.2)   &    3.421                 \\
$m_s$              &  0.180 \ (0.050)   &  0.148          \\
$m_d/m_s$          &  0.050 \ (0.015)   &  0.0589        \\
$Q^{-2}$           &  0.00203 \ (0.00020)  &  0.00201                \\
$M_{\tau}$         &  1.777 \ (0.0018)   &  1.776         \\
$M_{\mu}$          & 0.10566 \ (0.00011)   & .1057           \\
$M_e \cdot 10^3$      &  0.5110 \ (0.00051) &  0.5110  \\
$V_{us}$         &  0.2205 \ (0.0026)      &  0.2205        \\
$V_{cb}$         & 0.03920 \ (0.0030)      &  0.0403           \\
$V_{ub}/V_{cb}$    &  0.0800 \ (0.02)    &  0.0691                 \\
$\hat B_K$          &  0.860 \ (0.08)    &  0.8703           \\
\mr
$B(b \rightarrow\ s \gamma) \ \cdot 10^{4}$  &  3.000 \ (0.47) &  2.995  \\
\mr
 TOTAL $\chi^2$ &   3.39  & \\
\br
\end{tabular}
\end{indented}
\end{table}
In [Bla\v{z}ek \etal (1999)] there is also a prediction for $\sin2\beta = 0.54$ which should be compared to the
present experimental value $0.727 \ (0.036)$.   The prediction for $\sin2\beta$ is off by 5 $\sigma$.  In
addition the present experimental value for $V_{ub}/V_{cb}$ is $0.086 \ (0.008)$, hence this fit (Table
\ref{t:u2fit}) is somewhat worse than before.    Both of these quantities are predictions due solely to the zeros
in the 11, 13 and 31 elements of the Yukawa matrices [Hall and Rasin (1993), Roberts \etal (2001), Kim \etal
(2004)].  These poor fits are remedied with the addition of a non-vanishing 13 / 31 element, i.e. $\xi \neq 0$.
In this case a good fit is obtained with one additional real parameter [Kim \etal (2004)].

\begin{table}
\caption{\label{t:u2fitpar} Input parameters at $M_G$ for fit in Table \ref{t:u2fit} [Bla\v{z}ek \etal (1999)].}
\begin{indented}
\item[]\begin{tabular}{@{}lc} \br (1/$\alpha_G, \, M_G, \, \epsilon_3$) = & ($24.52, \, 3.05 \cdot 10^{16}$
GeV,$\, -4.08$\%), \makebox[1.8em]{ }\\
($\lambda, \,\epsilon, \, \sigma, \, \tilde \epsilon, \, \rho, \, \epsilon^\prime, \, \xi$ ) = &
($ 0.79, \, 0.045, \, 0.84, \, 0.011, \,  0.043,\,  0.0031,\, 0.00$),\\
($\Phi_\sigma, \, \Phi_\epsilon = \Phi_{\tilde \epsilon}, \, \Phi_\rho$) = & ($0.73, \, -1.21, \, 3.72$)rad,
\makebox[6.6em]{ }\\
($m_{16}, \, M_{1/2}, \, A_0, \, \mu(M_Z)$) = & ($1000,\, 300, \, -1437, \,
110$) GeV,\\
($(m_{H_d}/m_{16})^2, \, (m_{H_u}/m_{16})^2, \, $tan$\beta$) = & ($2.22,\, 1.65, \, 53.7$) \\
\end{tabular}
\end{indented}
\end{table}

\subsection{Neutrino masses \label{sec:neutrinos}}

The combined data from all neutrino experiments can be fit by the hypothesis of neutrino oscillations with the
neutrino masses and mixing angles given by
\begin{eqnarray}
\Delta m^2_{atm} = & |m_3^2 - m_2^2| \approx 3 \times 10^{-3} \; {\rm eV}^2 & \\
& \sin 2\theta_{atm} \approx 1 &  \nn \\
\Delta m^2_{sol} = & |m_2^2 - m_1^2| \approx 7 \times 10^{-5} \; {\rm eV}^2 & \nn \\
& 0.8 < \sin 2\theta_{sol} < 1 & \nn
\end{eqnarray}
For recent theoretical analyses of the data, see [Barger \etal (2003), Maltoni \etal (2003), Gonzales-Garcia and
Pe\~{n}a-Garay (2003)]. This so-called bi-large neutrino mixing is well described by the PMNS mixing matrix
\begin{eqnarray} \left( \begin{array}{c} \nu_e \\ \nu_\mu \\ \nu_\tau \end{array}
\right) \approx & \left( \begin{array}{ccc}
c_{sol} &  s_{sol} & 0 \\ -s_{sol}/\sqrt{2} & c_{sol}/\sqrt{2} & 1/\sqrt{2} \\
-s_{sol}/\sqrt{2} & c_{sol}/\sqrt{2} & -1/\sqrt{2} \end{array} \right) \; \left(
\begin{array}{c} \nu_1 \\ \nu_2 \\ \nu_3 \end{array} \right) . & \nonumber
\end{eqnarray}
which takes mass eigenstates into flavor eigenstates.   The 1-3 mixing angle satisfies $\sin\theta_{1 3} < 0.2$
at 3 $\sigma$ [Maltoni \etal (2003)].

Using the See-Saw mechanism [Yanagida (1979), Glashow (1979), Gell-Mann \etal (1979), Mohapatra and Senjanovic
(1980)], neutrino masses are given in terms of two completely independent $3 \times 3$ mass matrices, i.e.  the
Dirac mass matrix $m_\nu$ and a Majorana mass matrix $M_N$ via the formula ${\cal M}_\nu = m_\nu^T \ M_N^{-1} \
m_\nu$. The smallness of neutrino masses is explained by the large Majorana mass scale, of order $10^{14} -
10^{15}$ GeV; very close to the GUT scale. In addition, the large mixing angles needed to diagonalize ${\cal
M}_\nu$ can be directly related to large mixing in $m_\nu$, in $M_N$ or in some combination of both.  Lastly, the
Dirac neutrino mass matrix $m_\nu$ is constrained by charged fermion masses in SO(10), but not in SU(5) where it
is completely independent.

The major challenge for theories of neutrino masses is to obtain two large mixing angles; as compared to charged
fermions where we only have small mixing angles in $V_{CKM}$.  There are several interesting suggestions for
obtaining large mixing angles in the literature.  [For recent reviews of models of neutrino masses, see
[Altarelli and Feruglio (2003), Altarelli \etal (2003), King (2003)].]
\begin{itemize}
\item Degenerate neutrinos and RG running [Mohapatra \etal (2003), Casas \etal (2003)],

It was shown that starting with three degenerate Majorana neutrinos and small mixing angles at a GUT scale, that
RG running can lead to bi-large neutrino mixing at low energies.

\item Minimal renormalizable SO(10) [Goh \etal (2003), Bajc \etal (2003)],

SO(10) with Higgs in the 10 and $\overline{126}$ representations can give predictable theories of fermion masses
with naturally large neutrino mixing angles.

\item Dominant Majorana neutrinos [King (1998,2000)],

It was shown that large neutrino mixing can be obtained via coupling to a single dominant right handed neutrino.

\item Minimal Majorana sector [Frampton \etal (2002), Raidal and Strumia (2003), Raby (2003)],

It was shown that a simple model with two right handed neutrinos can accommodate bi-large neutrino mixing with
only one CP violating phase.   In such a theory, CP violating neutrino oscillations measured in low energy
accelerator experiments are correlated with the matter -- anti-matter asymmetry obtained via leptogenesis.

\item Lopsided charged lepton and down quark matrices   [Lola and Ross (1999), Nomura and Yanagida (1999),
Albright and Barr (2000(a,b),2001), Altarelli \etal (2000), Barr and Dorsner (2003)].

In SU(5) (or even in some SO(10) models) the down quark mass matrix is related to the transpose of the charged
lepton mass matrix.  A large left-handed $\mu - \tau$ mixing angle is thus directly related to a large
right-handed  $s - b$ mixing angle.   Whereas right-handed quark mixing angles are not relevant for CKM mixing,
the large left-handed charged lepton mixing angle can give large $\nu_\mu - \nu_\tau$ mixing.
\end{itemize}
Let us now consider the last two mechanisms in more detail.

\subsubsection{SU(5) $\times$ U(1) flavor symmetry}

One popular possibility has the large $\nu_\mu - \nu_\tau$ mixing in the Dirac charged Yukawa matrix with
\begin{eqnarray}     &  Y_e  = \left( \begin{array}{ccc} \lambda^5 &
\lambda^4 & \lambda^2 \\ \lambda^{3} & \lambda^{2} & 1 \\
\lambda^3 & \lambda^{2} & 1 \end{array} \right) \lambda^4 & = Y_d^T .
\end{eqnarray}
The neutrino Dirac Yukawa matrix and the Majorana matrix are given by
\begin{eqnarray}     &  Y_\nu  = \left( \begin{array}{ccc} \lambda^3 &
\lambda & \lambda^2 \\ \lambda & 0 & 1 \\
\lambda & 0 & 1 \end{array} \right) &
\end{eqnarray}
\begin{eqnarray}     &  M_N  = \left( \begin{array}{ccc} \lambda^2 &
1 & \lambda \\ 1 & 0 & 0 \\
\lambda & 0 & 1 \end{array} \right) &
\end{eqnarray}
where we use the results of [Altarelli \etal (2000)].  The light neutrino mass matrix is given by the standard
See-Saw formula. We obtain:
\begin{eqnarray}     &  {\cal M}_\nu  = U_e^{tr} \ \left( \begin{array}{ccc} \lambda^4 &
\lambda^2 & \lambda^2 \\ \lambda^{2} & 1 & 1 \\
\lambda^2 & 1 & 1 \end{array} \right) \ U_e \ v_u^2/M  . &
\end{eqnarray}
where $v_u$ is the VEV of the Higgs doublet giving mass to the up quarks,  $M$ is the heavy Majorana mass scale
and {\em all entries in each matrix are specified up to order one coefficients}.  $U_e$ is the mixing matrix
taking left-handed charged leptons into the mass eigenstate basis.   It is given by
\begin{eqnarray}
 \left( \begin{array}{ccc} m_e^2 & 0 & 0 \\
                            0   &  m_\mu^2 & 0 \\
                            0 &  0 &  m_\tau^2  \end{array} \right) = &  U_e^{tr} \ ( Y_e \ Y_e^\dagger ) \ U_e^* \
                            \langle H_d \rangle^2 &
\end{eqnarray}
where
\begin{eqnarray}     &  Y_e \ Y_e^\dagger = \left( \begin{array}{ccc} \lambda^4 &
\lambda^2 & \lambda^2 \\ \lambda^{2} & 1 & 1 \\
\lambda^2 & 1 & 1 \end{array} \right)  \end{eqnarray} {\em up to order one coefficients}.   It is clear that
without order one coefficients, the neutrino mixing matrix is the identity matrix.   Hence the arbitrary order
one coefficients are absolutely necessary to obtain bi-large neutrino mixing.\footnote{I thank G. Altarelli,
private communication, for emphasizing this point.} Hierarchical neutrino masses with $m_3 \gg m_2 \gg m_1$ and
bi-large neutrino mixing can naturally be obtained [Altarelli and Feruglio (2003)].

Of course, different U(1) charge assignments for all the fields can lead to other experimentally acceptable
solutions to the solar neutrino problem.  For a review and further references, see [Altarelli and Feruglio
(2003)].

\subsubsection{$SO_{10} \times [SU_2 \times U_1^n]_{FS}$ model}
Within the context of the $SO_{10} \times [SU_2 \times U_1^n]_{FS}$ model, bi-large neutrino mixing is naturally
obtained using the mechanism of the minimal two Majorana neutrino sector.  The Dirac neutrino mass is fixed once
charged fermion masses and mixing angles are fit.  It is given by the formula:
\begin{eqnarray}
Y_{\nu} =&  \left(\begin{array}{ccc}  0 & - \epsilon' \ \omega & {3 \over 2} \ \epsilon \ \xi \ \omega  \\
      \epsilon'  \ \omega &  3 \ \tilde \epsilon \  \omega & {3 \over 2} \ \epsilon \ \omega \\
       - 3 \ \epsilon \ \xi \ \sigma   & - 3 \ \epsilon \ \sigma & 1 \end{array} \right) \; \lambda & \nn
 \end{eqnarray} with $\omega \;\; =  \;\; 2 \, \sigma/( 2 \,
\sigma - 1)$ and the Dirac neutrino mass matrix given by $ m_\nu \equiv Y_\nu \frac{v}{\sqrt{2}} \sin\beta $. Of
course, all the freedom is in the Majorana neutrino sector.  The FGY ansatz [Frampton \etal (2002)] is obtained
with the following Majorana neutrino sector [Raby (2003)]:
\begin{eqnarray} W_{neutrino} = & \frac{\overline{16}}{\hat M}  \left( N_1 \ \tilde \phi^a \ 16_a \ + \
N_2 \ \phi^a \ 16_a \ + \ N_3 \ \theta \ 16_3 \right) &  \nn \\ & + \frac{1}{2} \left( S_1 \ N_1^2 \ + \ S_2 \
N_2^2 \right)  & \nonumber
\end{eqnarray} where \{$N_i,  \; i = 1,2,3$\} are SO(10) and SU(2) - flavor singlets.  In this version of the
theory, the symmetric two index tensor flavon field $S^{a b}$ is replaced by an SU(2) doublet $\tilde \phi^a$
such that $S^{a b} \equiv \tilde \phi^a \ \tilde \phi^b / \hat M $.   Note, since the singlet $N_3$ has no large
Majorana mass, it gets a large Dirac mass by mixing directly with $\bar \nu_3$ at the GUT scale.  Thus $\bar
\nu_3$ is removed from the See-Saw mechanism and we effectively have only two right-handed neutrinos taking part.

Integrating out the heavy neutrinos we obtain the light neutrino mass matrix given by \bea {\cal M}_\nu = &
U_e^{tr} \; [ \ D^{tr} \ \hat M_N^{-1} \ D \ ] \; U_e & \eea where  $U_e$ is the unitary matrix diagonalizing the
charge lepton mass matrix and \bea  D^{tr}  \equiv  \left(
\begin{array}{cc}
a & 0  \\
a^\prime  & b  \\
0 & b^\prime
\end{array} \right), \;\;\;\;\; &  \hat M_N \equiv  \left(
\begin{array}{cc} \langle S_1 \rangle & 0
\\ 0 & \langle S_2 \rangle \end{array} \right) &  \eea
with
\begin{eqnarray} b \equiv & \epsilon^\prime \ \omega \ \lambda \ (M_2/\langle \phi^1 \rangle) \
\frac{\hat M}{ v_{16}} \frac{v \sin\beta}{\sqrt{2}} &  \\
b^\prime \equiv & - 3 \ \epsilon \ \xi \ \sigma \ \lambda \ (M_2/\langle \phi^1 \rangle) \ \frac{\hat M }{
v_{16}} \frac{v
\sin\beta}{\sqrt{2}} . & \nonumber \\
a \equiv & - \epsilon^\prime \ \omega \ \lambda \ (M_1/\langle \tilde \phi^2 \rangle) \ \frac{\hat M }{ v_{16}}
\frac{v
\sin\beta}{\sqrt{2}} & \nn \\
a^\prime \equiv & ( -\epsilon^\prime \ \xi^{-1} + 3 \ \tilde \epsilon ) \ \omega \ \lambda \ (M_1/\langle \tilde
\phi^2 \rangle) \ \frac{\hat M }{v_{16}} \frac{v \sin\beta}{\sqrt{2}} . & \nonumber \end{eqnarray} We obtain \beq
b \sim b^\prime \eeq naturally, since $\epsilon^\prime \ \sim \ \epsilon \ \xi$.  In addition we can accommodate
\beq a \sim a^\prime \eeq with minor fine-tuning O(1/10) since $\epsilon^\prime \ \xi^{-1} \ \sim \ \tilde
\epsilon.$ Note, this is the Frampton-Glashow-Yanagida ansatz [Frampton \etal (2002)] with a bi-large neutrino
mixing matrix obtained naturally in a \ss GUT.

\subsection{Flavor Violation \label{sec:flavorviol}}

Quarks and leptons come in different flavors:  up, down, charm, strange, top, bottom;  electron, muon, tau.  We
observe processes where bottom quarks can decay into charm quarks or up quarks. Hence quark flavors (for quarks
with the same electric charge) are interchangeable. In the standard model, this is parametrized by the CKM mixing
matrix. We now have direct evidence from neutrino oscillation experiments showing that tau, muon and electron
numbers are not separately conserved.  Yet, we have never observed muons changing into electrons.   In
supersymmetric theories there are many more possible ways in which both lepton and quark flavors can change. This
is because scalar quarks and leptons carry the flavor quantum numbers of their \ss partners.  Thus flavor
violation in the scalar sector can lead to flavor violation in the observed fermionic sector of the theory.  This
gives rise to the SUSY flavor problem. We consider two examples here:  $\mu \rightarrow e \gamma$ or $B_s
\rightarrow \mu^+ \mu^-$. We show why \gs and/or neutrino masses can cause enhanced flavor violation beyond that
of the standard model. In this section we consider several ways to solve the SUSY flavor problem.

\begin{figure}[t!] \vspace*{0.50in} \hspace*{0.80in}
\begin{center}
%\hspace*{-1.00in}
\begin{minipage}{4.00in}
\epsfig{file=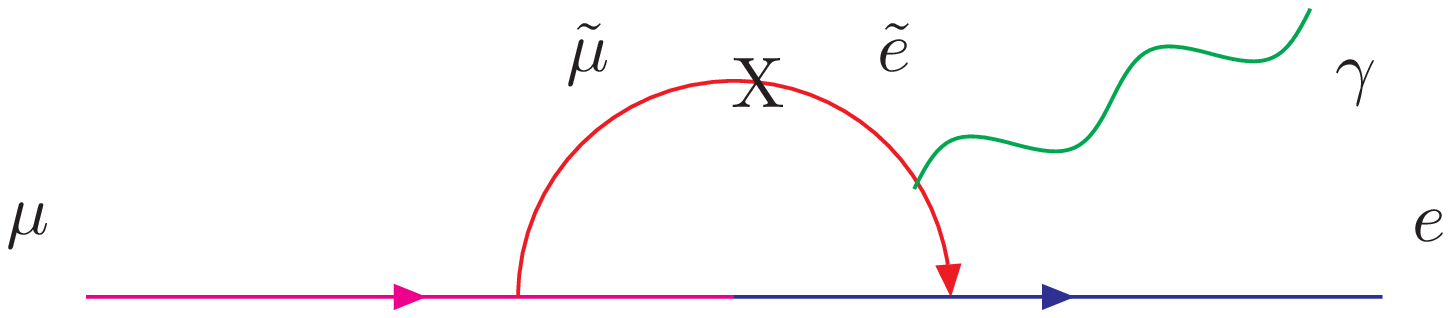,width=4.00in}
\end{minipage}
\end{center}
%\vspace*{-0.00in}
\begin{center} \caption{\label{fig:muegamma} The one loop contribution to the process $\mu \rightarrow e \gamma$
proportional to an off-diagonal scalar muon - electron mass term in the charged lepton mass eigenstate basis.}
\end{center}
\end{figure}

But first let us illustrate the problem with a few of the dominant examples [Gabbiani \etal (1996)].  In the
first column of Table \ref{t:flavorviol} we present four flavor violating observables with their experimental
bounds.  In the second and third columns we present the bounds on flavor violating scalar mass corrections \beq
\delta_{i j}^f \equiv \Delta {m^2_{i j}}^f/\bar m^2 \eeq where $i, j = 1,2,3$ are family indices, $\Delta {m^2_{i
j}}^f$ is an off-diagonal scalar mass insertion for $f =$ \{quark, lepton\} flavor (treated to lowest non-trivial
order in perturbation theory) in the flavor diagonal basis for quarks and leptons.  $\bar m$ is the average
squark or slepton mass squared.  The subscripts $LL$ refer to left-handed squark or slepton mass insertions.
There are separate limits on $RR$ and $LR$ mass insertions [Gabbiani \etal (1996)] which are not presented here.
The only difference between the second and third columns is the fiducial value of $\bar m^2$. Clearly as the mean
squark or slepton mass increases, the fine tuning necessary to avoid significant flavor violation is dramatically
reduced.  As seen in Fig. \ref{fig:muegamma}, the amplitude for $\mu \rightarrow e \gamma$ is proportional to
$\delta_{1 2}^e$ and is suppressed by $1/\bar m^2$, since it is an effective dimension 5 operator.

\begin{table}
\caption{\label{t:flavorviol}  Some constraints from the non-observation of flavor violation on squark, slepton
and gaugino masses [Gabbiani \etal (1996)].  For the electron electric dipole moment we use the relation $d_N^e
\sim 2 (100 /m_{\tilde l}({\rm GeV}))^2 sin\Phi_{A,B} \times 10^{-23} {\rm e \ cm}$.}
\begin{indented}
\item[] \hspace*{-1.00in} \begin{tabular}{@{}lcc} \br Observable & Experimental bound (1) & Experimental bound (2) \\
$B(\mu \rightarrow e \gamma) < 1.2 \times 10^{-11}$ & $|(\delta_{12}^l)_{LL}| <
2.1 \times 10^{-3} (m_{\tilde l}({\rm GeV})/100)^2$ &  $|(\delta_{12}^l)_{LL}| <  0.8 \ (m_{\tilde l}({\rm TeV})/2)^2$ \\
$\Delta m_{K} < 3.5 \times 10^{-12}$ MeV  & $\sqrt{|Re(\delta_{12}^d)^2_{LL}|} < 1.9 \times 10^{-2} (m_{\tilde
q}({\rm GeV})/500)$
& $\sqrt{|Re(\delta_{12}^d)^2_{LL}|} < 7.6 \times 10^{-2} \ (m_{\tilde q}({\rm TeV})/2)$ \\
$\epsilon_{K} < 2.28 \times 10^{-3}$   & $\sqrt{|Im(\delta_{12}^d)^2_{LL}|} < 1.5 \times 10^{-3} (m_{\tilde
q}({\rm GeV})/500)$ &
$\sqrt{|Re(\delta_{12}^d)^2_{LL}|} < 6.0 \times 10^{-3} \ (m_{\tilde q}({\rm TeV})/2)$  \\
$d_N^e  < 4.3 \times 10^{-27} {\rm e \ cm} $ & $sin\Phi_{A,B} < 4 \times 10^{-4} \times (m_{\tilde l}({\rm
GeV})/100)^2$ & $sin\Phi_{A,B}  <
0.16 \times (m_{\tilde l} \ ({\rm TeV})/2)^2$ \\
\end{tabular}
\end{indented}
\end{table}

\subsubsection{The Origin of flavor violation in \ss theories}

There are three possible ways to avoid large flavor violation.
\begin{enumerate}
\item  Having squarks and sleptons, with the same standard model gauge charges, be degenerate and, in addition,
the cubic scalar interactions proportional to the Yukawa matrices.

\item  Alignment of squark and slepton masses with quark and lepton masses.

The fermion and scalar mass matrices are ``aligned" when, in the basis where fermion masses are diagonal, the
scalar mass matrices and cubic scalar interactions are approximately diagonal as well.

\item  Heavy first and second generation squarks and sleptons.
\end{enumerate}
The CMSSM (or mSUGRA) is an example of the first case.  It has a universal scalar mass $m_0$ and tri-linear
scalar interactions proportional to Yukawa matrices.   These initial conditions correspond to a symmetry limit
[Hall \etal (1986)] -- dubbed minimal flavor violation [Ciuchini \etal (1998)]-- where the only flavor violation
occurs in the CKM matrix at the messenger scale for \ss breaking, or in this case, the Planck scale.
Gauge-mediated \ss breaking, where squarks and sleptons obtain soft \ss breaking masses via standard model gauge
interactions, is another example of the first case (for a review, see [Giudice and Rattazzi (1999)].  In this
case the messenger mass is arbitrary. Finally, within the context of perturbative heterotic string theory,
dilaton \ss breaking gives universal scalar masses at the string scale. For moduli \ss breaking, on the other
hand, scalar masses depend on modular weights, whose values are very model dependent.

Abelian flavor symmetries can be used to align quark (lepton) and the corresponding squark (slepton) mass
matrices, but they still require one of the above mechanisms for obtaining degenerate scalar masses at zeroth
order in symmetry breaking. Non-abelian symmetries, on the other hand, can both align fermion and scalar mass
matrices and guarantee the degeneracy of the scalar masses at zeroth order.

Finally, since the most stringent limits from flavor violating processes come from the lightest two families, if
the associated squarks and sleptons are heavy these processes are suppressed [Dimopoulos and Giudice (1995)]. A
natural mechanism for obtaining this inverted scalar mass hierarchy with the first and second generation scalars
heavier than the third was discussed by [Bagger \etal (1999,2000)].

If the messenger scale for \ss breaking is above the GUT scale or even above the See-Saw scale for neutrino
masses then squark and slepton masses can receive significant flavor violating radiative corrections due to this
beyond the standard model physics [Hall \etal (1986), Georgi (1986), Borzumati and Masiero (1986), Leontaris
\etal (1986), Barbieri \etal(1995a,b), Hisano \etal (1995,1996)].  In addition, if the fermion mass hierarchy is
due to flavor symmetry breaking using a Froggatt-Nielsen mechanism, then upon integrating out the heavy
Froggatt-Nielsen sector new flavor violating soft \ss breaking terms may be induced [Dimopoulos and Pomarol
(1995), Pomarol and Dimopoulos (1995)]. Hence the low energy MSSM is sensitive to physics at short distances.
This is both a problem requiring natural solutions and a virtue leading to new experimentally testable
manifestations of \ss theories.

For example in \ss GUTs, color triplet Higgs  fields couple quarks to leptons.   As a consequence flavor mixing
in the quark sector can, via loops, cause flavor mixing, proportional to up quark Yukawa couplings, in the lepton
sector. This comes via off-diagonal scalar lepton masses in the basis where charged lepton masses are diagonal.
While the one loop contribution of charm quarks give a branching ratio $Br(\mu \rightarrow e \gamma) \sim
10^{-15}$ [Hall \etal (1986)], the top quark contribution leads to very observable rates [Barbieri \etal (1995a)]
near the experimental bounds. Moreover, new experiments will soon test these results.  In addition, experimental
evidence for neutrino oscillations makes it clear that the lepton sector has its own intrinsic flavor violation.
In the standard model, these effects are suppressed by extremely small ($<$ eV) neutrino masses. In \ss however,
flavor violation in the (s)neutrino sector leads, again via loops, to eminently observable mixing in the charged
(s)lepton sector [Hisano \etal (1995,1996)].  There are a large number of papers in the literature which try to
use low energy neutrino oscillation data in an attempt to predict rates for lepton flavor violation.   However a
bottom-up approach is fraught with the problem that low energy oscillation data cannot completely constrain the
neutrino sector [Casas and Ibarra (2001), Lavignac \etal (2001,2002)].  It has been shown that neutrino
oscillation data and $l_i \rightarrow l_j \ \gamma$ measurements can nevertheless provide complementary
information on the See-Saw parameter space [Davidson and Ibarra (2001), Ellis \etal (2002)]. In a recent
analysis, it was shown that lepton flavor violation can constrain typical \ss SO(10) theories [Masiero \etal
(2003)]. Finally, it is important to note, that the same physics can lead to enhanced contributions to flavor
conserving amplitudes such as the anomalous magnetic moment of the neutrino ($a_\mu$) [Chattopadhyay and Nath
(1996), Moroi (1996)] and the electric dipole moments of the electron ($d^e_e$) and neutron ($d^n_e$) [Dimopoulos
and Hall (1995), Hisano and Tobe (2001), Demir \etal (2003)].  Moreover, the rates for these flavor violating
processes increase with $\tan\beta$.

\begin{figure}[t!] \vspace*{0.50in} \hspace*{0.80in}
\begin{center}
%\hspace*{-1.00in}
\begin{minipage}{4.00in}
\epsfig{file=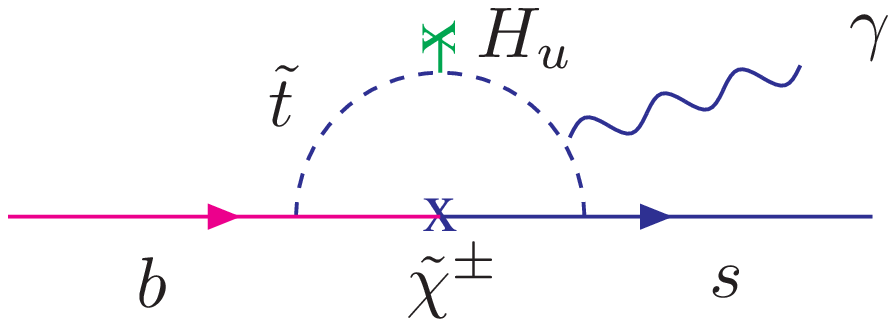,width=3.50in}
\end{minipage}
\end{center}
%\vspace*{-0.00in}
\begin{center} \caption{\label{fig:bsgamma} The one loop chargino contribution to the
process $B \rightarrow X_s \gamma$ proportional to $\lambda_t$ and $\tan\beta$.}
\end{center}
\end{figure}

Let us now consider flavor violating hadronic interactions at large $\tan\beta$.  We focus on a few important
examples, in particular, the processes $B \rightarrow X_s \gamma$, $B \rightarrow X_s \ l^+ \ l^-$
forward-backward asymmetry and $B_s \rightarrow \mu^+ \mu^-$.  For a more comprehensive study, see [Hall \etal
(1994), Hempfling (1994), Carena \etal (1994), Bla\v{z}ek \etal (1995), Chankowski and Pokorski (1997), Misiak
\etal (1998), Huang and Yan (1998), Huang \etal (1999), Hamzaoui \etal (1999), Babu and Kolda (2000), Chankowski
and Slawianowska (2001), Carena \etal (2001), Bobeth \etal (2001), Huang \etal (2001), Dedes \etal (2001),
Isidori and Retico (2001), Buras \etal (2002,2003), Dedes and Pilaftsis (2003)].   The standard model
contribution to $B \rightarrow X_s \gamma$ has significant uncertainties, but the calculated branching ratio is
consistent with the latest experimental data. Supersymmetry contributions are typically divided into two
categories, i.e. the contributions contained in a two Higgs doublet model and then the rest of the \ss spectrum.
The charged Higgs contribution has the same sign as the standard model contribution and thus increases the
predicted value for the branching ratio. This spoils the agreement with the data and thus a lower limit on the
charged Higgs mass is obtained. In the minimal flavor \ss scenario, the additional \ss contribution is dominated
by the chargino loop (Fig. \ref{fig:bsgamma}).  The sign of this term depends on the sign of $\mu$. For $\mu > 0$
(this defines my conventions) the chargino contribution is the opposite sign of the standard model contribution
to the coefficient $C_7$ of the magnetic moment operator $O_7 \sim \bar s_L \ \Sigma_{\mu \nu} \ b_R \ F^{\mu
\nu}$.   Moreover, this contribution is proportional to $\tan\beta$.  For small or moderate values of $\tan\beta$
the \ss correction to $C_7$ is small and for $\mu > 0$ it tends to cancel the charged Higgs and standard model
contributions.   This is in the right direction, giving good agreement with the data.  For $\mu < 0$ the
agreement with the data gets worse and can only work for large Higgs and squark masses, so that the overall \ss
contribution is small.    On the other hand, for $\mu > 0$ and large $\tan\beta \sim 50$ there is another
possible solution with the total \ss contribution to $C_7$ equal to twice the standard model contribution but
with opposite sign.  In this case $C_7^{total} = C_7^{SM} + C_7^{\ss} \approx - C_7^{SM}$ and good fits to the
data are obtained [Bla\v{z}ek and Raby (1999)]. Although the sign of $C_7$ is not observable in $B_s \rightarrow
X_s \gamma$, it can be observed by measuring the forward - backward asymmetry in the process $B \rightarrow X_s \
l^+ \ l^-$ [Huang and Yan (1998), Huang \etal (1999), Lunghi \etal (2000), Ali \etal (2002), Bobeth \etal (2003)]
where forward (backward) refers to the positive lepton direction with respect to the B flight direction in the
rest frame of the di-lepton system.\footnote[1]{I thank K. Tobe for pointing out this possibility to me.}

\subsubsection{Flavor violating Higgs couplings at large $\tan\beta$}

The MSSM has two Higgs doublets which at tree level satisfy the Glashow - Weinberg condition for natural flavor
conservation.  Up quarks get mass from $H_u$ and down quarks and charged leptons get mass from $H_d$.  Thus when
the fermion mass matrices are diagonalized (and neglecting small neutrino masses) the Higgs couplings to quarks
and leptons are also diagonal.  However, this is no longer true once \ss is broken and radiative corrections are
considered.   In particular,  for large values of $\tan\beta$ the coupling of $H_u$ to down quarks, via one loop
corrections, results in significant flavor violating vertices for neutral and charged Higgs.  These one loop
corrections to the Higgs couplings contribute to an effective Lagrangian (Eqn. \ref{eq:effhiggs}) [Bla\v{z}ek
\etal (1995), Chankowski and Pokorski (1997)].   The chargino contribution (Fig. \ref{fig:delta_mb}) is
proportional to the square of the up quark Yukawa matrix, which is not diagonal in the diagonal down quark mass
basis.  As a result, at one loop order, the down quark mass matrix is no longer diagonal (Eqn. \ref{eq:md}). This
leads to $\tan\beta$ enhanced corrections to down quark masses and to CKM matrix elements [Bla\v{z}ek \etal
(1995)].  Upon re-diagonalizing the down quark mass matrix we obtain the effective flavor violating Higgs - down
quark Yukawa couplings given in Eqns. \ref{eq:higgscoupling1} and \ref{eq:higgscoupling2} [Chankowski and
Pokorski (1997), Babu and Kolda (2000), Chankowski and Slawianowska (2001), Bobeth \etal (2001), Huang \etal
(2001), Dedes \etal (2001), Isidori and Retico (2001), Buras \etal (2002,2003), Dedes and Pilaftsis (2003)].
\begin{eqnarray}
{\cal {L}}_{eff}^{ddH}= &  -\bar{d}_{Li} \ \lambda_{di}^{diag} \ d_{Ri} \ H_d^{0*} & \label{eq:effhiggs}
\\ & -\bar{d}_{Li} \ \Delta \lambda_d^{ij} \ d_{Rj} \ H_d^{0*} & \nonumber \\ & -\bar{d}_{Li} \ \delta
\lambda_d^{ij} \ d_{Rj} \ H_u^0 + {\rm h.c.} & \nonumber
\end{eqnarray}
\begin{eqnarray}
m_{d}^{Diagonal} &=& V_d^L \ \left[\lambda_d^{diag} +\Delta \lambda_d + \delta \lambda_d \tan\beta  \right] \
V_d^{R\dagger} \ \frac{v\cos\beta}{\sqrt{2}} \label{eq:md}
\end{eqnarray}
\begin{eqnarray}
{\cal {L}}_{FV}^{i\neq j} =& -\frac{1}{\sqrt{2}}\bar{d'}_i \left[
F^h_{ij} \ P_R+ F^{h*}_{ji} \ P_L \right] d'_j \ h & \label{eq:higgscoupling1}   \\
& -\frac{1}{\sqrt{2}}\bar{d'}_i \left[ F^H_{ij} \ P_R + F^{H*}_{ji} \ P_L \right] d'_j \ H &
\nonumber \\
& -\frac{i}{\sqrt{2}}\bar{d'}_i \left[ F^A_{ij} \ P_R + F^{A*}_{ji} \ P_L \right] d'_j \ A , & \nonumber
\end{eqnarray}
where
\begin{eqnarray}
F^h_{ij} &\simeq& \delta \lambda_d^{ij}  (1+\tan^2\beta) \cos\beta \ \cos(\alpha-\beta), \label{eq:higgscoupling2} \\
F^H_{ij} &\simeq& \delta \lambda_d^{ij}  (1+\tan^2\beta) \cos\beta \ \sin(\alpha-\beta), \nonumber \\
F^A_{ij} &\simeq& \delta \lambda_d^{ij} (1+\tan^2\beta) \ \cos\beta . \nonumber
\end{eqnarray}

\begin{figure}[t!] \vspace*{0.50in} \hspace*{0.80in}
\begin{center}
\hspace*{-1.00in}
\begin{minipage}{4.00in}
\epsfig{file=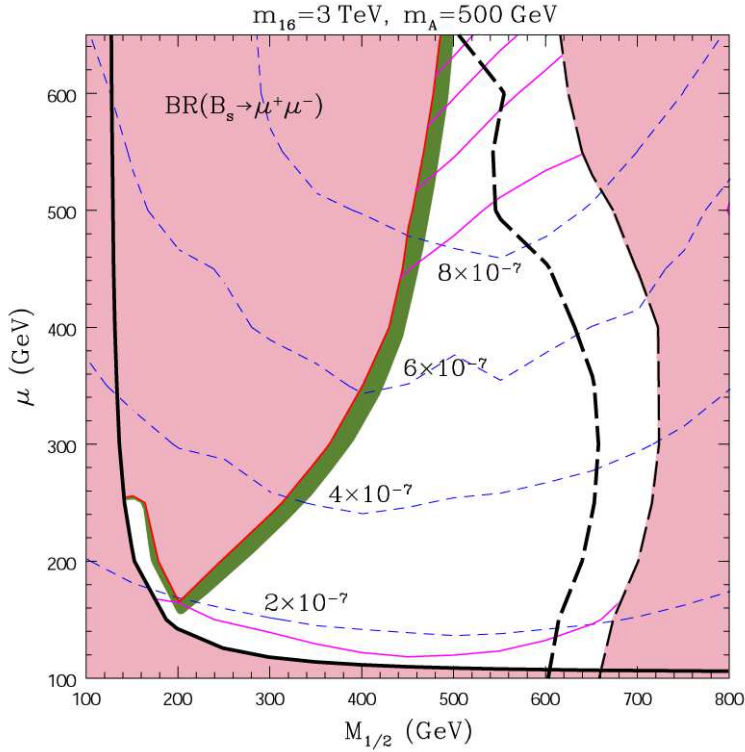,width=4.00in}
\end{minipage}
\end{center}
%\vspace*{-0.00in}
\begin{center} \caption{\label{fig:bsmm} The dashed (blue) lines are contours for constant branching ratio
$B(B_s \rightarrow \mu^+ \mu^-)$ as a function of $\mu, \ M_{1/2}$ for $m_{16} = 3$ TeV.  The green shaded region
is consistent with the recent WMAP data for dark matter abundance of the neutralino LSP. The light shaded region
in the lower left hand corner (separated by the solid line) is excluded by chargino mass limits, while the light
shaded region in the upper left (right side) is excluded by a cosmological dark matter abundance which is too
large (Higg mass which is too light).}
\end{center}
\end{figure}

This leads to $\tan\beta$ enhanced flavor violating couplings for the neutral Higgs bosons.  For example, the
branching ratio $B(B_s \rightarrow \mu^+ \mu^-)$ is proportional to $\tan\beta^4$ and inversely proportional to
the fourth power of the CP odd Higgs mass $m_A$ [Babu and Kolda (2000), Chankowski and Slawianowska (2001),
Bobeth \etal (2001), Huang \etal (2001), Dedes \etal (2001), Isidori and Retico (2001), Buras \etal (2002,2003),
Dedes and Pilaftsis (2003)], since the contributions of the two CP even Higgs bosons approximately cancel [Babu
and Kolda (2000)]. The present D0 and CDF bounds constrain $m_A \geq 250$ GeV for $\tan\beta \sim 50$, although
this result is somewhat model dependent [Derm\' \i \v sek \etal (2003)].  Note the D0 and CDF bounds from the
Tevatron Run 2 now give $B(B_s \rightarrow \mu^+ \mu^-) < 1.2 \times 10^{-6}$ (CDF) [Lin (2003)] at 95\% CL and
$< 1.6 \times 10^{-6}$ (D0) [Kehoe (2003)]. An order of magnitude improvement will test large $\tan\beta$ \ss for
$m_A$ up to $\sim 500$ GeV [Derm\' \i \v sek \etal (2003)] (see Fig. \ref{fig:bsmm}).

\section{\ss Dark Matter \label{sec:darkmatter}}

The two most popular dark matter candidates are axions or the LSP of SUSY.   Both are well motivated cold dark
matter candidates.   There have been several recent studies of SUSY dark matter in light of the recent WMAP data.
For these analyses and references to earlier works, see [Roszkowski \etal (2003), Ellis \etal (2003c,d,e,f),
Chattopadhyay \etal (2003)]. These calculations have been performed with different assumptions about soft \ss
breaking parameters, assuming the CMSSM boundary conditions at $M_G$ [Ellis \etal (2003bc,d,f), Chattopadhyay
\etal (2003)] or arbitrary low energy scalar masses [Ellis \etal (2003d)].  Soft \ss breaking outside the realm
of the CMSSM has also been considered.  For example, soft breaking with non-universal Higgs masses have been
analyzed recently by [Ellis \etal (2003a,b), Roszkowski \etal (2003)].  In the latter case, the soft \ss breaking
parameters consistent with SO(10) Yukawa unification were studied.

In the limit of large squark and slepton masses, it is important to have efficient mechanisms for dark matter
annihilation. Most recent studies have focused on the neutralino LSP in the limit of large $\tan\beta$ and/or the
focus point limit.  In both cases there are new mechanisms for efficient neutralino annihilation. For large
$\tan\beta \geq 40$ neutralino annihilation via direct s-channel neutral Higgs boson exchange dominates
[Roszkowski \etal (2001), Ellis \etal (2001a,b)]. In this limit the CP even and odd Higgs bosons have large
widths, due to their larger coupling to bottom quarks and $\tau$ leptons. In the focus point limit, on the other
hand, the neutralino LSP is a mixed Higgsino - gaugino state. Thus it has more annihilation channels than the
pure bino LSP case, valid for values of the universal scalar mass $m_0 < $ TeV [Feng \etal (2000c,2001)].

Direct detection [Ellis \etal (2003b,e), Roszkowski \etal (2003), Munoz (2003)] and/or indirect detection [Baer
and Farrill (2003), de Boer \etal (2003)] of neutralino dark matter has also been considered.  In fact, [de Boer
\etal (2003)] suggests that some indirect evidence for \ss dark matter already exists.

\section{Open questions \label{sec:open}}

It is beyond the scope of this review to comment on many other interesting topics affected by supersymmetric
theories.  Several effective mechanisms for generating the matter- anti-matter asymmetry of the universe have
been suggested, including the Affleck-Dine mechanism [Affleck and Dine (1985)], which is purely a supersymmetric
solution, or leptogenesis [Fukugita and Yanagida (1986)], which is not necessarily supersymmetric.   There have
also been many interesting studies of inflation in a \ss context.   Finally, we have only made passing reference
to superstring theories and \ss breaking mechanisms or fermion masses there.

Simple "naturalness" arguments would lead one to believe that SUSY should have already been observed.  On the
other hand,  ``focus point" or ``minimal SO(10) SUSY" regions of soft \ss breaking parameter space extend to
significantly heavier squark and slepton masses {\em without giving up on ``naturalness."} In both the ``focus
point" and ``mSO$_{10}$SM" regions of parameter space we expect a light Higgs with mass of order $114 - 120$ GeV.
Both ameliorate the \ss flavor problem with heavy squark and slepton masses.  They are nevertheless both
surprisingly consistent with cosmological dark matter abundances.  In addition, we have shown that the
``mSO$_{10}$SM" satisfies Yukawa coupling unification with an inverted scalar mass hierarchy.   Thus one finds
first and second generation squarks and sleptons with mass of order several TeV, while gauginos and third
generation squarks and sleptons are much ligher.   In addition, it requires large values of $\tan\beta \sim 50$
resulting in enhanced flavor violation.

\gs are the most natural extensions of the standard model, and thus they are the new ``standard model" of
particle physics. The mSUGRA (or CMSSM) boundary conditions at the GUT scale provide excellent fits to precision
low energy electroweak data. \ss GUTs, besides predicting gauge coupling unification, also provide a framework
for resolving the gauge hierarchy problem and understanding fermion masses and mixing angles, including
neutrinos. It also gives a natural dark matter candidate, and a framework for leptogenesis and inflation.

BUT there are two major challenges with any supersymmetric theory.  We do not know how \ss is spontaneously
broken or the origin of the $\mu$ term.    We are thus unable to predict the \ss particle spectrum, which makes
\ss searches very difficult.   Nevertheless, ``naturalness" arguments always lead to some light \ss sector,
observable at the LHC, a light Higgs, with mass less than O(135 GeV), or observable flavor violating rates beyond
that of the standard model.  Assuming \ss particles are observed at the LHC, then the fun has just begun.   It
will take many years to prove that it is really supersymmtry.   Assuming \ss is established, a \ss desert from
$M_Z$ to $M_G$ (or $M_N$) becomes highly likely.   Thus precision measurements at the LHC or a Linear Collider
will probe the boundary conditions at the very largest and fundamental scales of nature.   With the additional
observation of proton decay and/or precise GUT relations for sparticle masses, \gs can be confirmed.   Hence with
experiments at TeV scale accelerators or in underground detectors for proton decay, neutrino oscillations or dark
matter, the fundamental superstring physics can be probed.   Perhaps then we may finally understand who ordered
three families. It is thus no wonder why the elementary particle physics community is {\em desperately seeking
SUSY}.

\section*{References}
\begin{harvard}
\item[] Affleck I and Dine M 1985
%``A New Mechanism For Baryogenesis,''
{\it Nucl. Phys. B} {\bf 249} 361

\item[] Albright C H and Barr S M 2000a {\it Phys. Rev. Lett.} {\bf 85} 244 [arXiv:hep-ph/0002155]

\item[] Albright C H and Barr S M 2000b
%``Construction of a minimal Higgs SO(10) SUSY GUT model,''
Phys.\ Rev.\ D {\bf 62}, 093008 (2000) [arXiv:hep-ph/0003251].

\item[] Albright C H and Barr S M 2001 {\it Phys. Rev. D}{\bf 64} 073010 [arXiv:hep-ph/0104294]

\item[] Ali A, Lunghi E, Greub C and Hiller G 2002
%``Improved model-independent analysis of semileptonic and radiative rare  B decays,''
{\it Phys. Rev. D} {\bf 66} 034002 [arXiv:hep-ph/0112300]

\item[] Allanach B C, King S F, Leontaris G K and Lola S 1997
%``Yukawa textures from family symmetry and unification,''
{\it Phys. Lett. B} {\bf 407} 275 [arXiv:hep-ph/9703361]

\item[] Altarelli G, Feruglio F and Masina I 2000 {\it JHEP} {\bf 0011} 040 [arXiv:hep-ph/0007254]

\item[] Altarelli G, Feruglio F 2001
%``SU(5) grand unification in extra dimensions and proton decay,''
{\it Phys. Lett. B}{\bf 511} 257 [arXiv:hep-ph/0102301]

\item[] Altarelli G and Feruglio F 2003
%``Phenomenology of neutrino masses and mixings,''
[arXiv:hep-ph/0306265]

\item[] Altarelli G, Feruglio F and Masina I 2003
%``Models of neutrino masses: Anarchy versus hierarchy,''
{\it JHEP} {\bf 0301} 035 [arXiv:hep-ph/0210342]

\item[] Alvarez-Gaume L, Claudson M and Wise M B 1982 {\it Nucl. Phys. B} {\bf 207} 96

\item[] Alvarez-Gaume L, Polchinski J and Wise M B 1983 {\it Nucl. Phys. B} {\bf 221} 495

\item[] Amaldi U, de Boer W and F\"urstenau H 1991 {\it Phys. Lett.} {\bf B260} 447

\item[] Ananthanarayan B, Lazarides G and Shafi Q 1991 {\it Phys. Rev. D} {\bf 44} 1613

\item[] Ananthanarayan B, Lazarides G and Shafi Q 1993 {\it Phys. Lett. B} {\bf 300} 245

\item[] Ananthanarayan B, Shafi Q and Wang X M 1994 {\it Phys. Rev. D} {\bf 50} 5980

\item[] Anderson G W and Casta\~{n}o D J 1995 {\it Phys. Lett. B} {\bf 347} 300

\item[] Anderson G W, Raby S, Dimopoulos S and Hall L J 1993
%``Precise predictions for m(t), V(cb), and tan Beta,''
{\it Phys. Rev. D} {\bf 47} 3702 [arXiv:hep-ph/9209250]

\item[] Anderson G, Raby S, Dimopoulos S, Hall L J and Starkman G D 1994 {\it Phys. Rev. D} {\bf 49} 3660
[arXiv:hep-ph/9308333]

\item[] Aoki S \etal (JLQCD) 2000 {\it Phys. Rev. D} {\bf 62} 014506

\item[] Aoki Y \etal (RBC) 2002 [arXiv:hep-lat/0210008]

\item[] Appelquist T, Dobrescu B A, Ponton E and Yee H U 2001
%``Proton stability in six dimensions,''
{\it Phys. Rev. Lett.}  {\bf 87} 181802 [arXiv:hep-ph/0107056]

\item[] Aranda A, Carone C D and Lebed R F 2000
%``U(2) flavor physics without U(2) symmetry,''
{\it Phys. Lett. B} {\bf 474} 170 [arXiv:hep-ph/9910392]

\item[] Arkani-Hamed N, Cheng H C and Hall L J 1996
%``A Supersymmetric Theory of Flavor with Radiative Fermion Masses,''
{\it Phys. Rev. D}{\bf 54} 2242 [arXiv:hep-ph/9601262]

\item[] Arkani-Hamed N, Dimopoulos S and Dvali G R 1998
%``The hierarchy problem and new dimensions at a millimeter,''
{\it Phys. Lett. B} {\bf 429} 263 [arXiv:hep-ph/9803315]

\item[] Aulakh C S, Bajc B, Melfo A, Senjanovic G and Vissani F 2003
%``The minimal supersymmetric grand unified theory,''
[arXiv:hep-ph/0306242]

\item[] Auto D, Baer H, Balazs C, Belyaev A, Ferrandis J and Tata X 2003 {\it JHEP} {\bf 0306}, 023
[arXiv:hep-ph/0302155]

\item[] Babu K S and Barr S M 1993
%``Natural suppression of Higgsino mediated proton decay in supersymmetric
%SO(10),''
{\it Phys. Rev. D} {\bf 48} 5354 [arXiv:hep-ph/9306242]

\item[] Babu K S and Barr S M 2002
%``Eliminating the d = 5 proton decay operators from SUSY GUTs,''
{\it Phys. Rev. D} {\bf 65} 095009 [arXiv:hep-ph/0201130]

\item[] Babu K S and Kolda C 2000 {\it Phys. Rev. Letters}{\bf 84} 228

\item[] Babu K S and Mohapatra R N 1995 {\it Phys. Rev. Lett.}{\bf 74} 2418

\item[] Babu K S, Pati J C and Wilczek F 2000 {\it Nucl. Phys. B} {\bf 566} 33

\item[] Babu K S and Strassler M J 1998 [arXiv:hep-ph/9808447]

\item[] Baer H, Balazs C, Belyaev A and O'Farrill J 2003
%``Direct detection of dark matter in supersymmetric models,''
{\it JCAP} {\bf 0309} 007 [arXiv:hep-ph/0305191]

\item[] Baer H, Balazs C, Mercadante P, Tata X and Wang Y 2001a
%``Viable supersymmetric models with an inverted scalar mass hierarchy at  the GUT scale,''
{\it Phys. Rev. D} {\bf 63} 015011 [arXiv:hep-ph/0008061]

\item[] Baer H, Balazs C, Brhlik M, Mercadante P, Tata X and Wang Y 2001b
%``Aspects of supersymmetric models with a radiatively driven inverted  mass hierarchy,''
{\it Phys. Rev. D} {\bf 64} 015002 [arXiv:hep-ph/0102156]

\item[] Baer H, Mercadante P and Tata X 2000
%``Calculable sparticle masses with radiatively driven inverted mass  hierarchy,''
{\it Phys. Lett. B} {\bf 475} 289 [arXiv:hep-ph/9912494]

\item[] Baer H and Ferrandis J 2001 {\it Phys. Rev. Lett.} {\bf 87} 211803

\item[] Baer H and Farrill J O 2003
 %``Probing Neutralino Resonance Annihilation via Indirect Detection of Dark
%Matter,''
[arXiv:hep-ph/0312350]

\item[] Bagger J, Dimopoulos S, Georgi H and Raby S 1984
%``Theories Of Fermion Masses,''
{\it Proceedings of the 5th Workshop on Grand Unification, Providence, RI, Apr 12-14, 1984}

\item[] Bagger J A, Feng J L and Polonsky N 1999
%``Naturally heavy scalars in supersymmetric grand unified theories,''
{\it Nucl. Phys. B} {\bf 563} 3 [arXiv:hep-ph/9905292]

\item[] Bagger J A, Feng J L, Polonsky N and Zhang R-J 2000 {\it Phys. Lett. B} {\bf 473} 264
[arXiv:hep-ph/9911255]

\item[] Bajc B, Perez P F and Senjanovic G 2002
%``Proton decay in minimal supersymmetric SU(5),''
{\it Phys. Rev. D} {\bf 66} 075005 [arXiv:hep-ph/0204311]

\item[] Bajc B, Senjanovic G and Vissani F 2003
%``b - tau unification and large atmospheric mixing: A case for non-canonical see-saw,''
{\it Phys. Rev. Lett.}{\bf 90} 051802 [arXiv:hep-ph/0210207]

\item[] Banks T 1988 {\it Nucl. Phys. B} {\bf 303} 172

\item[] Banks T and Kaplunovsky V 1983 {\it Nucl. Phys. B} {\bf 211} 529

\item[] Barbieri R, Dvali G R and Hall L J 1996
%``Predictions From A U(2) Flavour Symmetry In Supersymmetric Theories,''
{\it Phys. Lett. B} {\bf 377} 76 [arXiv:hep-ph/9512388].

\item[] Barbieri R, Ferrara S and Savoy C A 1982
%``Gauge Models With Spontaneously Broken Local Supersymmetry,''
{\it Phys. Lett. B} {\bf 119} 343

\item[] Barbieri R and Giudice G F 1988 {\it Nucl. Phys. B} {\bf 306} 63

\item[] Barbieri R and Hall L J 1997
%``A grand unified supersymmetric theory of flavor,''
{\it Nuovo Cim. A} {\bf 110} 1 [arXiv:hep-ph/9605224]

\item[] Barbieri R, Hall L J, Raby S and Romanino A 1997a {\it Nucl. Phys. B} {\bf 493} 3

\item[] Barbieri R, Hall L J and Romanino A 1997b
%``Consequences of a U(2) flavour symmetry,''
{\it Phys. Lett. B} {\bf 401} 47 [arXiv:hep-ph/9702315]

\item[] Barbieri R, Hall L J and Romanino A 1999
%``Precise tests of a quark mass texture,''
{\it Nucl. Phys. B} {\bf 551} 93 [arXiv:hep-ph/9812384]

\item[] Barbieri R, Hall L J and Strumia A 1995a {\it Nucl. Phys. B}{\bf 445} 219 [arXiv:hep-ph/9501334]

\item[] Barbieri R, Hall L J and Strumia A 1995b
%``Hadronic flavor and CP violating signals of superunification,''
{\it Nucl. Phys. B} {\bf 449} 437 [arXiv:hep-ph/9504373]

\item[] Barbieri R and Strumia A 1998 {\it Phys. Lett. B} {\bf 433} 63 hep-ph/9801353

\item[] Barger V D, Berger M S and Ohmann P 1993 {\it Phys. Rev. D} {\bf 47} 1093 [arXiv:hep-ph/9209232]

\item[] Barger V, Marfatia D and Whisnant K 2003
%``Progress in the physics of massive neutrinos,''
{\it Int. J. Mod. Phys. E} {\bf 12} 569 [arXiv:hep-ph/0308123]

\item[] Barr S M and Dorsner I 2003 {\it Phys. Lett. B} {\bf 556} 185 [arXiv:hep-ph/0211346]

\item[] Berezhiani Z G 1983 {\it Phys. Lett. B} {\bf 129} 99

\item[] Berezhiani Z G 1985 {\it Phys. Lett. B} {\bf 150} 177

\item[] Berezhiani Z 1998
%``Unified picture of the particle and sparticle masses in SUSY GUT,''
{\it Phys. Lett. B} {\bf 417} 287 [arXiv:hep-ph/9609342]

\item[] Berezhiani Z and Rossi A 2001 {\it Nucl. Phys. B} {\bf 594} 113

\item[] Berezinsky V, Bottino A, Ellis J R, Fornengo N, Mignola G and Scopel S 1996
%``Searching for relic neutralinos using neutrino telescopes,''
{\it Astropart. Phys.}  {\bf 5}, 333 [arXiv:hep-ph/9603342]

\item[] Binetruy P, Lavignac S and Ramond P 1996
%``Yukawa textures with an anomalous horizontal abelian symmetry,''
{\it Nucl. Phys. B}{\bf 477} 353 [arXiv:hep-ph/9601243]

\item[] Bla\v{z}ek T, Carena M, Raby S and Wagner C 1997a
 %``A global chi**2 analysis of electroweak data (including fermion masses  and
%mixing angles) in SO(10) SUSY GUTs,''
{\it Nucl. Phys. Proc. Suppl. A}  {\bf 52} 133 [arXiv:hep-ph/9608273]

\item[] Bla\v zek T, Carena M, Raby S and Wagner C E M 1997b {\it Phys. Rev. D} {\bf 56} 6919
%``A global chi**2 analysis of electroweak data in SO(10) SUSY GUTs,''
[arXiv:hep-ph/9611217]

\item[] Bla\v{z}ek T, Derm\' \i \v sek R and Raby S 2002a {\it Phys. Rev. Lett.} {\bf 88} 111804

\item[] Bla\v{z}ek T, Derm\' \i \v sek R and Raby S 2002b {\it Phys. Rev. D} {\bf 65} 115004
[arXiv:hep-ph/0201081]

\item[] Bla\v{z}ek T, Pokorski S and Raby S 1995 {\it Phys. Rev. D}{\bf 52} 4151

\item[] Bla\v{z}ek T and Raby S 1999
%``b $\to$ s gamma with large tan(beta) in MSSM analysis constrained by a  realistic SO(10) model,''
{\it Phys. Rev. D} {\bf 59} 095002 [arXiv:hep-ph/9712257]

\item[] Bla\v{z}ek T, Raby S and Tobe K 1999 {\it Phys. Rev. D} {\bf 60} 113001 [arXiv:hep-ph/9903340]

\item[] Bla\v{z}ek T, Raby S and Tobe K 2000 {\it Phys. Rev. D} {\bf 62} 055001 [arXiv:hep-ph/9912482]

\item[] Bobeth C, Ewerth T, Kruger F and Urban J 2001
%``Analysis of neutral Higgs-boson contributions to the decays anti-B/s  $\to$ l+ l- and anti-B $\to$ K l+ l-,''
{\it Phys. Rev. D} {\bf 64} 074014 [arXiv:hep-ph/0104284]

\item[] Bobeth C, Gambino P, Gorbahn M and Haisch U 2003
%``Complete NNLO QCD Analysis of B $\to$ X_s l^+ l^- and Higher Order Electroweak Effects,''
[arXiv:hep-ph/0312090]

\item[] Borzumati F and Masiero A 1986 {\it Phys. Rev. Lett.}{\bf 57} 961

\item[] Brodsky S J, Ellis J, Hagelin J S and Sachradja C 1984 {\it Nucl. Phys. B}{\bf 238} 561

\item[] Buchmuller W 2001 hep-ph/0107153

\item[] Buchmuller W, Di Bari P and Plumacher M 2003
%``The neutrino mass window for baryogenesis,''
{\it Nucl. Phys. B} {\bf 665} 445 [arXiv:hep-ph/0302092]

\item[] Buras A J, Chankowski P H, Rosiek J and Slawianowska L 2002
%``Correlation between Delta M(s) and B/(s,d)0 $\to$ mu+ mu- in  supersymmetry at large tan(beta),''
{\it Phys. Lett. B} {\bf 546} 96 [arXiv:hep-ph/0207241]

\item[] Buras A J, Chankowski P H, Rosiek J and Slawianowska L 2003
%``Delta(M(d,s)), B/(d,s)0 $\to$ mu+ mu- and B $\to$ X/s gamma in supersymmetry at large tan(beta),''
{\it Nucl. Phys. B}{\bf 659} 3 [arXiv:hep-ph/0210145]

\item[] Carena M \etal 1994 {\it Nucl. Phys. B}{\bf 426} 269

\item[] Carena M \etal 1995 {\it Phys. Lett. B}{\bf 355} 209

\item[] Carena M, Garcia D, Nierste U and Wagner C E M 2001
%``b $\to$ s gamma and supersymmetry with large tan(beta),''
{\it Phys. Lett. B} {\bf 499} 141 [arXiv:hep-ph/0010003]

\item[] Carena M, Quiros M and Wagner C E M 1996 {\it Nucl. Phys. B}{\bf 461} 407

\item[] Carone C D, Hall L J and Murayama H 1996
%``$(S_3)~3$ flavor symmetry and $p\to K~0 e~+$,''
{\it Phys. Rev. D} {\bf 53} 6282 [arXiv:hep-ph/9512399]

\item[] Carone C D and Lebed R F 1999 {\it Phys. Rev. D} {\bf 60} 096002

\item[] Casas J A, Espinosa J R and Navarro I 2003
 ``Large mixing angles for neutrinos from infrared fixed points,''
 {\it JHEP}{\bf 0309} 048 [hep-ph/0306243]

\item[] Casas J A, Espinosa J R, Quiros M and Riotto A 1995 {\it Nucl. Phys. B}{\bf 436} 3; Erratum 1995 {\it
Nucl. Phys. B}{\bf 439} 466

\item[] Casas J A and Ibarra A 2001 {\it Nucl. Phys. B} {\bf 618} 171 [arXiv:hep-ph/0103065]

\item[] Chamseddine A H, Arnowitt R and Nath P 1982
%``Locally Supersymmetric Grand Unification,''
{\it Phys. Rev. Lett.}  {\bf 49} 970

\item[] Chankowski P H, Ellis J R and Pokorski S 1997 {\it Phys. Lett. B} {\bf 423} 327 [arXiv:hep-ph/9712234]

\item[] Chankowski P H, Ellis J R, Olechowski M and Pokorski S 1999 {\it Nucl. Phys. B} {\bf 544} 39
[arXiv:hep-ph/9808275]

\item[] Chankowski P H and Pokorski S 1997, in {\it Perspectives on supersymmetry}, Kane, G.L. (ed.), p. 402,
World Scientific
%``Supersymmetric loop effects,''
[arXiv:hep-ph/9707497]

\item[] Chankowski P H and Slawianowska L 2001
%``B0/d,s $\to$ mu- mu+ decay in the MSSM,''
{\it Phys. Rev. D}{\bf 63} 054012 [arXiv:hep-ph/0008046]

\item[] Chattopadhyay U, Corsetti A and Nath P 2003
%``WMAP constraints, SUSY dark matter and implications for the direct
%detection of SUSY,''
{\it Phys. Rev. D} {\bf 68} 035005 [arXiv:hep-ph/0303201]

\item[] Chattopadhyay U and Nath P 1996
%``Probing supergravity grand unification in the Brookhaven g-2 experiment,''
{\it Phys. Rev. D} {\bf 53} 1648 [arXiv:hep-ph/9507386]

\item[] Choudhury S R and Gaur N 1999
%``Dileptonic decay of B/s meson in SUSY models with large tan(beta),''
{\it Phys. Lett. B}{\bf 451} 86 [arXiv:hep-ph/9810307]

\item[] Chen M C and Mahanthappa K T 2003 {\it Phys. Rev. D}{\bf 68} 017301

\item[] Chen M C and Mahanthappa K T 2003
%``Fermion masses and mixing and CP-violation in SO(10) models with family  symmetries,''
[arXiv:hep-ph/0305088]

\item[] Claudson M, Wise M and Hall L J 1982 {\it Nucl. Phys. B}{\bf 195} 297

\item[] Chadha S and Daniel M 1983 {\it Nucl. Phys. B}{\bf 229} 105

\item[] Ciuchini M, Degrassi G, Gambino P and Giudice G F 1998
%``Next-to-leading {QCD} corrections to B $\to$ X/s gamma in supersymmetry,''
{\it Nucl. Phys. B} {\bf 534} 3 [arXiv:hep-ph/9806308]

\item[] Contino R, Pilo L, Rattazzi R and Trincherini E 2002
%``Running and matching from 5 to 4 dimensions,''
{\it Nucl. Phys. B} {\bf 622} 227 [arXiv:hep-ph/0108102]

\item[] Davidson S and Ibarra A 2001 {\it JHEP}{\bf 0109} 013 [arXiv:hep-ph/0104076]

\item[] Davier M, Eidelman S, Hocker A and Zhang Z 2003
 %``Updated estimate of the muon magnetic moment using revised results from e+
%e- annihilation,''
{\it Eur. Phys. J. C} {\bf 31} 503 [arXiv:hep-ph/0308213]

\item[] de Boer W, Herold M, Sander C and Zhukov V 2003
%``Indirect evidence for the supersymmetric nature of dark matter from the combined data on galactic positrons, antiprotons and gamma rays,''
[arXiv:hep-ph/0309029]

\item[] de Boer W and Sanders C 2003 [arXiv:hep-ph/0307049]

\item[] de Carlos B and Casas J A 1993 {\it Phys. Lett. B} {\bf 309} 320

\item[] Dedes A, Dreiner H K and Nierste U 2001 [arXiv:hep-ph/0108037]

\item[] Dedes A and Pilaftsis A 2003
%``Resummed effective Lagrangian for Higgs-mediated FCNC interactions in the CP-violating MSSM,''
{\it Phys. Rev. D}{\bf 67} 015012 [arXiv:hep-ph/0209306]

\item[] Degrassi G, Heinemeyer S, Hollik W, Slavich P and Weiglein G 2003
%``Towards high-precision predictions for the MSSM Higgs sector,''
{\it Eur. Phys. J. C} {\bf 28} 133 [arXiv:hep-ph/0212020]

\item[] Demir D, Lebedev O, Olive K A, Pospelov M and Ritz A 2003
%``Electric dipole moments in the MSSM at large tan(beta),''
[arXiv:hep-ph/0311314]

\item[] Derm\' \i \v sek R 2001 Published in *Dubna 2001, Supersymmetry and unification of fundamental
interactions* 284-286 [arXiv:hep-ph/0108249]

\item[] Derm\' \i \v sek R, Mafi A and Raby S 2001 {\it Phys. Rev. D} {\bf 63} 035001 [arXiv:hep-ph/0007213]

\item[] Derm\' \i \v sek R and Raby S 2000
%``Fermion masses and neutrino oscillations in SO(10) SUSY GUT with  D(3) x U(1) family symmetry,''
{\it Phys. Rev. D} {\bf 62} 015007 [arXiv:hep-ph/9911275]

\item[] Derm\' \i \v sek R, Raby S, Roszkowski L and Ruiz De Austri R 2003
%``Dark matter and B/s $\to$ mu+ mu- with minimal SO(10) soft SUSY  breaking,''
{\it JHEP} {\bf 0304} 037 [arXiv:hep-ph/0304101]

\item[] Dimopoulos S 1983 {\it Phys. Lett. B}{\bf 129} 417

\item[] Dimopoulos S, Dine M, Raby S and Thomas S 1996 {\it Phys. Rev. Lett.} {\bf 76} 3494
[arXiv:hep-ph/9601367]

\item[] Dimopoulos S and Georgi H 1981 {\it Nucl. Phys. B} {\bf 193} 150

\item[] Dimopoulos S and Giudice G F 1995
%``Naturalness constraints in supersymmetric theories with nonuniversal soft terms,''
{\it Phys. Lett. B} {\bf 357} 573 [arXiv:hep-ph/9507282]

\item[] Dimopoulos S and Hall L J 1995
%``Electric dipole moments as a test of supersymmetric unification,''
{\it Phys. Lett. B} {\bf 344} 185 [arXiv:hep-ph/9411273]

\item[] Dimopoulos S, Hall L J and Raby S 1992 {\it Phys. Rev. Lett.} {\bf 68} 1984; ibid., {\it Phys. Rev. D}
{\bf 45} 4192

\item[] Dimopoulos S and Pomarol A 1995
%``Nonunified sparticle and particle masses in unified theories,''
{\it Phys. Lett. B} {\bf 353} 222 [arXiv:hep-ph/9502397]

\item[] Dimopoulos S and Raby S 1981 {\it Nucl. Phys. B} {\bf 192} 353

\item[] Dimopoulos S and Raby S 1983 {\it Nucl. Phys. B} {\bf 219} 479

\item[] Dimopoulos S, Raby S, and Wilczek F 1981 {\it Phys. Rev. D} {\bf 24} 1681

\item[] Dimopoulos S, Raby S and Wilczek F 1982 {\it Phys. Lett.} {\bf 112B} 133

\item[] Dimopoulos S, Raby S, and Wilczek F 1991 {\it Phys. Today} {\bf 44N10} 25

\item[] Dimopoulos S and Sutter D W 1995
%``The Supersymmetric flavor problem,''
{\it Nucl. Phys. B} {\bf 452} 496 [arXiv:hep-ph/9504415]

\item[] Dine M, Fischler W and Srednicki M 1981 {\it Nucl. Phys. B} {\bf 189} 575

\item[] Dine M, Fischler W 1982 {\it Phys. Lett. B} {\bf 110} 227

\item[] Dine M, Leigh R and Kagan A 1993 {\it Phys. Rev. D}{\bf 48} 4269

\item[] Dine M and Nelson A E 1993 {\it Phys. Rev. D} {\bf 48} 1277 [arXiv:hep-ph/9303230]

\item[] Dine M, Nelson A E and Shirman Y 1995 {\it Phys. Rev. D} {\bf 51} 1362 [arXiv:hep-ph/9408384]

\item[] Dine M, Nelson A E, Nir Y and Shirman Y 1996 {\it Phys. Rev. D} {\bf 53} 2658 [arXiv:hep-ph/9507378]

\item[] Dine M, Nir Y and Shadmi Y 2002
%``Product groups, discrete symmetries, and grand unification,''
{\it Phys. Rev. D} {\bf 66} 115001 [arXiv:hep-ph/0206268]

\item[] Dixon L J, Harvey J A, Vafa C and Witten E 1985
%``Strings On Orbifolds,''
{\it Nucl. Phys. B}{\bf 261} 678

\item[] Dixon L J, Harvey J A, Vafa C and Witten E 1986
%``Strings On Orbifolds. 2,''
{\it Nucl. Phys. B}{\bf 274} 285 (1986)

\item[] Dreiner H K, Murayama H and Thormeier M 2003
%``Anomalous Flavor U(1)_X for Everything,''
[arXiv:hep-ph/0312012]

\item[] Dudas E, Pokorski S and Savoy C A 1995 {\it Phys. Lett. B}{\bf 356} 45

\item[] Dudas E, Pokorski S and Savoy C A 1996 {\it Phys. Lett. B}{\bf 369} 255

\item[] Einhorn M B and Jones D R 1982 {\it Nucl. Phys. B} {\bf 196} 475

\item[] Elwood J K, Irges N, and Ramond P 1997 {\it Phys. Lett. B}{\bf 413} 322 [arXiv:hep-ph/9705270]

\item[] Elwood J K, Irges N, and Ramond P 1998 {\it Phys. Rev. Lett.}{\bf 81} 5064 [arXiv:hep-ph/9807228]

\item[] Ellis J, Falk T, Ganis G, Olive K A and Srednicki M 2001b {\it Phys. Lett. B}{\bf 510} (236)

\item[] Ellis J R, Hisano J, Raidal M and Shimizu Y  2002
%``A new parametrization of the seesaw mechanism and applications in
%supersymmetric models,''
{\it Phys. Rev. D} {\bf 66} 115013 [arXiv:hep-ph/0206110]

\item[] Ellis J, Kelly S and Nanopoulos D V 1991 {\it Phys. Lett. B} {\bf 260} 131

\item[] Ellis J, Nanopoulos D V and Rudaz S 1982 {\it Nucl. Phys. B} {\bf 202} 43

\item[] Ellis J, Nanopoulos D V and Olive K A 2001a {\it Phys. Lett. B}{\bf 508} (65)

\item[] Ellis J R, Enqvist K, Nanopoulos D V and Zwirner F 1986
 %``Aspects Of The Superunification Of Strong, Electroweak And Gravitational
%Interactions,''
{\it Nucl. Phys. B} {\bf 276} 14

\item[] Ellis J R, Falk T, Olive K A and Santoso Y 2003a
%``Exploration of the MSSM with non-universal Higgs masses,''
Nucl.\ Phys.\ B {\bf 652}, 259 (2003) [arXiv:hep-ph/0210205].

\item[] Ellis J R, Ferstl A, Olive K A and Santoso Y 2003b
 %``Direct detection of dark matter in the MSSM with non-universal Higgs
%masses,''
{\it Phys. Rev. D} {\bf 67} 123502 [arXiv:hep-ph/0302032]

\item[] Ellis J R, Olive K A, Santoso Y and Spanos V C 2003c
%``Supersymmetric dark matter in light of WMAP,''
{\it Phys. Lett. B} {\bf 565} 176 [arXiv:hep-ph/0303043]

\item[] Ellis J R, Olive K A, Santoso Y and Spanos V C 2003d
%``Phenomenological constraints on patterns of supersymmetry breaking,''
Phys.\ Lett.\ B {\bf 573}, 162 (2003) [arXiv:hep-ph/0305212].

\item[] Ellis J R, Olive K A, Santoso Y and Spanos V C 2003e
%``High-energy constraints on the direct detection of MSSM neutralinos,''
[arXiv:hep-ph/0308075]

\item[] Ellis J R, Olive K A, Santoso Y and Spanos V C 2003f
%``Likelihood analysis of the CMSSM parameter space,''
[arXiv:hep-ph/0310356]

\item[] Ellis J R, Ridolfi G and Zwirner F 1991
%``Radiative Corrections To The Masses Of Supersymmetric Higgs Bosons,''
{\it Phys. Lett. B} {\bf 257} 83

\item[] Espinosa J R and Zhang R J (2000a)
%MSSM LIGHTEST CP-EVEN HIGGS BOSON MASS TO O(ALPHA_S ALPHA_T): THE EFFECTIVE POTENTIAL APPROACH
{\it JHEP}{0003} 026

\item[] Espinosa J R and Zhang R J (2000b)
%COMPLETE TWO-LOOP LEADING CORRECTIONS TO THE MASS OF THE LIGHTEST
%CP-EVEN HIGGS BOSON IN THE MINIMAL SUPERSYMMETRIC STANDRAD MODEL
{\it Nucl. Phys. B}{\bf 586} 3

\item[] Eyal G 1998
%``Models of supersymmetric U(2) x U(1) flavor symmetry,''
{\it Phys. Lett. B} {\bf 441} 191 [arXiv:hep-ph/9807308]

\item[] Faraggi A E and Pati J C 1998 {\it Nucl. Phys. B} {\bf 526} 21

\item[] Farrar G and Fayet P 1978 {\it Phys. Lett.} {\bf B76} 575

\item[] Fayet P and Ferrara S 1977
%``Supersymmetry,''
{\it Phys. Rept.}  {\bf 32} 249

\item[] Feng J L and Moroi T 2000 {\it Phys. Rev. D} {\bf 61} 095004 [arXiv:hep-ph/9907319]

\item[] Feng J L, Matchev K T and Moroi T 2000a {\it Phys. Rev. Lett.} {\bf 84} 2322 [arXiv:hep-ph/9908309]

\item[] Feng J L, Matchev K T and Moroi T 2000b {\it Phys. Rev. D} {\bf 61} 075005 [arXiv:hep-ph/9909334]

\item[] Feng J L, Matchev K T and Wilczek F 2000c {\it Phys. Lett. B} {\bf 482} 388 [arXiv:hep-ph/0004043]

\item[] Feng J L, Matchev K T and Wilczek F 2001 {\it Phys. Rev. D}{\bf 63} (045024)

\item[] Feng J L, Matchev K T 2001 {\it Phys. Rev. D} {\bf 63} 095003 [arXiv:hep-ph/0011356]

\item[] Frampton P H, Glashow S L and Yanagida T 2002 {\it Phys. Lett. B}{\bf 548} 119
%``Cosmological sign of neutrino CP violation,''
[arXiv:hep-ph/0208157]

\item[] Frampton P H and Kephart T W 1995a {\it Phys. Rev. D}{\bf 51} 1

\item[] Frampton P H and Kephart T W 1995b {\it Int. J. Mod. Phys. A} {\bf 10} 4689

\item[] Frampton P H and Kong O C W 1996
%``Horizontal symmetry for quark and squark masses in supersymmetric SU(5),''
{\it Phys. Rev. Lett.}  {\bf 77} 1699 [arXiv:hep-ph/9603372]

\item[] Frampton P H and Rasin A 2000
%``Nonabelian discrete symmetries, fermion mass textures and large  neutrino mixing,''
{\it Phys. Lett. B} {\bf 478} 424 [arXiv:hep-ph/9910522]

\item[] Fritzsch H and Minkowski P 1975 {\it Ann. Phys.} {\bf 93} 193

\item[] Froggatt C D and Nielsen H B 1979 {\it Nucl. Phys. B}{\bf 147} 277

\item[] Fukugita M and Yanagida T 1986 {\it Phys. Lett. B} {\bf 174} 45
%``Baryogenesis Without Grand Unification,''

\item[] Gabbiani F, Gabrieli E, Masiero A and Silvestrini L 1996 {\it Nucl. Phys. B}{\bf 477} 321
%``A complete analysis of FCNC and CP constraints in general SUSY extensions of the standard model,''
[arXiv:hep-ph/9604387]

\item[] Gavela M B, King S F, Sachrajda C T, Martinelli G, Paciello M L, and Taglienti B 1989 {\it Nucl. Phys. B}
{\bf 312} 269

\item[] Gell-Mann M, Ramond P and Slansky R 1979 {\it Supergravity}, [Sept. 27-29 1979], eds. D. Freedman et al.,
(North Holland, 1980, Amsterdam)

\item[] Georgi H 1975 Particles and Fields, Proceedings of the APS Div. of Particles and Fields, ed C. Carlson,
p. 575

\item[] Georgi G 1986
%``The Flavor Problem,''
{\it Phys. Lett. B} {\bf 169} 231

\item[] Georgi H and Glashow S L 1974 {\it Phys. Rev. Lett.} {\bf 32} 438

\item[] Georgi H and Jarlskog C 1979 {\it Phys. Lett. B} {\bf86} 297

\item[] Georgi H, Quinn H R and Weinberg S 1974
%``Hierarchy Of Interactions In Unified Gauge Theories,''
{\it Phys. Rev. Lett.}  {\bf 33} 451

\item[] Georgi H and Nanopoulos D V 1979 {\it Nucl. Phys. B} {\bf 159} 16

\item[] Girardello L and Grisaru M T 1982 {\it Nucl. Phys. B} {\bf 194} 65

\item[] Giudice G F, Notari A, Raidal M, Riotto A and Strumia A 2003
%``Towards a complete theory of thermal leptogenesis in the SM and MSSM,''
[arXiv:hep-ph/0310123]

\item[] Giudice G F and Rattazzi R 1999 {\it Phys. Rept.} {\bf 322} 419 [arXiv:hep-ph/9801271]

\item[] Glashow S L 1979 {\it Quarks and Leptons, Carg`ese}, [July 9-29,1979], eds. M. L´evy, et al., (Plenum,
1980, New York), p. 707

\item[] Goh H S, Mohapatra R N and Ng S P 2003
%``Minimal SUSY SO(10), b tau unification and large neutrino mixings,''
{\it Phys. Lett. B} {\bf 570} 215 [arXiv:hep-ph/0303055]

\item[] Gonzalez-Garcia M C and Pena-Garay C 2003
%``Three-neutrino mixing after the first results from K2K and KamLAND,''
{\it Phys. Rev. D} {\bf 68} 093003 [arXiv:hep-ph/0306001]

\item[] Goto T and Nihei T 1999 {\it Phys. Rev. D} {\bf 59} 115009 [arXiv:hep-ph/9808255]

\item[] Grinstein B 1982 {\it Nucl. Phys. B} {\bf 206} 387

\item[] Haber H E and Hempfling R 1993 {\it Phys. Rev. D}{\bf 48} 4280

\item[] Haber H E, Hempfling R and Hoang A H 1997
 %``Approximating the radiatively corrected Higgs mass in the minimal
%supersymmetric model,''
{\it Z. Phys. C} {\bf 75} 539 [arXiv:hep-ph/9609331]

\item[] Hagiwara K, Martin A D, Nomura D and Teubner T 2003
%``Predictions for g-2 of the muon and alpha_QED(M_Z^2),''
[arXiv:hep-ph/0312250]

\item[] Hall L J, Kostelecky V A and Raby S 1986
%``New Flavor Violations In Supergravity Models,''
{\it Nucl. Phys. B }{\bf 267} 415

\item[] Hall L J, Lykken J and Weinberg S 1983
%``Supergravity As The Messenger Of Supersymmetry Breaking,''
{\it Phys. Rev. D} {\bf 27} 2359

\item[] Hall L J and Murayama H 1995 {\it Phys. Rev. Lett.}{\bf 75} 3985

\item[] Hall L J and Nomura Y 2001
%``Gauge unification in higher dimensions,''
{\it Phys. Rev. D} {\bf 64} 055003 [arXiv:hep-ph/0103125]

\item[] Hall L J and Nomura Y 2002a
%``Gauge coupling unification from unified theories in higher dimensions,''
{\it Phys. Rev. D} {\bf 65} 125012 [arXiv:hep-ph/0111068]

\item[] Hall L J and Nomura Y 2002b
%``A complete theory of grand unification in five dimensions,''
{\it Phys. Rev. D}{\bf 66} 075004 [arXiv:hep-ph/0205067]

\item[] Hall L J and Randall L 1990 {\it Phys. Rev. Lett.}{\bf 65} 2939

\item[] Hall L J and Rasin A 1993
%``On the generality of certain predictions for quark mixing,''
{\it Phys. Lett. B} {\bf 315} 164 [arXiv:hep-ph/9303303]

\item[] Hall L J, Rattazzi R and Sarid U 1994 {\it Phys. Rev. D}{\bf 50} 7048

\item[] Hamzaoui C, Pospelov M and Toharia M 1999 {\it Phys. Rev. D}{\bf 59} 095005

\item[] Hara Y, Itoh S, Iwasaki Y and Yoshie T 1986 {\it Phys. Rev. D}{\bf 34} 3399

\item[] Harvey J, Ramond P and Reiss D B 1980 {\it Phys. Lett. B} {\bf 92} 309

\item[] Harvey J, Ramond P and Reiss D B 1982 {\it Nucl. Phys. B} {\bf 199} 223

\item[] Hebecker A and March-Russell J 2002
%``Proton decay signatures of orbifold GUTs,''
{\it Phys. Lett. B}{\bf 539} 119 [arXiv:hep-ph/0204037]

\item[] Hempfling R 1994 {\it Phys. Rev. D}{\bf 49} 6168

\item[] Hisano J, Moroi T, Tobe K, Yamaguchi M and Yanagida T 1995
%``Lepton flavor violation in the supersymmetric standard model with seesaw induced neutrino masses,''
{\it Phys. Lett. B} {\bf 357} 579 [arXiv:hep-ph/9501407]

\item[] Hisano J, Moroi T, Tobe K and Yamaguchi M 1996
%``Lepton-Flavor Violation via Right-Handed Neutrino Yukawa Couplings in Supersymmetric Standard Model,''
{\it Phys. Rev. D} {\bf 53} 2442 [arXiv:hep-ph/9510309]

\item[] Hisano J and Tobe K 2001
%``Neutrino masses, muon g-2, and lepton-flavour violation in the  supersymmetric see-saw model,''
{\it Phys. Lett. B} {\bf 510} 197 [arXiv:hep-ph/0102315]

\item[] Huang C S and Yan Q S 1998
%``B $\to$ X/s tau+ tau- in the flipped SU(5) model,''
{\it Phys. Lett. B}{\bf 442} 209 [arXiv:hep-ph/9803366]

\item[] Huang C S, Liao W and Yan Q S (1999)
%``The promising process to distinguish supersymmetric models with large
%tan(beta) from the standard model: B $\to$ X/s mu+ mu-,''
{\it Phys. Rev. D}{\bf 59} 011701 [arXiv:hep-ph/9803460]

\item[] Huang C S, Liao W, Yan Q S and Zhu S H 2001 {\it Phys. Rev. D}{\bf 63} 114021 [Erratum: Phys. Rev. D64
(2001), 059902]

\item[] Iba\~{n}ez L E 1982
%``Locally Supersymmetric SU(5) Grand Unification,''
{\it Phys. Lett. B} {\bf 118} 73

\item[] Iba\~{n}ez L E and Lopez C 1983
%``N=1 Supergravity, The Breaking Of SU(2) X U(1) And The Top Quark Mass,''
{\it Phys. Lett. B} {\bf 126} 54

\item[] Iba\~{n}ez L E and Lopez C 1984 {\it Nucl. Phys. B} {\bf 233} 511

\item[] Iba\~{n}ez L and Ross G G 1981 {\it Phys. Lett. B} {\bf 105} 439

\item[] Iba\~{n}ez L and Ross G G 1982 {\it Phys. Lett. B} {\bf 110} 215

\item[] Iba\~{n}ez L and Ross G G 1992 ``Perspectives on Higgs physics" Kane G L (ed.) 229-275
[arXiv:hep-ph/9204201]

\item[] Inoue K, Kakuto A, Komatsu H and Takeshita S 1982 {\it Prog. Theor. Phys.} {\bf 67} 1889

\item[] Irges N, Lavignac S and Ramond P 1998
%``Predictions from an anomalous U(1) model of Yukawa hierarchies,''
{\it Phys. Rev. D}{\bf 58} 035003 [arXiv:hep-ph/9802334]

\item[] Isidori G and Retico A 2001 [arXiv:hep-ph/0110121]

\item[] Jeannerot R, Khalil S and Lazarides G 2001
%``Leptogenesis in smooth hybrid inflation,''
{\it Phys. Lett. B }{\bf 506} 344 [arXiv:hep-ph/0103229]

\item[] Jung C K 2002 Brookhaven National Lab preprint, UNO 02-BNL

\item[] Kakizaki M and Yamaguchi M 2002
%``U(1) flavor symmetry and proton decay in supersymmetric standard model,''
{\it JHEP}{\bf 0206} 032 [arXiv:hep-ph/0203192]

\item[] Kane G L, Kolda C F, Roszkowski L and Wells J D 1994
%``Study of constrained minimal supersymmetry,''
{\it Phys. Rev. D} {\bf 49} 6173 [arXiv:hep-ph/9312272]

\item[] Kane G L, Nelson B D, Wang T T and Wang L T (2003)
%``Phenomenology and theory of possible light Higgs bosons,''
[arXiv:hep-ph/0304134]

\item[] Kaplan D B and Schmaltz M 1994 {\it Phys. Rev. D}{\bf 49} 3741
%``Flavor unification and discrete nonAbelian symmetries,''
[arXiv:hep-ph/9311281]

\item[] Kawamura Y 2001a
%``Triplet-doublet splitting, proton stability and extra dimension,''
{\it Prog. Theor. Phys.}{\bf 105} 999 [arXiv:hep-ph/0012125]

\item[] Kawamura Y 2001b
%``Split multiplets, coupling unification and extra dimension,''
{\it Prog. Theor. Phys.}{\bf 105} 691 [arXiv:hep-ph/0012352]

\item[] Kehoe R 2003 http://www-clued0.fnal.gov/~kehoe/physics/fermiwc.pdf

\item[] Kim H D and Raby S 2003
%``Unification in 5D SO(10),''
{\it JHEP} {\bf 0301} 056 [arXiv:hep-ph/0212348]

\item[] Kim H D, Raby S and Schradin L (2004) [arXiv:hep-ph/0401169]

\item[] King S F 1998
%``Atmospheric and solar neutrinos with a heavy singlet,''
{\it Phys. Lett. B} {\bf 439} 350 [arXiv:hep-ph/9806440]

\item[] King S F 2000
%``Large mixing angle MSW and atmospheric neutrinos from single  right-handed neutrino dominance and U(1) family symmetry,''
{\it Nucl. Phys. B }{\bf 576} 85 [arXiv:hep-ph/9912492]

\item[] King S F 2003
%``Neutrino mass models,''
[arXiv:hep-ph/0310204]

\item[] King S F and Ross G G 2003
%``Fermion masses and mixing angles from SU(3) family symmetry and unification,''
{\it Phys. Lett. B}{\bf 574} 239 [arXiv:hep-ph/0307190]

\item[] Kitano R and Mimura Y 2001 {\it Phys. Rev. D}{\bf 63} 016008

\item[] Klebanov I R and Witten E 2003
%``Proton decay in intersecting D-brane models,''
{\it Nucl. Phys. B} {\bf 664} 3 [arXiv:hep-th/0304079]

\item[] Langacker P and Luo M 1991 {\it Phys. Rev. D} {\bf 44} 817

\item[] Lavignac S, Masina I and Savoy C A 2001 {\it Phys. Lett. B}{\bf 520} 269 [arXiv:hep-ph/0106245]

\item[] Lavignac S, Masina I and Savoy C A 2002 {\it Nucl. Phys. B}{\bf 633} 139 [arXiv:hep-ph/0202086]

\item[] LEP2 (2003), Barate R {\it et al.}  [ALEPH Collaboration],
%``Search for the standard model Higgs boson at LEP,''
{\it Phys. Lett. B}{\bf 565} 61 [arXiv:hep-ex/0306033]

\item[] Leontaris G K, Tamvakis K and Vergados J D 1986 {\it Phys. Lett. B}{\bf 171} 412

\item[] Leurer M, Nir Y and Seiberg N 1993 {\it Nucl. Phys. B}{\bf 398} 319

\item[] Leurer M, Nir Y and Seiberg N 1994 {\it Nucl. Phys. B}{\bf 420} 319468

\item[] Lin, C-J S 2003 (CDF Collaboration) Beauty 2003 Conference, Pittsburgh PA

\item[] Lola S and Ross G G 1999
%``Neutrino masses from U(1) symmetries and the Super-Kamiokande data,''
{\it Nucl. Phys. B} {\bf 553} 81 [arXiv:hep-ph/9902283]

\item[] Lucas V and Raby S 1996 {\it Phys. Rev. D} {\bf 54} 2261 [arXiv:hep-ph/9601303]

\item[] Lucas V and Raby S 1997 {\it Phys. Rev. D} {\bf 55} 6986 [arXiv:hep-ph/9610293]

\item[] Lunghi E, Masiero A, Scimemi I and Silvestrini L 2000
%``B $\to$ X/s l+ l- decays in supersymmetry,''
{\it Nucl. Phys. B} {\bf 568} 120 [arXiv:hep-ph/9906286]

\item[] Maekawa N 2001 {\it Prog. Theor. Phys.}{\bf 106} 401

\item[] Maltoni M, Schwetz T, Tortola M A and Valle J W F (2003)
%``Status of three-neutrino oscillations after the SNO-salt data,''
{\it Phys. Rev. D}{\bf 68}, 113010 [arXiv:hep-ph/0309130]

\item[] Marciano W J and Senjanovic G 1982 {\it Phys. Rev. D} {\bf 25} 3092

\item[] Masiero A, Nanopoulos D V, Tamvakis K and Yanagida T 1982 {\it Phys. Lett. B} {\bf 115} 380

\item[] Masiero A, Vempati S K and Vives O 2003
%``Seesaw and lepton flavour violation in SUSY SO(10),''
{\it Nucl. Phys. B} {\bf 649} 189 [arXiv:hep-ph/0209303]

\item[] Matalliotakis D and Nilles H P 1995 {\it Nucl. Phys. B}{\bf 435} 115

\item[] Melnikov K and Vainshtein A 2003
%``Hadronic light-by-light scattering contribution to the muon anomalous
%magnetic moment revisited,''
[arXiv:hep-ph/0312226]

\item[] Misiak M, Pokorski S and Rosiek J 1998
%``Supersymmetry and FCNC effects,''
{\it Adv. Ser. Direct. High Energy Phys.}{\bf 15} 795 [arXiv:hep-ph/9703442]

\item[] Mohapatra R N, Parida M K and Rajasekaran G 2003 [arXiv:hep-ph/0301234]

\item[] Mohapatra R N and Senjanovic G 1980 {\it Phys. Rev. Lett.}{\bf 44} 912

\item[] Moroi T 1996
%``The Muon Anomalous Magnetic Dipole Moment in the Minimal Supersymmetric Standard Model,''
{\it Phys. Rev. D} {\bf 53} 6565 [Erratum-ibid.\ D {\bf 56}, 4424 (1997)] [arXiv:hep-ph/9512396]

\item[] Muon (g - 2) Collaboration, G. W. Bennett et al. 2002 {\it Phys. Rev. Lett.}{\bf 89} 101804,
Erratum-ibid. {\bf 89} (2002) 129903

\item[] Munoz C 2003
 %``Theoretical Predictions for the Direct Detection of Supersymmetric Dark
%Matter,''
[arXiv:hep-ph/0312321]

\item[] Murayama H, Olechowski M and Pokorski S 1996 {\it Phys. Lett. B}{\bf 371} 57

\item[] Murayama H and Pierce A 2002 {\it Phys. Rev. D} {\bf 65} 055009 [arXiv:hep-ph/0108104]

\item[] Nath P and Arnowitt R 1997
%``Non-universal soft SUSY breaking and dark matter,''
{\it Phys. Rev. D} {\bf 56} 2820 [arXiv:hep-ph/9701301]

\item[] Nilles H P, Srednicki M and Wyler D 1983
%``Weak Interaction Breakdown Induced By Supergravity,''
{\it Phys. Lett. B} {\bf 120} 346

\item[] Nir Y and Seiberg N 1993 {\it Phys. Lett. B}{\bf 309} 337

\item[] Nomura Y 2002
%``Strongly coupled grand unification in higher dimensions,''
{\it Phys. Rev. D} {\bf 65} 085036 [arXiv:hep-ph/0108170]

\item[] Nomura Y, Smith D R and Weiner N 2001
%``GUT breaking on the brane,''
{\it Nucl. Phys. B} {\bf 613} 147 [arXiv:hep-ph/0104041]

\item[] Nomura Y and Yanagida T 1999
%``Bi-maximal neutrino mixing in SO(10)(GUT),''
{\it Phys. Rev. D} {\bf 59} 017303 [arXiv:hep-ph/9807325]

\item[] Okada Y, Yamaguchi M and Yanagida T 1991
 %``Upper Bound Of The Lightest Higgs Boson Mass In The Minimal Supersymmetric
%Standard Model,''
{\it Prog. Theor. Phys.}  {\bf 85} 1

\item[] Olechowski M and Pokorski S 1988 {\it Phys. Lett. B}{\bf 214} 393

\item[] Olechowski M and Pokorski S 1995 {\it Phys. Lett. B}{\bf 344} 201

\item[] Ovrut B A and Wess J 1982
%``A New Mechanism For Supersymmetry Breaking,''
{\it Phys. Lett. B} {\bf 112} 347

\item[] Paschos E A 2003
%``Leptogenesis,''
[arXiv:hep-ph/0308261]

\item[] Pati J C and Salam A 1973a {\it Phys. Rev. D} {\bf 8} 1240

\item[] Pati J C and Salam A 1973b
%``Is Baryon Number Conserved?,''
{\it Phys. Rev. Lett.}{\bf 31} 661

\item[] Pati J C and Salam A 1974
%``Lepton Number As The Fourth Color,''
{\it Phys. Rev. D}{\bf 10} 275

\item[] Pierce D \etal 1997 {\it Nucl. Phys. B}{\bf 491} 3

\item[] Pokorski S 1990 {\it Nucl. Phys. B} {\bf 13} (Proc. Supp.) 606

\item[] Polchinski J and Susskind L 1982 {\it Phys. Rev. D} {\bf 26} 3661

\item[] Polonsky N and Pomarol A 1995 {\it Phys. Rev. D}{\bf 51} 6532

\item[] Polony J 1977 Budapest preprint KFKI-1977-93 (unpublished)

\item[] Pomarol A and Dimopoulos S 1995
%``Superfield derivation of the low-energy effective theory of softly broken supersymmetry,''
{\it Nucl. Phys. B} {\bf 453} 83 [arXiv:hep-ph/9505302]

\item[] Pomarol A and Tommasini D 1996 {\it Nucl. Phys. B} {\bf 466} 3 [arXiv:hep-ph/9507462]

\item[] Pouliot P and Seiberg N 1993 {\it Phys. Lett. B}{\bf 318} 169

\item[] Raby S 1995 Published in Trieste HEP Cosmology 1994:0126-180
%``Introduction to theories of fermion masses,''
[arXiv:hep-ph/9501349]

\item[] Raby S 2001 Proceedings of Higgs \& SUSY, March 2001, Orsay, France
http://www.lal.in2p3.fr/actualite/conferences/higgs2001/slides/; Published in *Dubna 2001, Supersymmetry and
unification of fundamental interactions* 168-176 [arXiv:hep-ph/0110203]

\item[] Raby S 2002a in K.~Hagiwara {\it et al.}  [Particle Data Group Collaboration] {\it Phys. Rev. D} {\bf 66}
010001

\item[] Raby S 2002b Published in the proceedings of SUSY'02  [arXiv:hep-ph/0211024]

\item[] Raby S 2003 {\it Phys. Lett. B}{\bf 561} 119

\item[] Raidal M and Strumia A 2003
%``Predictions of the most minimal see-saw model,''
{\it Phys. Lett. B} {\bf 553} 72 [arXiv:hep-ph/0210021]

\item[] Rattazzi R and Sarid U 1996 {\it Phys. Rev. D}{\bf 53} 1553

\item[] Roberts R G, Romanino A, Ross G G and Velasco-Sevilla L 2001
%``Precision test of a fermion mass texture,''
{\it Nucl. Phys. B} {\bf 615} 358 [arXiv:hep-ph/0104088]

\item[] Romanino A and Strumia A 2000 {\it Phys. Lett. B} {\bf 487} 165 [arXiv:hep-ph/9912301]

\item[] Ross G G and Velasco-Sevilla L 2003 {\it Nucl. Phys. B}{\bf 653} 3

\item[] Roszkowski L, Ruiz de Austri, R and Nihei T 2001
%``New cosmological and experimental constraints on the CMSSM,''
{\it JHEP }{\bf 0108} 024 [arXiv:hep-ph/0106334]

\item[] Sakai N 1981 {\it Z. Phys.} {\bf C11} 153

\item[] Sakai N and Yanagida T 1982 {\it Nucl. Phys.} {\bf B197}, 533

\item[] Shafi Q and Ananthanarayan B 1991 {\it ICTP Summer School lectures}

\item[] Shafi Q and Tavartkiladze Z 2000 {\it Phys. Lett. B} {\bf 487} 145

\item[] Strumia A 1999 hep-ph/9904247

\item[] Tobe K and Wells J D 2003 {\it Nucl. Phys. B} {\bf 663} 123 [arXiv:hep-ph/0301015]

\item[] Weinberg S 1982 {\it Phys. Rev.} {\bf D26} 287

\item[] Weisskopf V 1939 {\it Phys. Rev.} {\bf 56} 72

\item[] Witten E 1981 {\it Nucl. Phys. B} {\bf 188} 513

\item[] Witten E 2002 Published in *Hamburg 2002, Supersymmetry and unification of fundamental interactions, vol.
1 472-491
%``Deconstruction, G(2) holonomy, and doublet-triplet splitting,''
[arXiv:hep-ph/0201018]

\item[] Yanagida T 1979 {\it Proceedings of the Workshop on the unified theory and the baryon number of the
universe}, [Feb. 13-14, 1979], eds. O. Sawada and A. Sugamoto (Tsukuba,1979) p.95

\item[] Zhang R J (1999)
%TWO-LOOP EFFECTIVE POTENTIAL CALCULATION OF THE LIGHTEST CP-EVEN HIGGS-BOSON MASS IN THE MSSM
{\it Phys. Lett. B}{\bf 447} 89

\end{harvard}

\end{document}